\documentclass[11pt]{report}

\usepackage{upennstyle} 
\usepackage{amsmath,amssymb,amsfonts}
\usepackage{graphicx}
\usepackage{subfiles}
\usepackage{stix}
\usepackage{algorithm}
\usepackage{algpseudocode} 
\usepackage{gensymb}
\usepackage{ragged2e}
\usepackage[nonamebreak,round]{natbib}
\usepackage{color,colortbl}
\usepackage{enumitem}

\usepackage{acronym}
\acrodef{AUC}{area under the curve}
\acrodef{BB}{Bachmann's Bundle}
\acrodef{CT}{Computed Tomography}
\acrodef{CV}{Conduction Velocity}
\acrodef{CA}{catheter ablation}
\acrodef{DT-MRI}{Diffusion Tensor Magnetic Resonance Imaging}
\acrodef{DE-MRI}{Delayed-Enhanced Magnetic Resonance Imaging}
\acrodef{ECG}{electrocardiogram}
\acrodef{EGM}{electrogram}
\acrodef{EAM}{electroanatomic mapping}
\acrodef{FIRM}{Focal Impulse and Rotor Modulation}
\acrodef{FPR}{false positive rate}
\acrodef{GPR}{Gaussian process regression}
\acrodef{ICP}{Iterative Closest Point}
\acrodef{LAT}{Local Activation Time}
\acrodef{LA}{Left Atrium}
\acrodef{LSPV}{Left Superior Pulmonary Vein}
\acrodef{LIPV}{Left Inferior Pulmonary Vein}
\acrodef{LAA}{Left Atrial Appendage}
\acrodef{LGE-MRI}{Late Gadolinium Enhancement Magnetic Resonance Imaging}
\acrodef{MV}{Mitral Valve}
\acrodef{MRI}{magnetic resonance imaging}
\acrodef{NSR}{normal sinus rhythm}
\acrodef{PV}{pulmonary vein}
\acrodef{PS}{Phase Singularity}
\acrodef{PVI}{Pulmonary Vein Isolation}
\acrodef{RSPV}{Right Superior Pulmonary Vein}
\acrodef{RIPV}{Right Inferior Pulmonary Vein}
\acrodef{RIGHT}{Rhythm ID Goes Head-to-head Trial}
\acrodef{ROI}{region of interest}

\graphicspath{{figure/}}
\title{CyberCardia: Patient-specific electrophysiological heart model for assisting left atrium arrhythmia ablation}
\author{Jiyue He}
\gradgroup{Electrical and Systems Engineering} 
\date{2023} 
\supervisor{Rahul Mangharam}
\supervisortitle{Professor of Electrical and Systems Engineering and Computer and Information Science Departments}
\gradchair{Troy Olsson, Professor of Electrical and Systems Engineering Department}
\committee{René Vidal, Director of the Innovation in Data Engineering and Science Initiative, Rachleff and Penn Integrates Knowledge University Professor of Electrical and Systems Engineering, Penn Medicine Radiology, and Computer and Information Science Departments}
\committee{Daniel Hashimoto, Director of Penn Computer Assisted Surgery and Outcomes Laboratory, Professor of Surgery at the Hospital of the University of Pennsylvania}
\committee{Arkady Pertsov, Professor of Pharmacology Department of Upstate Medical University}

\authorlegal{Jiyue He}

\acknowledgement{
Special thanks to my parents for their unconditional support.

Sincere thanks to my advisor Professor Rahul Mangharam for giving me an extraordinary opportunity to pursue a Doctor of Philosophy, and providing me precious guidance in both research and life.

Many thanks to my collaborators (names are in alphabetical order):
\vspace{-10pt}
\begin{itemize}[leftmargin=0pt,itemsep=-10pt]
\item[] Arkady Pertsov, Professor, Upstate Medical University. 
\item[] Caroline Roney, Lecturer, Queen Mary University of London, United Kingdom. 
\item[] Elizabeth Cherry, Professor, Georgia Institute of Technology. 
\item[] Eric Toolan, Technician, Biosense Webster Inc. 
\item[] Erik Pernod, Director, InfinyTech3D, France.
\item[] Flavio Fenton, Professor, Georgia Institute of Technology. 
\item[] Hugo Talbot, PhD, Simulation Open Framework Architecture Consortium, France.
\item[] Jackson Liang, MD, Professor, University of Michigan. 
\item[] John Bullinga, MD, EP Director, Penn Presbyterian Medical Center. 
\item[] Katie Walsh, MD, Hospital of the University of Pennsylvania. 
\item[] Kuk Jin Jang, Postdoctoral Researcher, University of Pennsylvania. 
\item[] Maxime Sermesant, Head of Computational Cardiology at Inria, Université Côte d'Azur, France.
\item[] Sanjay Dixit, MD, EP Director, Professor, Hospital of the University of Pennsylvania. 
\item[] Steven Niederer, Professor, King’s College London, United Kingdom. 
\item[] Zirui Zang, PhD candidate, University of Pennsylvania.
\end{itemize}
}

\abstract{
Atrial arrhythmia can be categorized into tachycardia, flutter, and fibrillation. Atrial fibrillation is a prevalent heart disease that results in weak and irregular contractions of the atria. It affects millions people worldwide and contributes to hundreds of thousands deaths annually \cite{AF2022,AF2021}. Cardiac ablation is among the most successful treatment options, involving the use of radio frequency energy to kill diseased cells or create lesion lines that obstruct abnormal activation waves. During the procedure, catheters are inserted into the left atrium to map the atrium geometry and record endocardium electrograms that are then converted into electroanatomical maps to pinpoint the arrhythmia source locations. 

However, identifying these sources is challenging. The electrograms are asynchronous and can be susceptible to noise. The spatial distribution of sampling sites is non-uniform, which leads to inaccurate maps. Identifying arrhythmia source locations is not a trivial task. Therefore, an ablation procedure often lasts from 3 to 6 hours, and arrhythmia recurrence within 12 months after first ablation is around 50\% \cite{Ganesan2013,Sultan2017}.

To address these challenges, we developed an integrated computational heart mode for clinical left atrium arrhythmia ablation. Our system takes in the left atrium geometry and electrograms, processes them to extract regional tissue properties, which are used to tune a heart model, creating a patient-specific whole-atrium model. With this model, we can simulate and detect arrhythmia sources, and provide ablation assistance. To build such a system, we investigated the fiber effects on atrial activation patterns. We developed a fast heart model tuning method which takes only a few seconds of computation time on a personal computer, enabling real-time assistance during the ablation procedure. We achieved high accuracy in simulating arrhythmias, which we validated on patient data.
} 

\begin{document}
\maketitle 
\setcounter{page}{2}

\makecopyright
\makeacknowledgement
\makeabstract

\tableofcontents

\clearpage \phantomsection \addcontentsline{toc}{chapter}{LIST OF TABLES} \begin{singlespacing} \listoftables \end{singlespacing}

\clearpage \phantomsection \addcontentsline{toc}{chapter}{LIST OF ILLUSTRATIONS} \begin{singlespacing} \listoffigures \end{singlespacing}

\begin{mainf}


\chapter{\MakeUppercase{Introduction}}
\label{chapter: introduction}

\section{Overview}
In this dissertation, we develop an integrated computational heart model for clinical left atrium arrhythmia ablation. Our approach is illustrated in Figure \ref{fig:our_approach}. During arrhythmia ablation surgery, the physician maneuvers catheters to acquire mapping data. This mapping data includes left atrium geometry and endocardium electrogram. The mapping data is processed into electroanatomical maps which inform local tissue properties. An iterative optimization process is run to tune heart model parameters according to these clinical data. Then this tuned heart model simulates patient-specific arrhythmia activations. The simulations are analyzed and converted into electroanatomical maps and arrhythmia source locations are detected, which will guide ablation. These 5 steps create a loop, that the physician continuously acquires new mapping data, and the heart model gets constant updates. More and more ablations are applied until atrial fibrillation is non-inducible.

\begin{figure}[!ht]
\centering
\includegraphics[width = 1\textwidth]{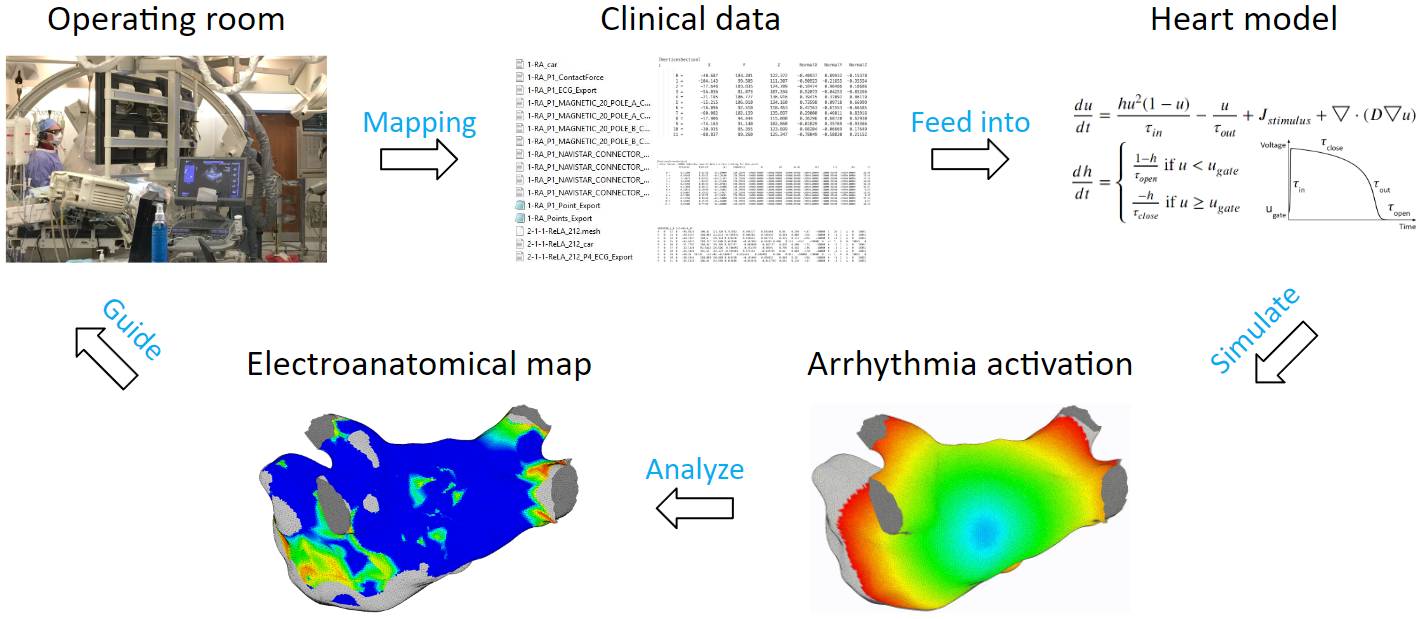}
\caption{Our approach of utilizing a heart model for assisting arrhythmia ablation. Our heart model is integrated into the clinical mapping system, where the patient data are processed and feed into the heart model. Then the heart model can simulate patient-specific arrhythmias. Processing the simulation will give electroanatomical maps, and arrhythmia source detection will provide potential ablation target locations. These information will help physician make ablation decisions.}
\label{fig:our_approach}
\end{figure}

To achieve such a system integration, we need to overcome several challenges: 1) The electrograms are asynchronous and can be susceptible to noise, and The spatial distribution of sampling sites is non-uniform. 2) Electrogram recordings can be a challenge to process. 3) Identifying arrhythmia source locations is not a trivial task. Challenge 1 can be solved with a heart model, which can provide synchronous, uniform, and noise free whole-atrium simulation. For challenge 2, we develop tools, user interfaces, and algorithms to process clinical and simulation data. For challenge 3, we develop various ablation guidance maps that help locate arrhythmia sources.

The key is to have the heart model tuned to patient-specific. However, the heart model requires myocardial fiber data, which is not available during the ablation procedure. Therefore, we start with investigating the fiber effects on atrium activation patterns (Chapter 2). The investigation conclusions encourage us to develop a patient-specific heart model that does not require fiber data (Chapter 3 and 4). Then, in Chapter 5 and 6, we explain the tools, user interfaces, and algorithms we developed for processing clinical data and generating ablation guidance maps. Lastly, we validated our heart model with patient data (Chapter 7).

\newpage

\section{What is left atrium arrhythmia}
Atrial arrhythmia is a heart disease that affects millions of Americans. During arrhythmia, the heart cannot sufficiently plumb blood. Causing hundreds of thousands of deaths each year \cite{AF2022,AF2021}. The symptoms includes fast heartbeat, chest pain, dizziness, shortness of breath and others.

While the underlying mechanism of atrial fibrillation is not clearly understood, it largely is due to the spatial variation in the conduction properties of the atrial myocardium. This causes the single wavefront propagating across the atria to being split into multiple wavefronts resulting in chaotic depolarization of the heart tissue and irregularly fast rhythm. Atrial fibrillation increases the risk of stroke and heart failure. If left untreated, it will become worse \cite{Wijffels}.

The main treatments are: medicines, surgery, and lifestyle changes. One of the most effective treatments for atrial fibrillation is catheter ablation. It is a minimally invasive surgery. This involves irreversible radiofrequency heating of the sources that initiate or perpetuate atrial fibrillation. The common current ablation protocol involves standard lesion locations for all patients. For example, in the \Ac{PVI} approach, ablation lesions encircle the pulmonary veins to prevent abnormal activations originated in the veins travel into the atria. 

Another common protocol involves \Ac{PVI} with additional ablation of non-pulmonary vein triggers and putative arrhythmia substrate \cite{Frankel}. While paroxysmal atrial fibrillation (i.e. spontaneous onset and termination) can be eliminated in 70-75\% of patients with a single ablation procedure, persistent atrial fibrillation can be eliminated in only about 50\% of patients with a single procedure \cite{AblationStats}. The reason is there are triggers other than pulmonary veins that cause fibrillation for the persistent atrial fibrillation. Electroanatomical mapping captures the heart geometry and tissue conductivity which are essential in identifying those trigger \cite{Santangeli}. 

\newpage

\section{Catheter ablation surgery}
Figure \ref{fig:operating_room} shows the left atrium catheter ablation operating room, where the electroanatomical maps are acquired. There is the patient table behind the glass, and multiple computer screens displaying the mapping system’s information. Which includes spatial data and temporal data. We also have X-ray and echocardiogram alongside the mapping system.

\begin{figure}[!ht]
\centering
\includegraphics[width = 1\textwidth]{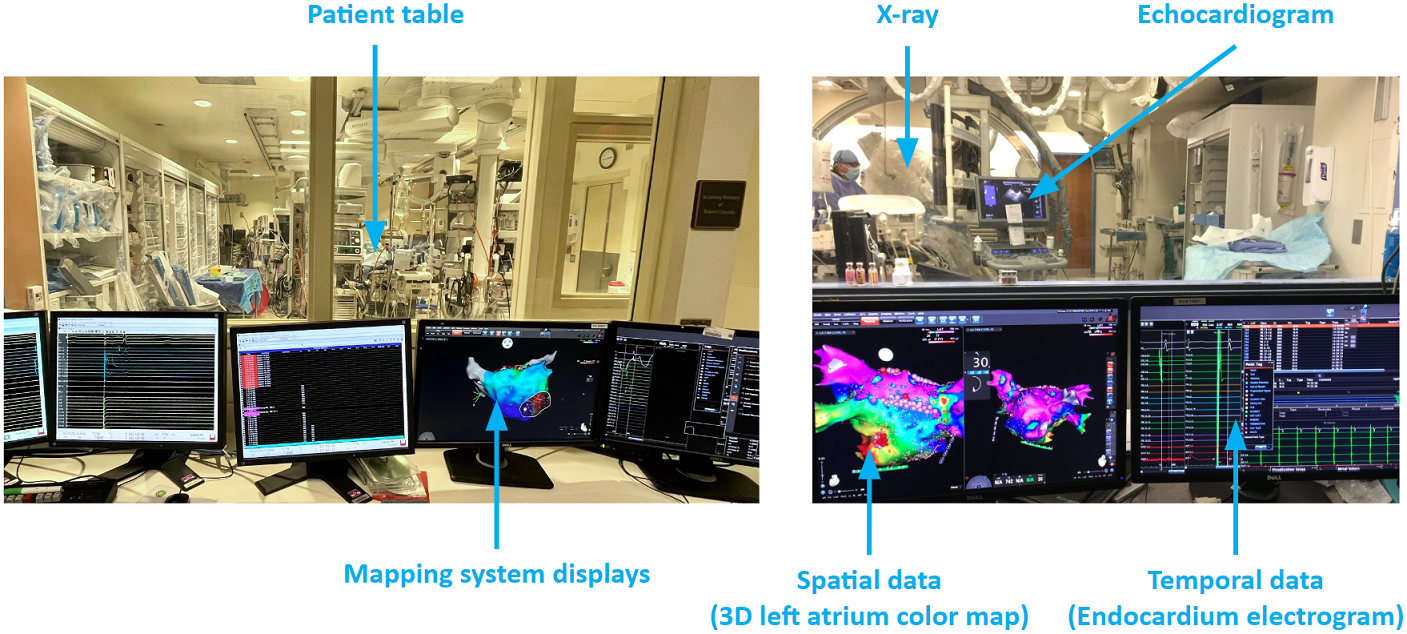}
\caption{Cardiac catheter ablation operating room.}
\label{fig:operating_room}
\end{figure}

Catheters are passed through a vein to reach the heart. They may be inserted in the groin, shoulder or neck. A 3D mapping system takes in electrograms from the catheters and creates colored maps to help the physician pinpoint the abnormal tissue, such map is called an electroanatomical map, as shown in Figure \ref{fig:ablation}. There, ablations created a line of block to the abnormal activations from the pulmonary veins, and a cluster of ablations stop abnormal activations from a focal source.

\begin{figure}[!ht]
\centering
\includegraphics[width = 1\textwidth]{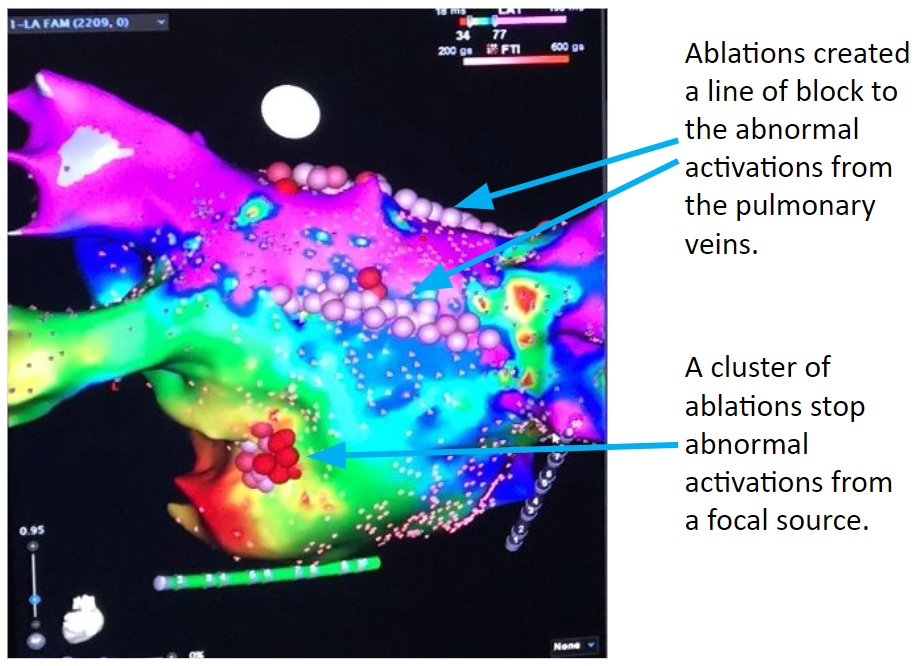}
\caption{Example of an electroanatomical map and some ablation lesions.}
\label{fig:ablation}
\end{figure}

\newpage

\section{Atrial arrhythmia ablation challenges}
A typical ablation procedure runs for 3 to 6 hours. As shown in Figure \ref{fig:ablation_procedure}. The first 1 to 2 hours is the preparation step: The patient is getting sedation anesthesia, physician or nurses connect the hardware wiring, technicians set up the mapping software, and the physician inserts catheters into the patient's left atrium. Then an initial map is acquired. From this initial map, the system can generate a voltage map, which displays scar tissue locations. Based on scar tissue locations, standard ablations such as pulmonary vein isolation is done first, this takes 1 to 2 hours. If that does not terminate arrhythmia, then additional ablations will be needed. This step will be iterations of mapping and ablation, which can take a few hours if the arrhythmia is complex. After all the arrhythmia sources are ablated, the procedure ends. Catheters are removed, and nurses clean up the operating room. This step takes about an hour. We developed an integrated system that provides assistance during the mapping and ablation steps.

\begin{figure}[!ht]
\centering
\includegraphics[width = 1\textwidth]{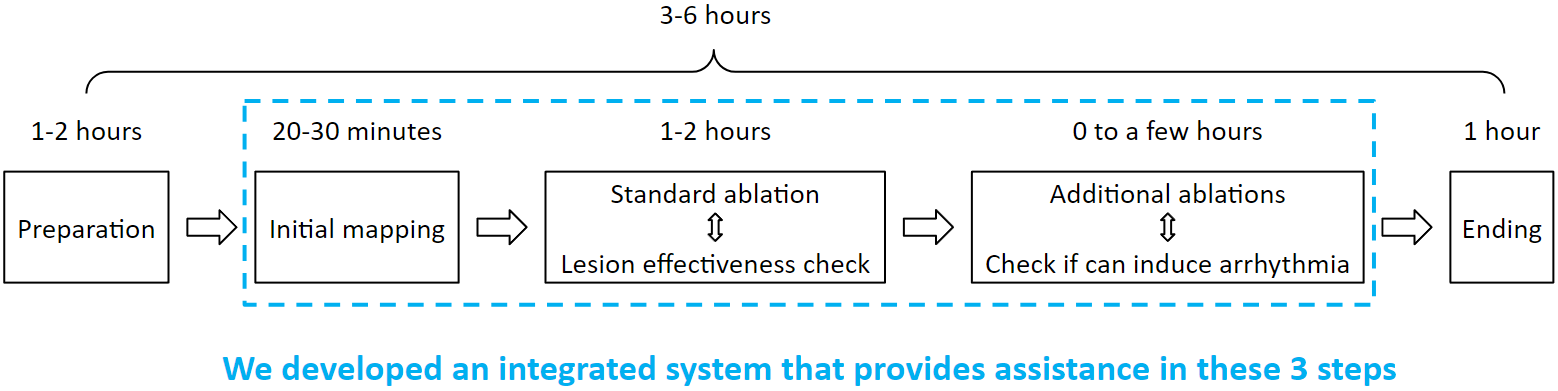}
\caption{The steps breakdown of a typical ablation procedure.}
\label{fig:ablation_procedure}
\end{figure}

Atrial arrhythmia ablation is challenging. Arrhythmia recurrence within 12 months is about 45\% \cite{Sultan2017}. There are mainly 3 challenges: 

\subsection{Challenge I: Data sampling is temporally asynchronous and spatially non-uniform}
A map usually consists of 1000-2500 electrogram samples. Each electrogram recording segment is 2.5 seconds long, and it is aligned with a reference timing. The reference timing is the ventricle beat from a surface electrode or a coronary sinus catheter electrode chosen by the physician. Figure \ref{fig:electrogram} shows examples of unipolar and bipolar electrogram. They are collected by maneuvering the catheter to different locations. Figure \ref{fig:recording_segment} shows the process of collecting electrograms around the endocardium. 

\begin{figure}[!ht]
\centerline{\includegraphics[width = 1\textwidth]{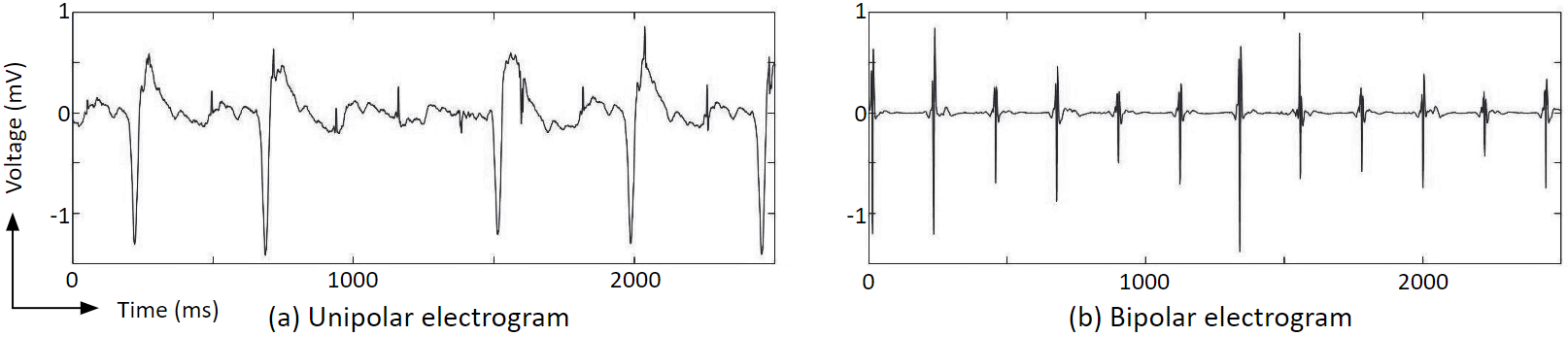}}
\caption{Examples of electrogram. (a) Unipolar electrogram. (b) Bipolar electrogram.}
\label{fig:electrogram}
\end{figure}

\begin{figure}[!ht]
\centerline{\includegraphics[width = 1\textwidth]{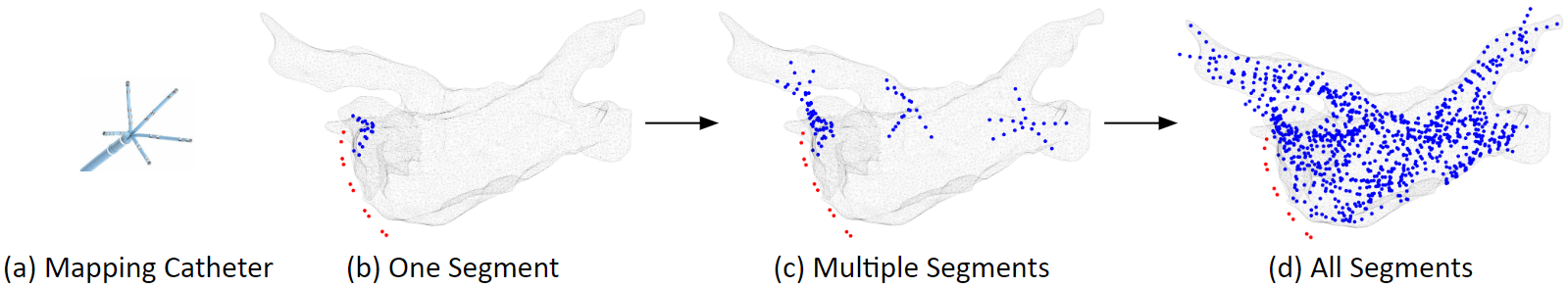}}
\caption{Electrogram sampling. In this example, the mapping catheter is the Pentaray catheter, which is a star-shape catheter that has 20 electrodes. The physician maneuvers the catheter to different locations of the endocardium tissue to record electrograms. The blue dots are the electrode locations of the Pentaray catheter, and the red dots are the electrode locations of the Decapolar catheter which is located at the coronary sinus. (Note: Some electrodes are located outside of the mesh, that is because the atrium is having small movements during the operation but the mesh is not updating in real time, rather, the mesh is fixed after its creation.)}
\label{fig:recording_segment}
\end{figure}

The temporally asynchronous problem is: Different regions may not represent the same rhythm. Sampling a map is done region by region, however, the pattern of the rhythm can change during the 10-20 minutes of mapping. The spatially non-uniform problem is: The sampling density is not uniform. Maneuvering and keeping the catheter stable at a location is technically challenging, as a result, some regions overlap while others disjoint. 

\subsection{Challenge II: The electrogram recordings can be a challenge to process}
\label{sec:electrogram_processing_challenge}
First, there can be bad electrodes, or poor electrode-tissue contact issues. Figure \ref{fig:catheter_twisted_spline}(a) shows an example of good catheter placement. (b), (c) and (d) show examples that the catheter splines are twisted, causing the electrodes collide with each other, making bad electrograms.  

\begin{figure}[!ht]
\centerline{\includegraphics[width = 1\textwidth]{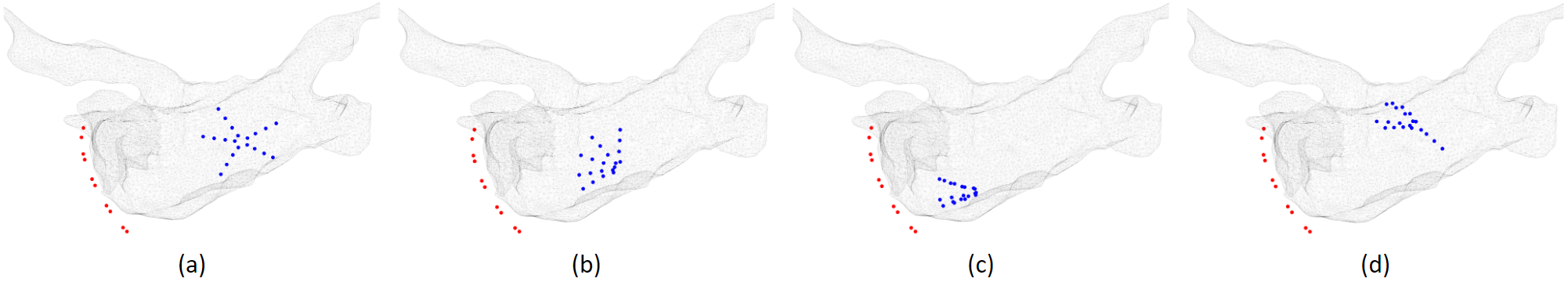}}
\caption{Catheter splines can be twisted, causing the electrodes collide with each other.}
\label{fig:catheter_twisted_spline}
\end{figure}

Next, there can be a lot of catheter movements. Figure \ref{fig:catheter_spatial_stability} shows examples of catheter movements. (a) shows an example of spatially stable catheter position. (b) shows an example of spatially unstable catheter position. The problem with unstable catheter position is that the electrogram recordings will no longer have any meaning, because they will not be from the same location.

\begin{figure}[!ht]
\centerline{\includegraphics[width = 1\textwidth]{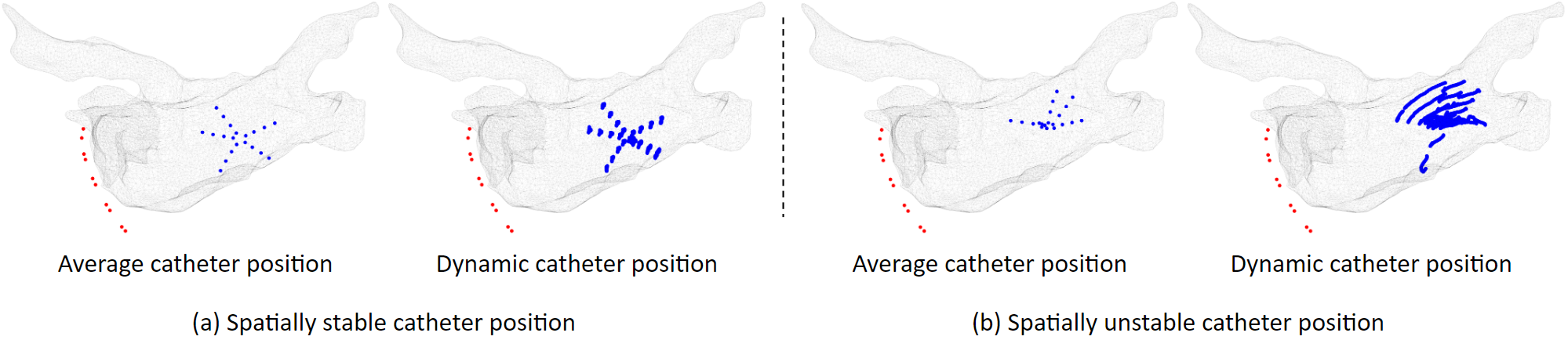}}
\caption{Catheter movements during electrogram recording. If the movement is too large, the recorded electrograms will no longer have meanings.}
\label{fig:catheter_spatial_stability}
\end{figure}

Next, ventricle activation signals can be large in the atrium electrograms causing difficulty in detecting atrial activations. Figure \ref{fig:far_field_interference} shows such an interference: The ventricle activations, or QRS waves in (b) are getting into the atrium electrograms in (a). We need to remove these interference to better detect atrial activations.

\begin{figure}[!ht]
\centering
\includegraphics[width=1\textwidth]{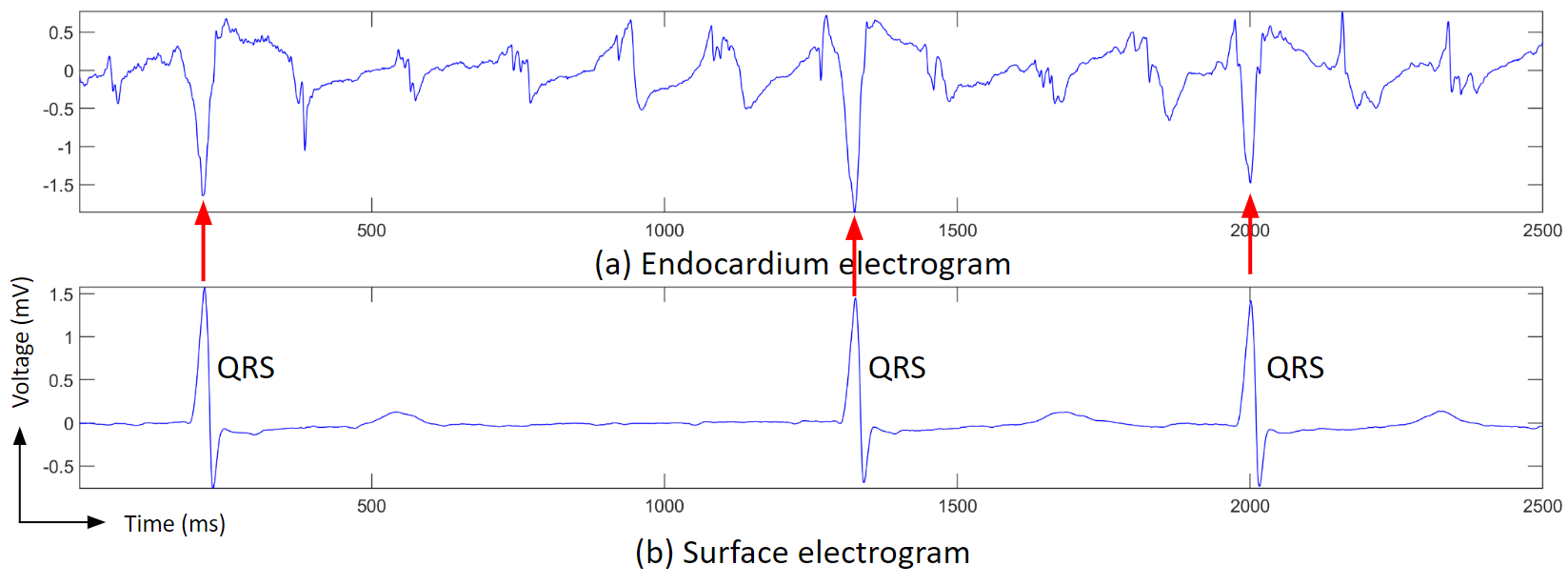}	
\caption{Far-field interference. The strong ventricle activation, or QRS, often get recorded in the atrium endocardium electrograms which will obscure atrium activation detection. (a) An example of endocardium electrogram that has ventricle activation interference marked with the red arrows. (b) An example of a surface electrogram. Here the electrogram is from surface lead "V5".}
\label{fig:far_field_interference}
\end{figure}

Next, for some electrograms, although they have clean and strong signal, they can be fractionated. These complex and fractionated electrograms may be a result of the electrode located at scar regions or located at a place that multiple activation waves collide with each other, so that the electrogram picks up multiple waves tightly together as shown in Figure \ref{fig:fractionated_egm}(a): the red portion consist of multiple activations with low amplitudes, it is hard to determine which activation time will be a better representation of the local activation time. 

\begin{figure}[!ht]
\centerline{\includegraphics[width = 1\textwidth]{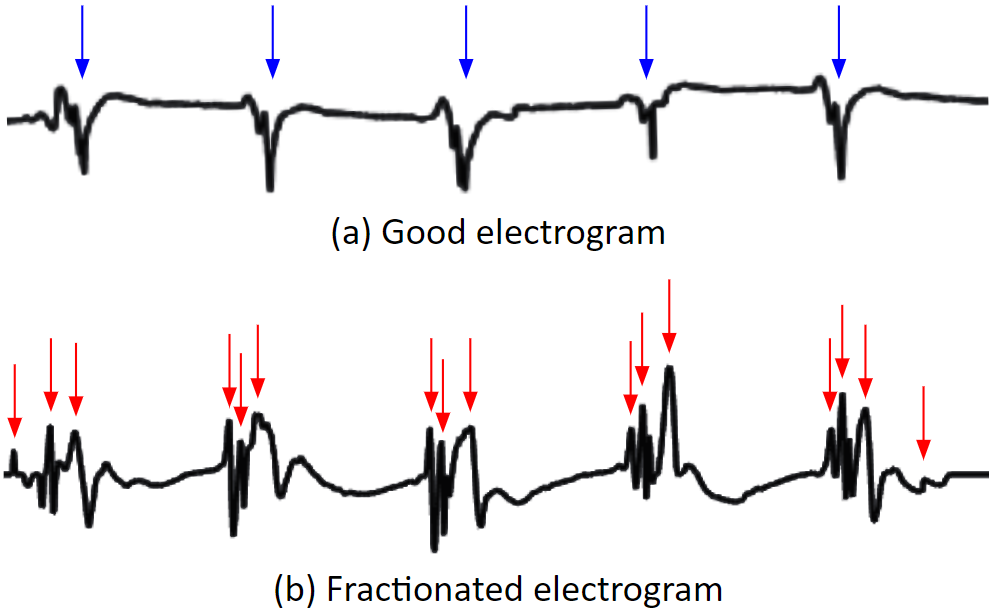}}
\caption{(a) Good electrogram. The activation time is easy to identify. (b) Electrogram is fractionated, making it difficult to find the local activation time. }
\label{fig:fractionated_egm}
\end{figure}

Next, bipolar electrogram has wave front direction dependency issue as shown in Figure \ref{fig:unipolar_bipolar}. When the bipolar orientation is parallel to the activation wave direction, it can record a strong signal. But when the bipolar orientation is perpendicular to the activation wave direction, it will not record any signal. This is a problem, because one of the most important arrhythmia sources, scar, is determined by voltage maps. Low voltage signifies scar tissue. But what if the low voltage is a result of bipolar orientation, then healthy tissue will be misclassified as scar.

\begin{figure}[!ht]
\centering
\includegraphics[width = 1\textwidth]{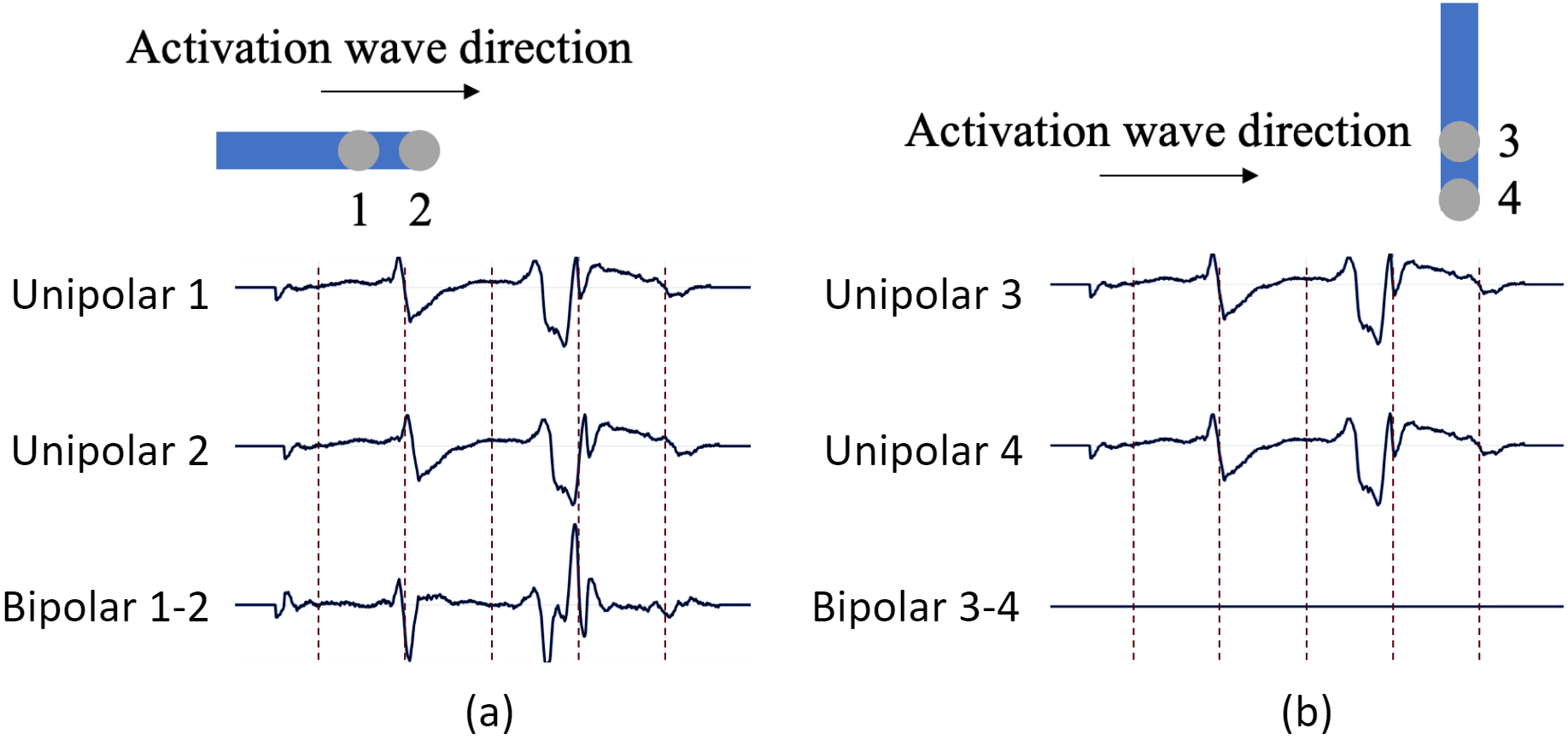}
\caption{Bipolar electrogram magnitude depends on activation wave direction. (a) Activation wave direction is along the bipolar orientation, which lead to the unipolar 2 electrogram is a slight delayed version of unipolar 1 electrogram, resulting a strong bipolar electrogram. (a) Activation wave direction is perpendicular to bipolar orientation, which lead to unipolar 3 and 4 electrograms the same, resulting a zero bipolar electrogram.}
\label{fig:unipolar_bipolar}
\end{figure}

Last but not least, interpolating data from electrode sample sites onto the left atrium mesh can also introduce noise. As shown in Figure \ref{fig:interpolation_limitation}, electrode samples are acquired at the locations of the blue dots. Then we need to interpolate them to the atrium mesh. Interpolation can introduce data outliers. The figure shows a portion of a voltage map. Magenta represents healthy tissue, yellow and green represents scar. We can see that in the yellow circle, there are small regions of scars which does not make too much sense, these can be interpolation errors.

\begin{figure}[!ht]
\centering
\includegraphics[width=0.5\textwidth]{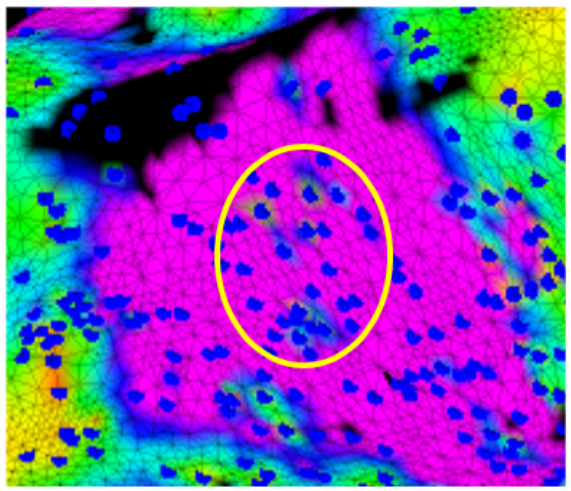}	
\caption{Clinical measurements from the blue dot locations need to be interpolate to the atrium mesh. Interpolation can introduce data outliers. The greenish spots within the yellow circle can be interpolation errors.}
\label{fig:interpolation_limitation}
\end{figure}

\subsection{Challenge III: Identify arrhythmia sources is difficult}
Identifying arrhythmia sources is difficult. The biggest question is where to ablate? As shown in Figure \ref{fig:difficulty_of_source_detection}, voltage maps can help us find out scar tissue distribution and re-entry circuit locations. But that depends on applying a correct voltage threshold. Activation maps can help us find out focal source locations. But that depends on choosing a correct time window of interest. Phase singularity tracking can help us find out rotor source locations. But tracking the rotor center is not a trivial task. Oftentimes, multiple arrhythmia sources appear simultaneously, perpetuating atrial fibrillation.

\begin{figure}[!ht]
\centering
\includegraphics[width=1\textwidth]{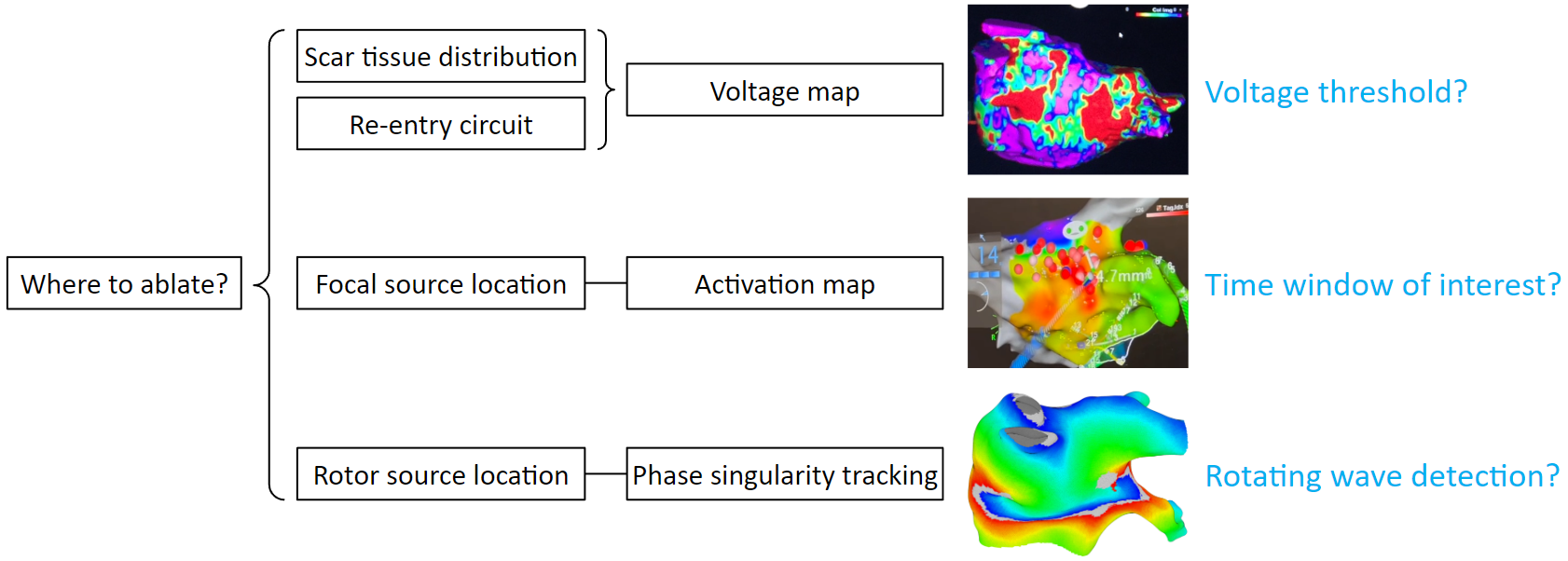}	
\caption{Difficulties of identifying arrhythmia source locations.}
\label{fig:difficulty_of_source_detection}
\end{figure}

\newpage

\section{Related work}
These many challenges inspired research in building a better ablation guidance system. Numerous research had shown applications of utilizing a heart model for assisting arrhythmia ablation. Among these research, we would like to mention 2 state of the arts implementations.

The first one is based on MRI scans. MRI scans provide patient-specific atrium geometry and scar tissue locations. Then dozens of simulations are run and analyzed to identify potential arrhythmia source locations. Simulations and virtual ablations are done iteratively until atrial fibrillation is non-inducible. Then export the ablation plan to guide physicians to do the real ablations \cite{Boyle2019}. 

It only requires \ac{MRI} data, and does not use electrograms. The heart model implements one set of parameters for healthy tissue, another set of parameters for scar. All steps can be completed within 2.5–5 days, and is done before the ablation procedure.

The other one is a simulation model reflective of individual anatomy, fiber orientation, fibrosis, and electrophysiology \cite{Lim2020}. It utilizes clinical voltage maps, activation maps, and computed tomography images. The myocardial fiber orientation for the heart model is from atlases. The model implements one set of parameters for healthy tissue, another set of parameters for scar. The time cost of loading and analyzing data is 48 minutes.

\newpage

\section{Our approach}
As illustrated in Figure \ref{fig:research_overview}, there are several clinical mapping systems built by different companies. They take measurements from the patient, interpolate the measurements onto the atrium mesh as colored maps to guide ablation. We integrated a heart model to the clinical mapping system. By doing so, we can reduce measurement noise, display more analysis, and provide clearer ablation guidance. We are the first to integrate a heart model to a clinical mapping system without adding new equipment or new procedures.

\begin{figure}[!ht]
\centering
\includegraphics[width = 1\textwidth]{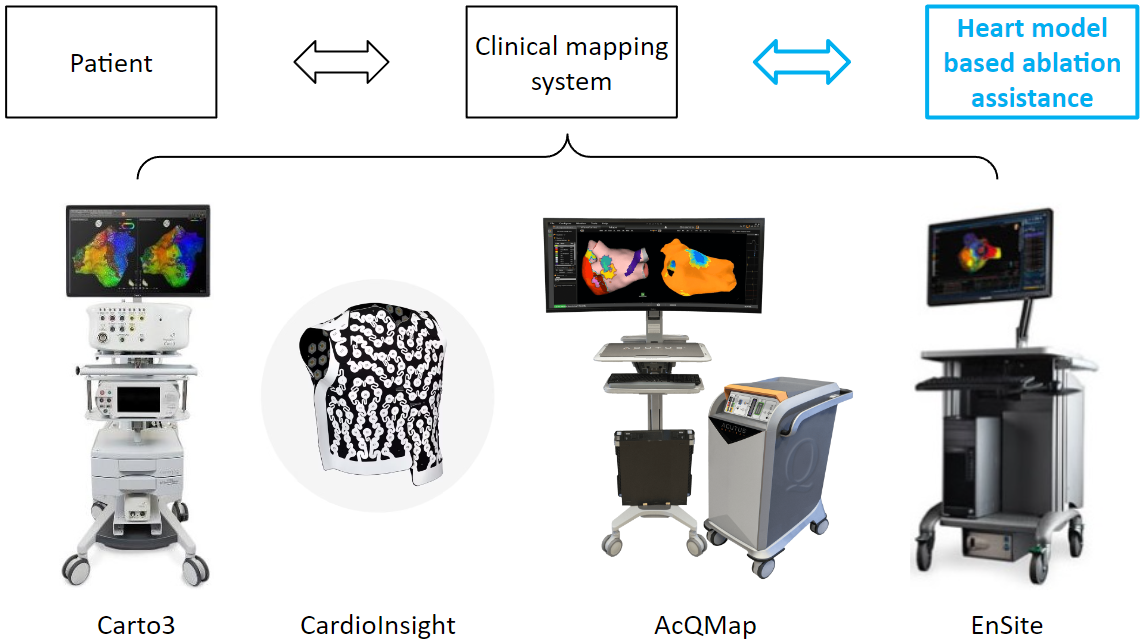}
\caption{There are several companies build clinical mapping systems, which take measurements from the patient and interpolate them to the atrium mesh to guide ablation. We integrated a heart model to the clinical mapping system without additional equipment or procedures. (Figures are from the companies website.)}
\label{fig:research_overview}
\end{figure}

Why did we integrate a heart model to the clinical system? Because having a heart model can provide better ablation guidance. As shown in Figure \ref{fig:why_a_heart_model}, the current ablation system can provide voltage maps, which help identify scar tissues. It can provide activation maps, which help locate focal sources. It can provide coherent maps, which help identify ablation block leakage. On the other hand, our integrated system can provide all of the above maps and additional custom maps. It can also provide arrhythmia simulations, and locate arrhythmia sources.

\begin{figure}[!ht]
\centering
\includegraphics[width = 1\textwidth]{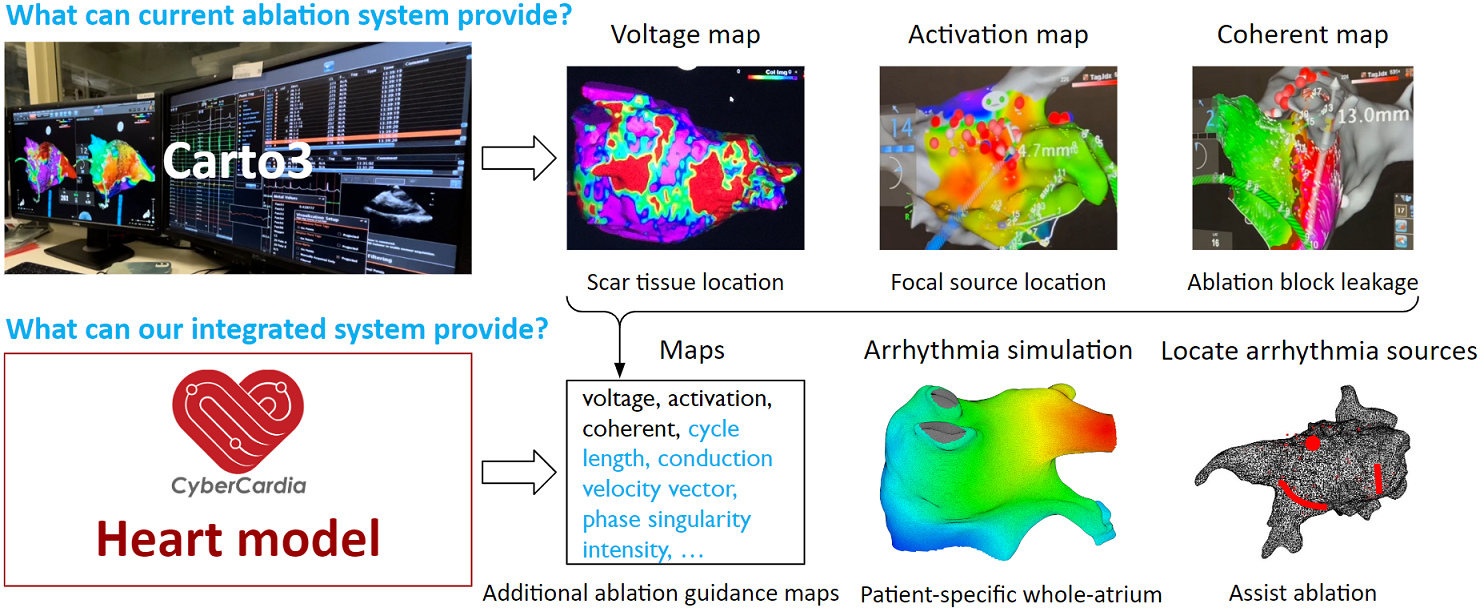}
\caption{Comparison between the current ablation system and our integrated system.}
\label{fig:why_a_heart_model}
\end{figure}

What makes our approach different? Our system is highly integrated into the current mapping system in the operating room. We only utilize patient data that is available during the ablation procedure. Our patient-specific tuning is more detailed: Parameters are tuned locally with continuous-values, rather than one value for healthy tissue and one value for scar. Our heart model can run in real-time: Tuning time is 15 seconds, simulation time for 1 heartbeat is 5 seconds on a personal computer, rather than hours or days on a supercomputer.

\newpage

\section{Our contributions}
With the many challenges to solve, we need to find the correct questions to ask. We asked ourselves 3 scientific questions. Question 1: What are the weaknesses of the current mapping system? We think the major ones are: 1) Electrogram recordings are temporally asynchronous and spatially non-uniform. 2) Limited available maps to help find arrhythmia source locations. 3) Patient data (atrium geometry and electrogram) contain noise.

Question 2: What can the mapping system benefit from a heart model? We believe: 1) A heart model can provide Patient-specific whole-atrium simulation which is synchronous and uniform, and noise free. 2) A heart model can produce ablation guidance maps for arrhythmia source detection. 3) A heart model can run ablation simulation before irreversibly create the lesions.

Question 3: How to integrate a heart model to the mapping system? We need to: 1) Process and clean up patient data. 2) Develop patient-specific model parameter tuning method. 3) Implement GPU parallel computing to enable real-time simulation.

Solving these questions, we have made several contributions. The core contributions are: 1) We investigated the fiber effects on activation patterns. 2) We developed a fast heart model tuning method. 3) We achieved high accuracy of simulating arrhythmias. In addition, we built several tools, user interfaces, and algorithms for processing clinical data, to remove atrium geometry defects and electrogram noise. We also developed new types of maps for assisting ablation. Lastly, we validated our system with patient data. These contributions turned into papers published in top journals and international conferences, details please refer to Appendix \ref{app:publication}.

\newpage

\chapter{\MakeUppercase{The effects of fiber organization on activation patterns}}
\label{chapter:fiber_effect}

\section{Overview}
To integrate a heart model to the ablation system, the heart model needs to be patient-specific. Therefore, we process clinical data and feed them into the heart model, so that we can tune the heart model’s parameters to make it patient-specific. As shown in Figure \ref{fig:fiber_not_available}, we have clinical data that can be transformed into the components of the heart model. But there is one heart model’s components that is not available: the myocardial fiber orientations. 

\begin{figure}[ht]
\centering
\includegraphics[width = 1\textwidth]{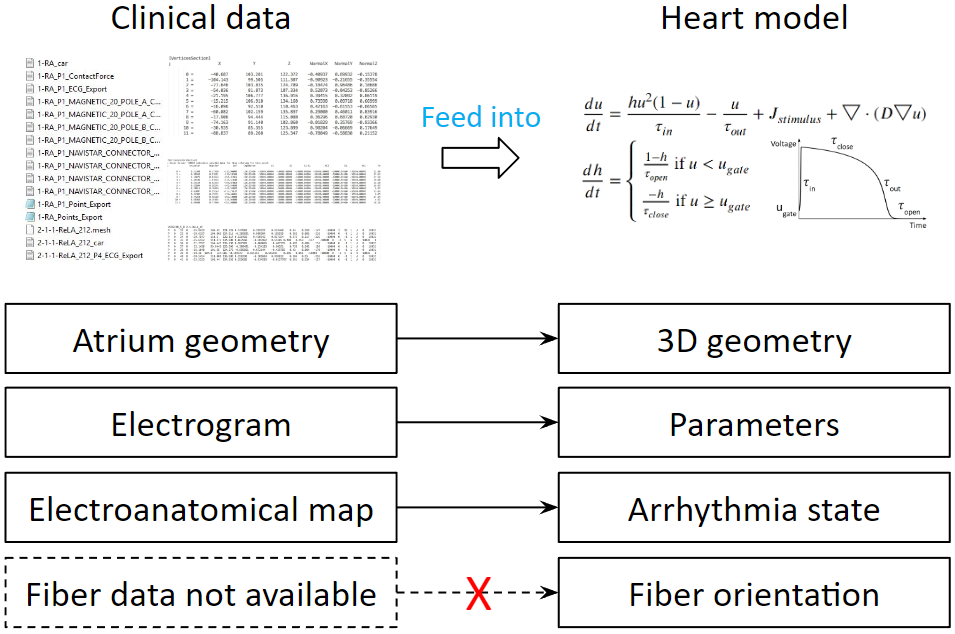}
\caption{Fiber data is not available during the ablation procedure.}
\label{fig:fiber_not_available}
\end{figure}

We do not want to introduce new equipment or new procedures to the current ablation system. Therefore, we would like to find a way to compensate for the lack of fiber data. To do that, we need to first learn the effects of fiber on activation patterns.

\newpage

\section{Slab tissue experiments of fiber effects}
We ran experiments on a slab tissue with different fiber organizations. 
We varied the fiber angle in between endocardium and epicardium ($\Delta\theta$), we also varied the activation wave direction ($\alpha$), creating a total of 50 scenarios: 10 $\Delta\theta$ scenarios, and each has 5 $\alpha$ scenarios as shown in Figure \ref{fig:angle_difference_vs_cv}.

\begin{figure}[!ht]
\centering
\includegraphics[width = 1\textwidth]{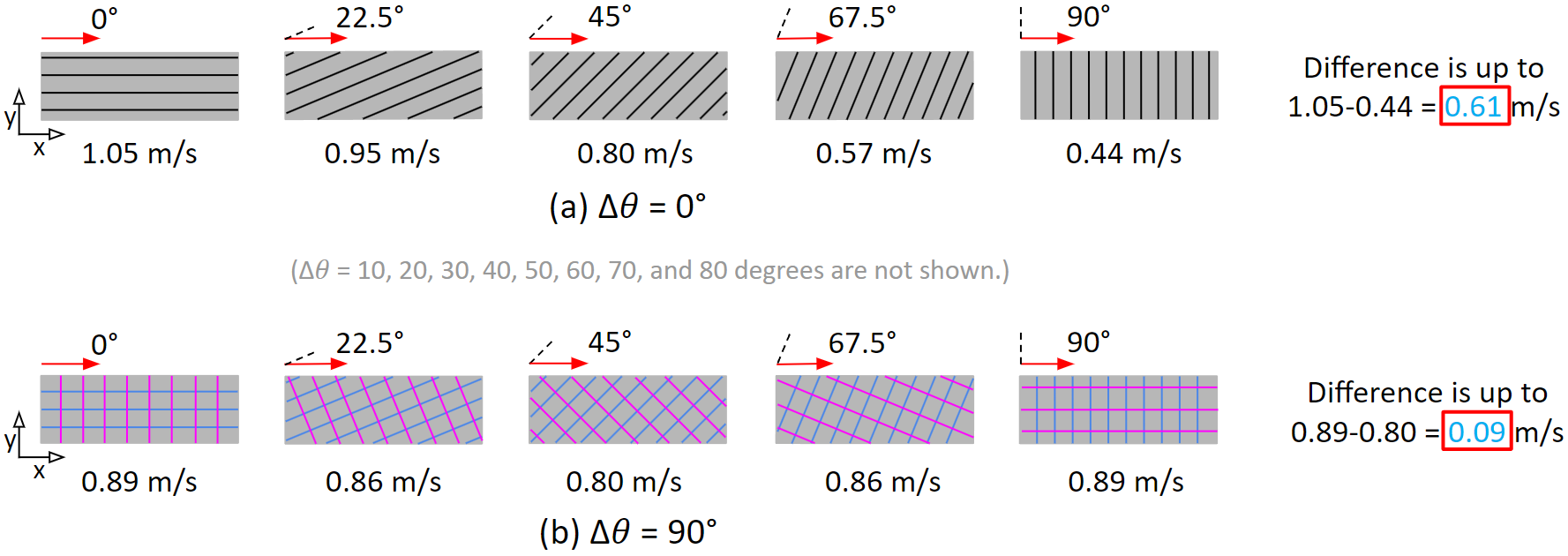}
\caption{Slab tissue experiments of 50 scenarios.}
\label{fig:angle_difference_vs_cv}
\end{figure}

As illustrated in Figure \ref{fig:angle_difference_vs_cv}(a), the endocardium and epicardium have the same fiber orientations, depending on the activation wave direction, conduction velocity can be a lot different: it varies from 1.05 m/s to 0.44 m/s, which is a difference of 0.61 m/s. As illustrated in Figure \ref{fig:angle_difference_vs_cv}(b), the endocardium and epicardium’s fiber are perpendicular to each other, the conduction velocity does not change much regardless of the activation wave direction: it varies from 0.89 m/s to 0.80 m/s, which is a difference of 0.09 m/s. We concluded that fiber has a stronger effect on conduction velocity when 
endocardium and epicardium fiber orientations are the same.

We than ran another set of experiments varying the tissue thickness. In these experiments, $\Delta\theta$ is set to 90 degrees, but the slab thickness has 5 different scenarios: 1.89 mm, 3.16 mm, 4.00 mm, 5.26 mm, 6.11 mm. And for each thickness scenario, there are 5 different $\alpha$ scenarios. We find that tissue thickness can also effect conduction velocity.

We ran yet another set of experiments varying the thickness ratio between endocardium epicardium. In these experiments, $\Delta\theta$ is set to 90 degrees. The total thickness is 4 mm, but the endocardium/epicardium thickness ratio has 5 scenarios: 10:10, 12:8, 14:6, 16:4, 18:2. Figures \ref{fig:thickness_ratio_vs_cv} are snap shots of activation waves at different times. We can see that if the layer thickness ratio is 10:10, there is no difference at the wavefront locations. But if the layer thickness ratio is 18:2, then the wavefront locations will be different.

\begin{figure}[!ht]
\centering
\includegraphics[width = 1\textwidth]{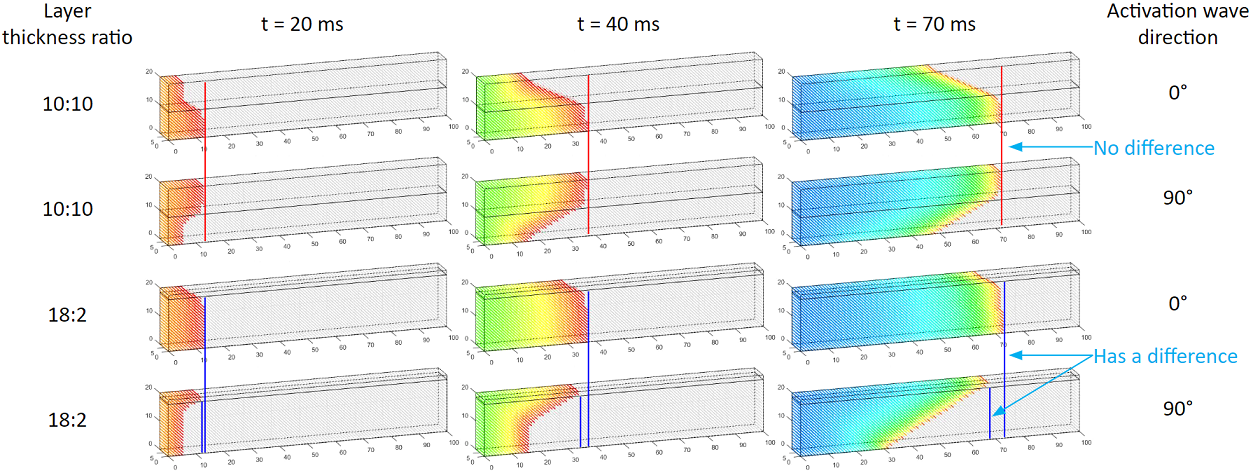}
\caption{Slab tissue experiments of varying the thickness ratio between endocardium and epicardium.}
\label{fig:thickness_ratio_vs_cv}
\end{figure}

A summary of the results of the above 3 sets of experiments are shown in Figure \ref{fig:slab_experiments_result}. It shows that conduction velocity varies more with: 1) Smaller $\Delta\theta$. This can be explained by the uniformity of fiber orientations between the endocardium and epicardium. If all fibers are in the same orientation, the effect of faster speed in longitudinal direction will be strong; if one layer of fiber in one orientation, the other layer of fiber in perpendicular orientation, then such effect will be reduced. 2) Thicker tissue. This can be explained by the influence between the 2 fiber layers: the thicker the tissue, the less influence from one layer to the other. 3) Larger difference between endocardium and epicardium thickness. A large difference in layer thickness will result in a dominant layer, which will have an effect similar to reducing $\Delta\theta$. This effect can be seen visually in Figure \ref{fig:thickness_ratio_vs_cv}.

\begin{figure}[!ht]
\centering
\includegraphics[width = 1\textwidth]{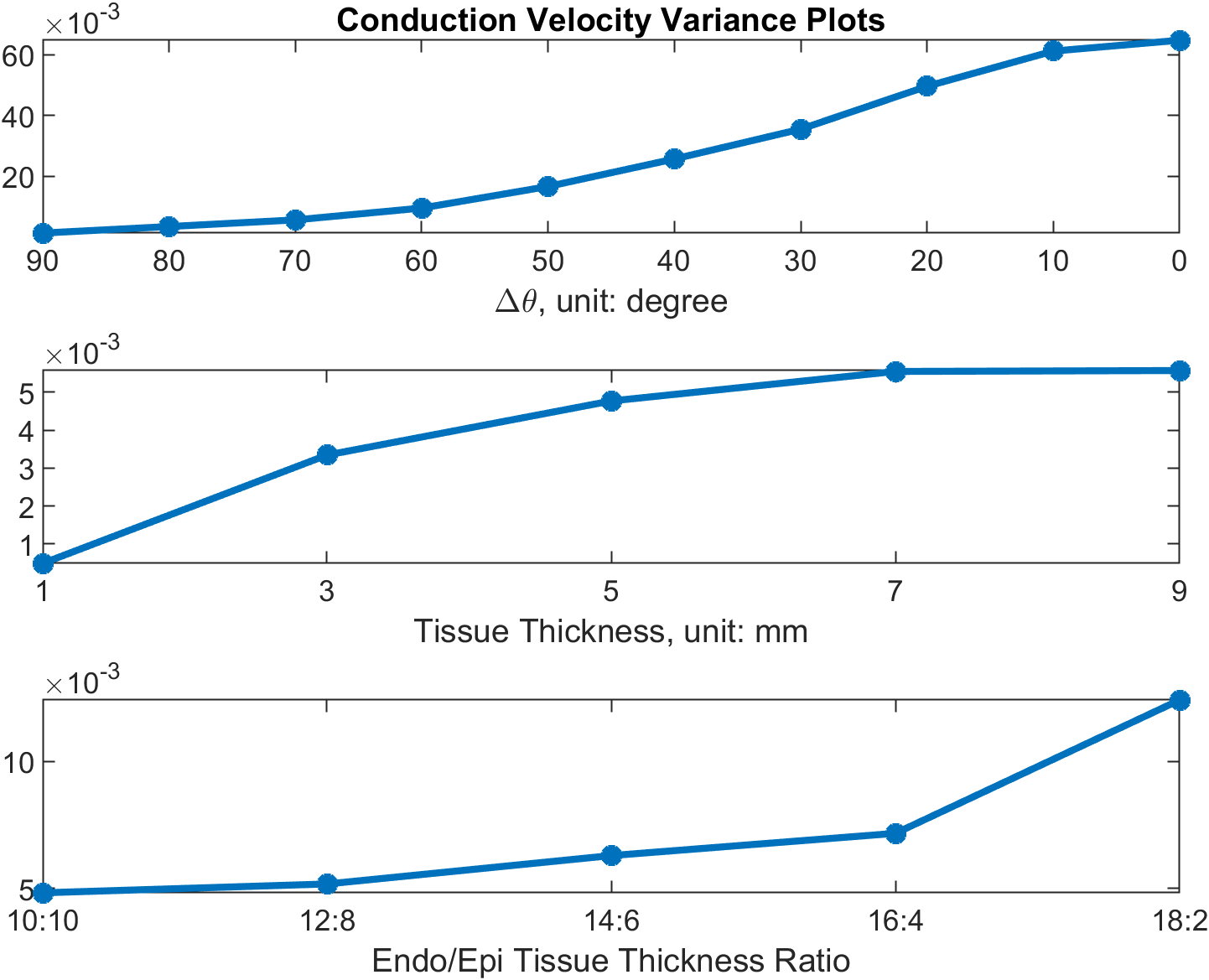}
\caption{Result summary of the effects of fiber organizations to conduction velocity.}
\label{fig:slab_experiments_result}
\end{figure}

\newpage

\section{Atrial fiber data analysis}
The observations of the slab tissue experiments show that fiber orientations could have a large effect on activation patterns. However, the left atrium is different from a slab tissue. Fiber organizations in the left atrium are much more complex. We want to know how different are the left atria fiber organizations. Is there a consistent fiber direction pattern based on left atrium anatomy location? Do different left atria have similar fiber direction pattern? To answer these questions, we analyze and compare 7 ex-vivo left atria fiber organizations. The fiber organizations of the atria is shown in Figure \ref{fig:fiber_data} (more examples of fibers are shown in Figure \ref{fig:fiber_data_7_atria} in the appendix). We do not observe consistent fiber organizations within a atrium.

\begin{figure}[!ht]
\centering
\includegraphics[width = 1\textwidth]{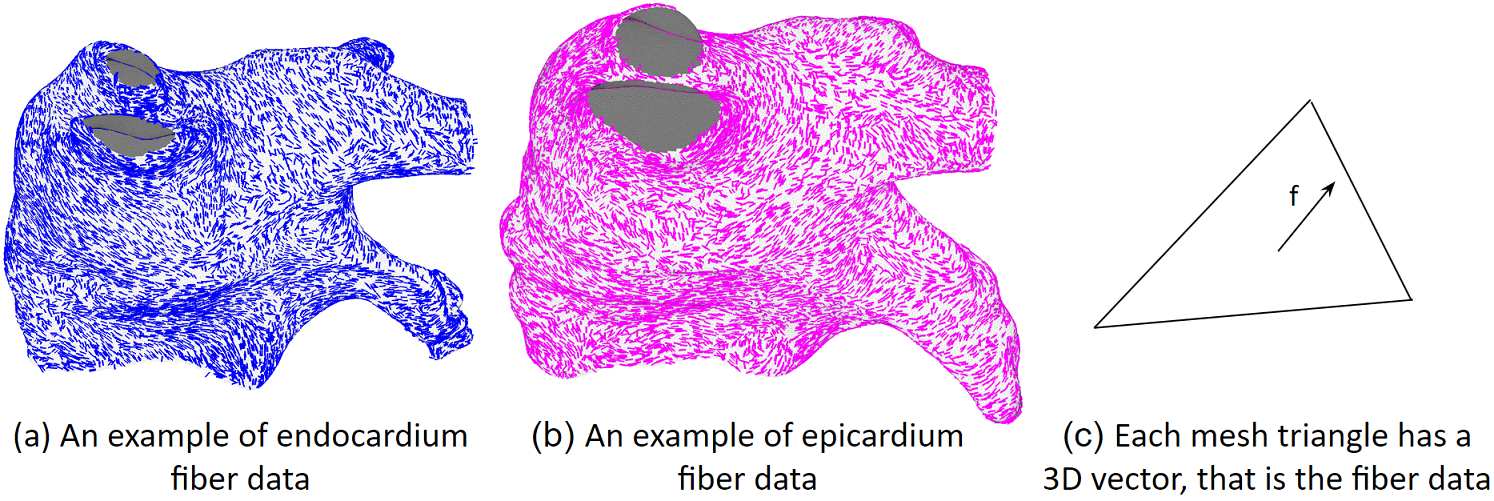}
\caption{An example of the fiber data of a left atrium.}
\label{fig:fiber_data}
\end{figure}

The angle difference between endocardial and epicardial fibers ($\Delta\theta$) indicates the degree of anisotropy of conductivity. We plot the $\Delta\theta$ maps in Figure \ref{fig:fiber_angle_difference}, where red represents small and blue represents large $\Delta\theta$. We observe that no large regions have the same $\Delta\theta$ value, and there is no regularity in the spatial distribution of the small and large $\Delta\theta$ regions in all atria tested. This also means that in all atria the fiber pattern of the endocardium is very different from that of the epicardium. 

\begin{figure}[!ht]
\centering
\includegraphics[width = 1\textwidth]{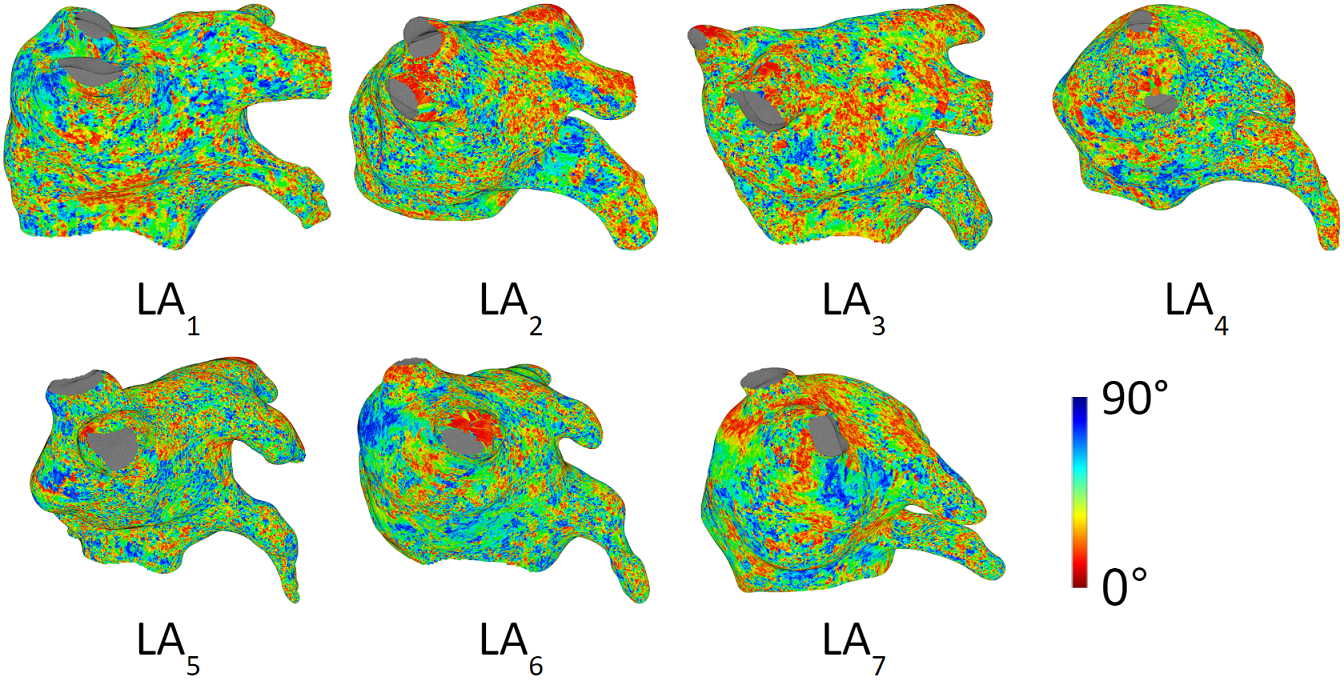}
\caption{Endocardium-epicardium fiber angle difference ($\Delta\theta$) maps. Red and blue colors represent small and large $\Delta\theta$, respectively. There is no consistent pattern for the red blue locations within and among different atria.}
\label{fig:fiber_angle_difference}
\end{figure}

To quantify our observations, we register the 7 fiber data onto the same geometry (LA$_1$) as shown in Figure \ref{fig:register_fibers_to_same_mesh}. (Details of fiber registration please refer to Appendix \ref{app:fiber_registration}.) Then we compute the correlations between different fiber data, and we also analyze the fiber angle differences between the two fiber layers. 

\begin{figure}[!ht]
\centering
\includegraphics[width = 1\textwidth]{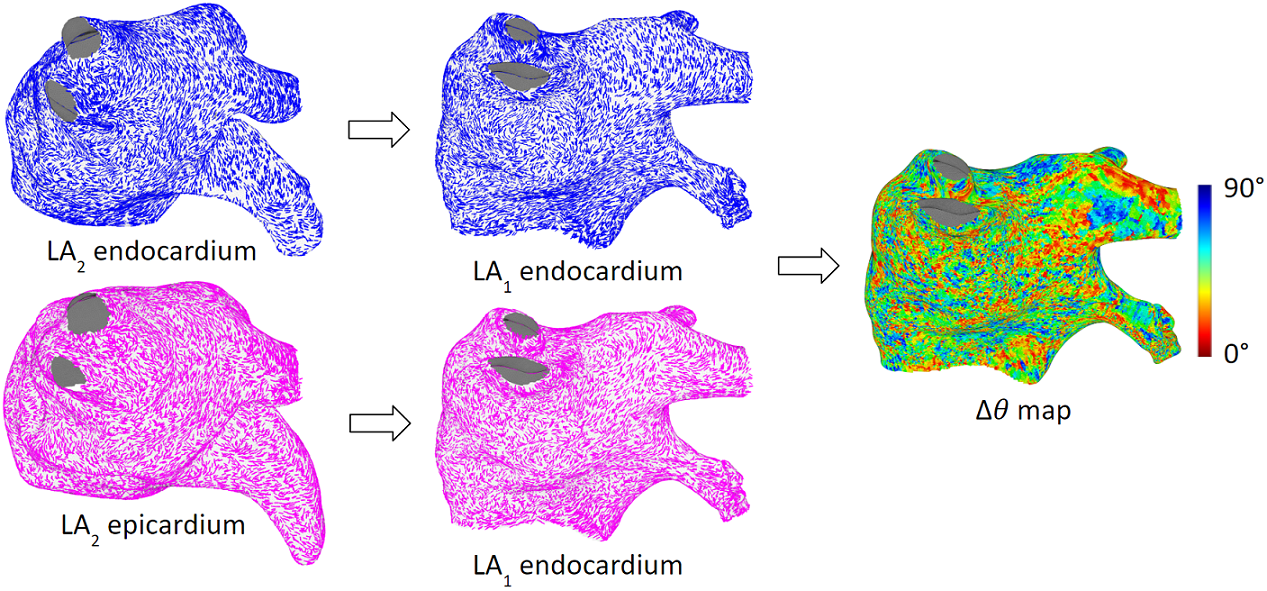}
\caption{Register fibers example: From LA$_2$ endocardium and epicardium onto LA$_1$ endocardium.}
\label{fig:register_fibers_to_same_mesh}
\end{figure}

Fibers are 3D vectors as shown in Figure \ref{fig:compare fiber}(a). It is inconvenient to directly compare 3D vectors, therefore, we make a reference frame transformation, so that the 3D fibers ($f_1$, $f_2$, and $f_3$) are represented as 1D values ($\theta_1$, $\theta_2$, and $\theta_3$) as shown in (b). The reference frame is a vector created by a fixed edge of each triangle (vector $e$ in (a)), and the transformation is to compute the angle between the fiber and that fixed edge follow the right hand rule of the triangle face normal vector ($\theta$ in (b)). (Note that it is a property of a triangular mesh, that the sequence of the 3 vertices of a triangle are in such a way that we can utilize the right-hand-rule to find out the triangular face normal vector that points outwards of the mesh.)

\begin{figure}[!ht]
\centering
\includegraphics[width = 0.8\textwidth]{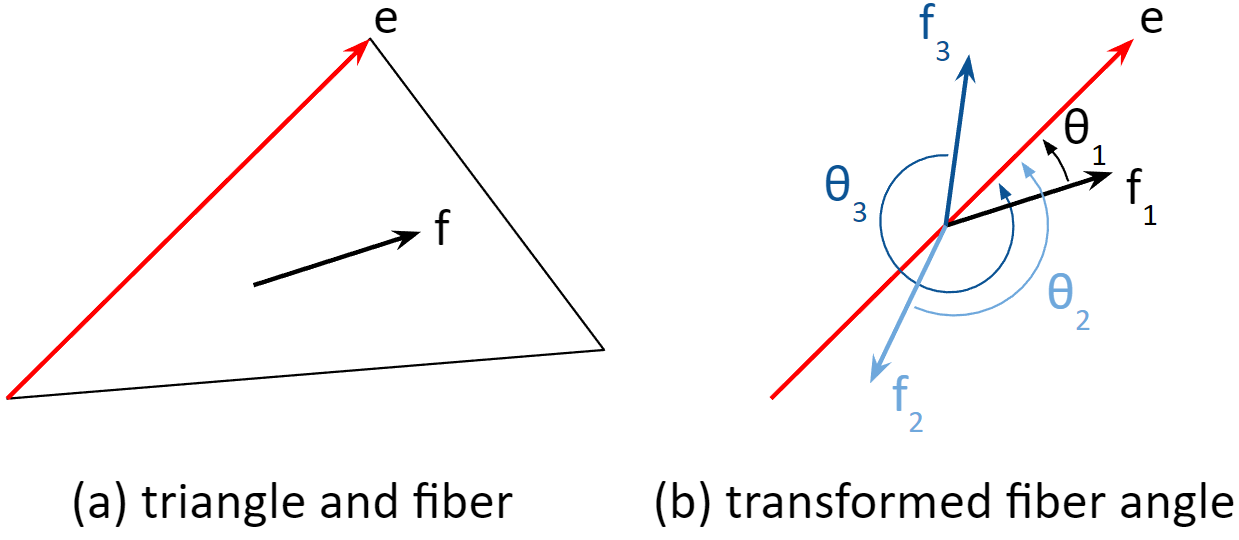}
\caption{Fiber data reference frame transformation. (a) Fiber $f$ resides on a triangle of the mesh. The reference vector $e$ created from a fixed edge, all the seven different fibers will reference this vector $e$. (b) The 3D fiber vector is transformed into a 1D value. Examples of fibers ($f_1$, $f_2$, and $f_3$) and their transformed values ($\theta_1$, $\theta_2$, and $\theta_3$).}
\label{fig:compare fiber}
\end{figure}

The $\Delta\theta$ maps of different atria registered on LA$_1$ is shown in Figure \ref{fig:delta theta map}. Table \ref{tb:delta theta percentage} shows the relative area occupied by smaller $\Delta\theta$ regions ($\Delta\theta \leq 45^{\circ}$) in different atria. Areas with smaller $\Delta\theta$ manifest higher effective anisotropy, which contributes to increased sensitivity of activation patterns to fiber organisation. 

\begin{figure}[!ht]
\centering
\includegraphics[width = 1\textwidth]{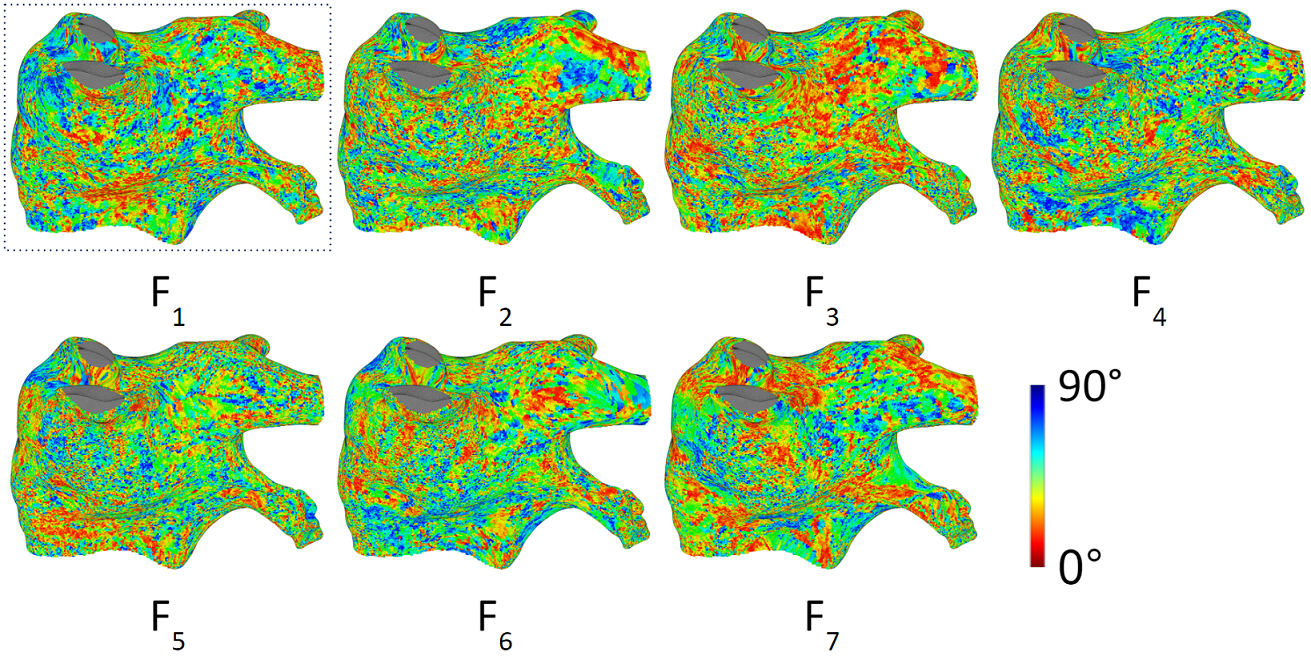}
\caption{Local differences between endocardial and epicardial fiber orientations ($\Delta\theta$ maps). The boxed F$_1$ fiber is on its intrinsic mesh (LA$_1$). F$_2$ to F$_7$ are fibers from LA$_2$ to LA$_7$ that registered on LA$_1$. Red and blue colors represent small and large $\Delta\theta$, respectively. Random local color variations indicate lack of large regions with the same $\Delta\theta$ value.}
\label{fig:delta theta map}
\end{figure}

\begin{table}[!ht]
    \centering
    \begin{tabular}{ c c c c c c c c}
    \hline
F1 & F2 & F3 & F4 & F5 & F6 & F7 \\ \hline
57\% & 56\% & 65\% & 55\% & 57\% & 53\% & 64\% \\ \hline 
    \end{tabular}
    \caption{$\Delta\theta$ area analysis. Percentage of the atrium area occupied by regions with $\Delta\theta \leq 45^{\circ}$ in different models.}
    \label{tb:delta theta percentage}
\end{table}

The correlations of fiber orientations are in the range between -0.12 and 0.18 (with the exception of auto-correlation). Low correlation indicates significant variation in fiber organization across different atria, which is consistent with earlier observations \cite{Ho2001, Pashakhanloo2016}.

\newpage

\section{Atrial experiments of fiber effects}
We found that fiber organization can have a relative large effect on conduction velocity in the slab tissue experiments. However, in the atrium, the fiber organizations are a lot different than in our slab tissue experiments. The question we have now is: Whether observations from slab tissue experiment still hold true in the left atrium?

To investigate fiber effects on activation patterns on the left atrium, we register 7 different patient’s fiber data onto the same left atrium. And run the same set of arrhythmias to observe the differences in activation patterns. This process is shown in Figure \ref{fig:fiber_effect_experiments}.

\begin{figure}[ht]
\centering
\includegraphics[width = 1\textwidth]{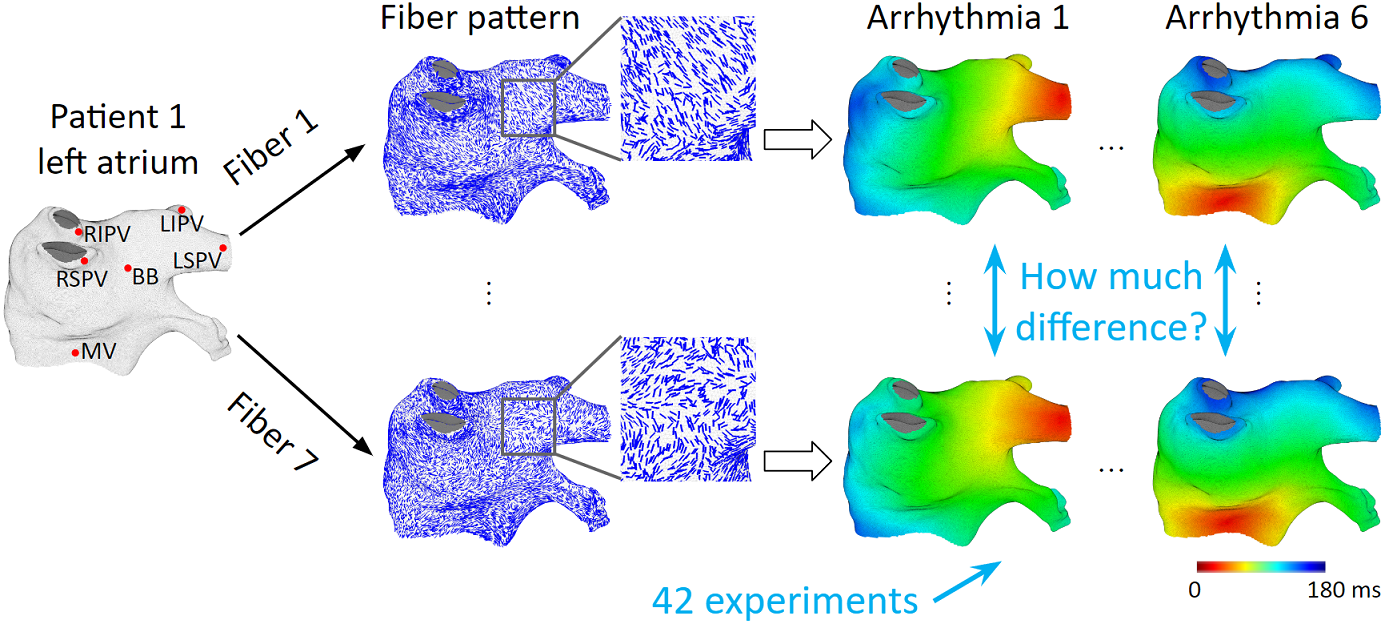}
\caption{Overview of the atrial experiments. Seven fiber organizations are registered onto left atrium 1, and each runs six focal arrhythmia simulations. For each arrhythmia, we compare the activation patterns resulted from different fibers.}
\label{fig:fiber_effect_experiments}
\end{figure}

\newpage

\section{Fiber's effect on local activation time}
\label{sec:42 simulations}
To test the effects of fiber organization on activation patterns, we generated 42 simulations: six pacing sites each with seven different fiber patterns. The six pacing sites are P1: \ac{RIPV}, P2: \ac{LIPV}, P3: \ac{RSPV}, P4: \ac{LSPV}, P5: \ac{MV}, and P6: \ac{BB}.

The resulting \ac{LAT} maps of the 42 experiments are shown in Figure \ref{fig:lat_maps_examples} (more details in Figure \ref{fig:lat maps} in the appendix). The first column shows the simulation results in the ground-truth model built upon the intrinsic left atrium geometry LA$_1$ and the fiber F$_{endo1}$ and F$_{epi1}$. Columns 2-7 are \ac{LAT} maps obtained in models in which the intrinsic fiber was substituted with other fibers. 

\begin{figure}[!ht]
\centering
\includegraphics[width = 1\textwidth]{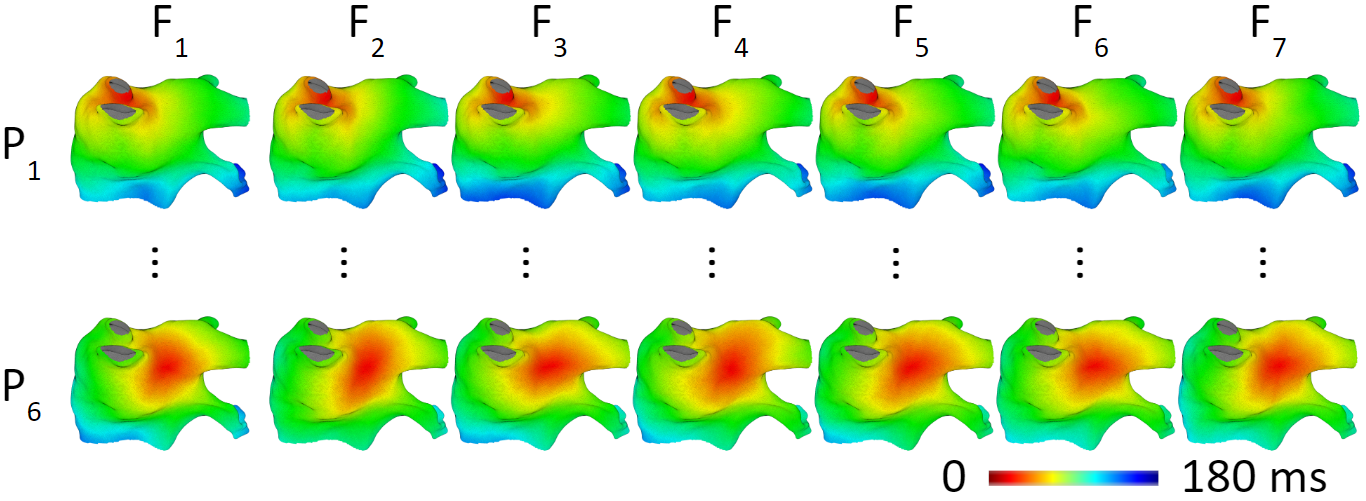}
\caption{\ac{LAT} maps produced by pacing from six different locations in the ground-truth (column F1) and chimeric models (columns F2 to F7). Color scales are normalized to each individual map. The maps in each row are similar to each other, demonstrating that despite significant differences in fiber patterns the activation pattern is not significantly affected.}
\label{fig:lat_maps_examples}
\end{figure}

The \ac{LAT} maps of each row are similar, indicating that differences in fiber organization do not significantly affect the activation pattern. To compare quantitatively the activation maps generated using the chimeric models with the ground-truth, we calculate \ac{LAT} errors (see Table \ref{tb:lat err abs}). The average \ac{LAT} error is 7.8 ms, which is relatively small compared to the time it takes for the activation to travel through the entire left atrium (4.3\% of 180 ms). The LAT correlation is also quite high ranging between 0.94 and 0.99 for different pacing sites and across all models. 

\begin{table}[!ht]
    \centering
    \begin{tabular}{ c | c c c c c c | c}
    \hline
         & F$_2$ & F$_3$ & F$_4$ & F$_5$ & F$_6$ & F$_7$ & \textbf{Avg}\\ \hline
P$_1$ & 4.6 & 6.1 & 5.5 & 4.2 & 5.7 & 5.0 & \textbf{5.2} \\ 
P$_2$ & 6.0 & 9.7 & 4.0 & 4.9 & 5.7 & 6.3 & \textbf{6.1} \\ 
P$_3$ & 9.0 & 11.5 & 7.1 & 8.8 & 10.4 & 12.5 & \textbf{9.9} \\ 
P$_4$ & 12.1 & 9.3 & 7.0 & 9.7 & 6.9 & 9.5 & \textbf{9.1} \\ 
P$_5$ & 13.8 & 7.5 & 13.6 & 3.8 & 6.7 & 7.8 & \textbf{8.9} \\ 
P$_6$ & 7.2 & 8.1 & 7.5 & 6.5 & 6.6 & 8.4 & \textbf{7.4} \\ \hline 
\textbf{Avg} & \textbf{8.8} & \textbf{8.7} & \textbf{7.5} & \textbf{6.3} & \textbf{7.0} & \textbf{8.3} & \textbf{7.8} \\ \hline 
    \end{tabular}
    \caption{Absolute LAT error (ms). The last row shows the average LAT errors across all pacing sites in different fiber models, the last column shows the average error for a given pacing site across different models. The overall average is 7.8 ms or 4.3\% of the LAT range (which is 180 ms).}
    \label{tb:lat err abs}
\end{table}

\newpage

\section{Fiber's effect on latest activation location}
Figure \ref{fig:latest activation time locations} shows the latest activation locations in the left atrium. Blue regions indicate where \ac{LAT} values are within 5 ms of the maximum LAT; The first column shows the results obtained in the ground-truth model with the intrinsic fiber pattern. Other columns are results obtained with different fiber organizations. 

\begin{figure*}[!ht]
\centering
\includegraphics[width = 1\textwidth]{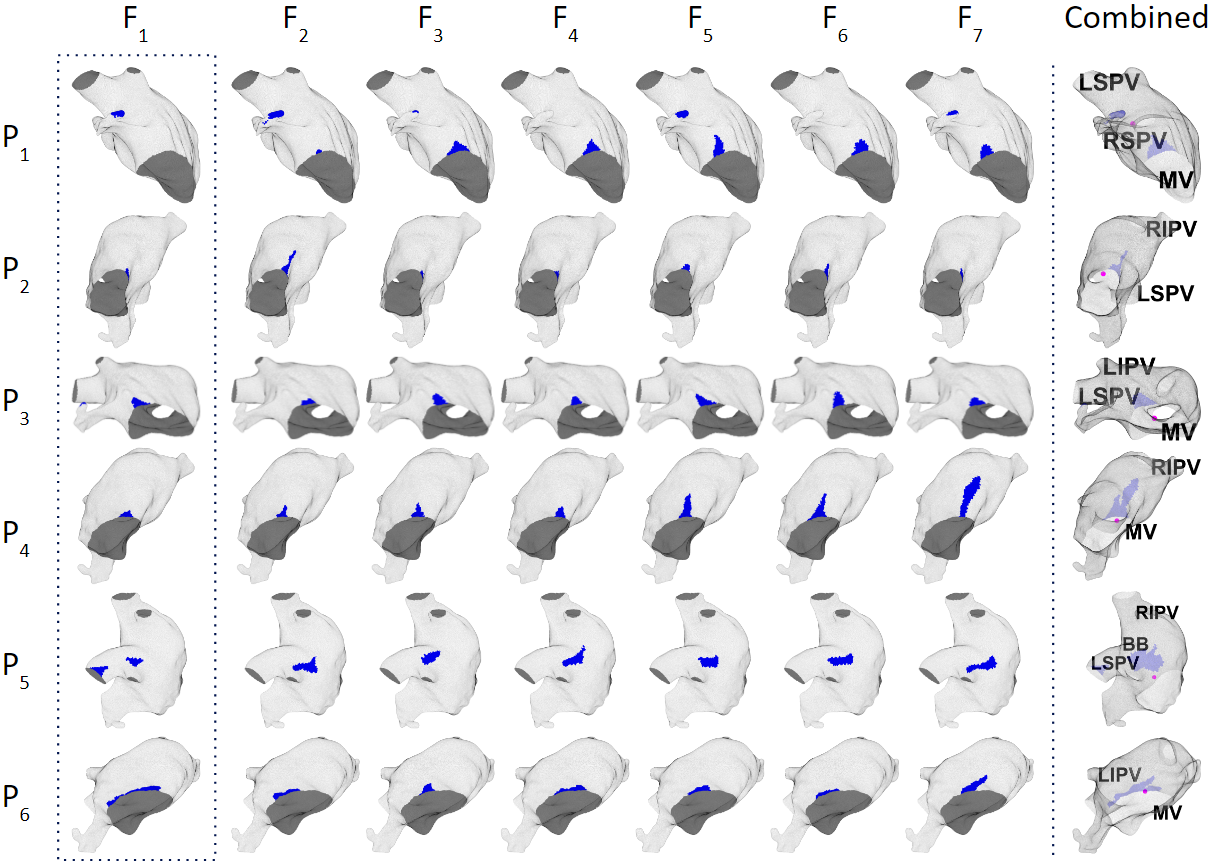}
\caption{The regions of the latest activation derived from the activation maps shown in Figure \ref{fig:lat maps}. The view angles are adjusted for each of pacing sites for better visualization. The first column (boxed) shows the results in the ground-truth model. Blue regions show the sites with the \ac{LAT} values within 5 ms from the maximum. The average maximum \ac{LAT} of the 42 scenarios is 180 ms. The last column shows the superposition of the latest activation regions for all fiber organizations. The magenta dots show the pacing sites on the back side of the atrium. For a given pacing site, the position of the blue regions are reasonably well preserved across different fiber organizations.}
\label{fig:latest activation time locations}
\end{figure*}

Notably, in most of cases, the regions of late activation have similar locations, indicating that different fiber organizations do not alter the start-to-end activation wave patterns significantly. For example, rows P$_3$ and P$_6$ have consistent latest activation locations regardless of fiber organization; rows P$_2$ and P$_4$ show some instances with dilated regions, and row P$_1$ has some different locations, but the clusters are still on the opposite side of the pacing site, which means the general activation pattern remains the same. 

\begin{table}[!ht]
    \centering
    \begin{tabular}{ c | c c c c c c | c}
    \hline
         & F$_2$ & F$_3$ & F$_4$ & F$_5$ & F$_6$ & F$_7$ & \textbf{Avg}\\ \hline
P$_1$ & 3.37 & 34.63 & 38.75 & 16.80 & 38.30 & 24.17 & \textbf{26.00} \\ 
P$_2$ & 6.95 & 1.86 & 1.73 & 8.02 & 1.38 & 3.17 & \textbf{3.85} \\ 
P$_3$ & 17.28 & 14.10 & 15.83 & 10.72 & 10.60 & 12.62 & \textbf{13.52} \\ 
P$_4$ & 2.95 & 3.16 & 4.24 & 6.13 & 6.32 & 17.12 & \textbf{6.65} \\ 
P$_5$ & 26.09 & 18.61 & 26.81 & 26.09 & 23.61 & 27.40 & \textbf{24.77} \\ 
P$_6$ & 9.72 & 4.68 & 5.23 & 2.49 & 8.33 & 5.78 & \textbf{6.04} \\ \hline 
\textbf{Avg} & \textbf{11.06} & \textbf{12.84} & \textbf{15.43} & \textbf{11.71} & \textbf{14.76} & \textbf{15.04} & \textbf{13.47} \\ \hline 
    \end{tabular}
    \caption{Latest activation location difference (mm). Latest activation location difference is the distance between the center of the blue area for F$_2$-F$_7$ and F$_1$ (the truth). The mean is 13.47 mm and the median is 9.43 mm, which are relatively small compared to the left atrium size, indicating that the latest activation locations for different fiber organizations are similar. Note that the values in rows P$_5$ and P$_6$ are very different, which demonstrates that the latest activation location differences depend on pacing location.}
    \label{tb:latest activation location difference}
\end{table}

The quantitative analysis is summarized in Table \ref{tb:latest activation location difference}. For each simulation illustrated in Figure \ref{fig:latest activation time locations} columns F$_2$-F$_7$, we compute the distance of the center of the blue area to the truth. For most scenarios, the distances are small (median: 9.43 mm; average: 13.47 mm) compared to the size of the left atrium mesh $M_{endo1}$ (84 mm $\times$ 97 mm $\times$ 88 mm). Notably, the average across different pacing sites  does not change significantly from model to model (see the last row in Table \ref{tb:latest activation location difference}). Yet, for some pacing sites, they are consistently better than for other pacing sites, for example, compare rows P$_5$ and P$_6$.

\newpage

\section{Fiber's local effect on activation propagation}
Although differences in fiber organization have little effect on the large scale, they can be observed locally. Figure \ref{fig:local effect} shows activation maps generated using models F$_2$ and F$_3$ paced at P$_2$. The activation propagates in the direction of the red-dashed arrow. In the F$_2$ row, the activation propagation slows down in the orange circled area (it has a smaller red region) because the fiber direction is perpendicular to the activation direction. In the F$_3$ row, because fiber direction is along the activation direction, it accelerates propagation and results in a larger red area. Such acceleration and slow-down effects occur throughout the left atrium. The cyan-circled area is another example: propagation is slower in the F$_2$ row than F$_3$ row, resulted in more blue area.

\begin{figure*}[!ht]
\centering
\includegraphics[width = 1\textwidth]{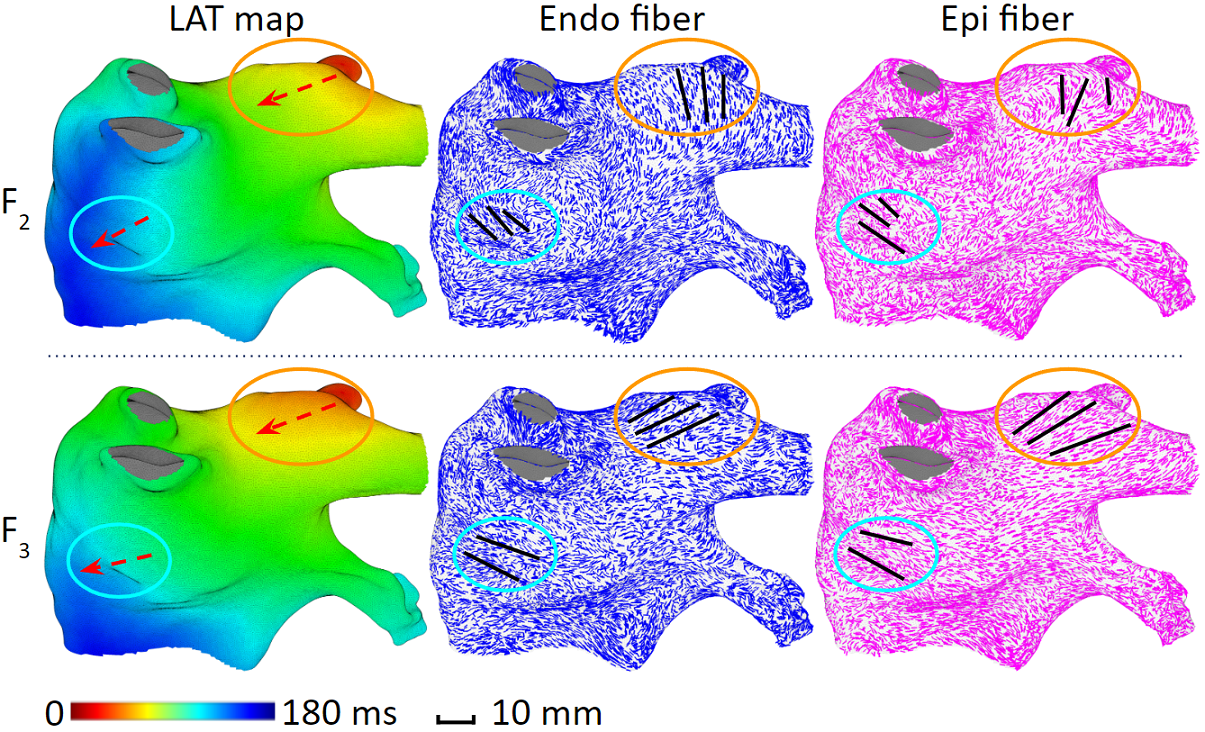}
\caption{Local effect of fiber organization. Upper and lower panels show LAT maps (left) and fiber organization (center and right) in models F$_2$ and F$_3$, respectively. Red arrows: propagation direction. Black lines: prevailing fiber direction in the areas of interest (orange and cyan ovals). The LAT gradient inside the orange oval is conspicuously greater in F$_2$ than in F$_3$ (red-green vs red-yellow). This is because the endocardial and epicardial fibers in F$_2$ are perpendicular to the activation direction, resulting in slower propagation. This is not the case for F$_3$ where propagation is parallel to fibers. Such local effects exist throughout the entire atrium, but they do not alter the overall activation pattern much.}
\label{fig:local effect}
\end{figure*}

The majority of simulations of atrial propagation use values of $r$ between 0.1 and 0.2. Some publications use $r = 0.11$ \cite{Coster2018, Aslanidi2011}, while some others use $r = 0.2$ \cite{Boyle2019}. To investigate the robustness of our conclusions, we performed the same experiment as described in Section \ref{sec:42 simulations} with $r = 0.1$ and $r = 0.5$, and the results are summarized in Table \ref{tb:r}. The small errors suggest that the anisotropy ratio is not the main factor responsible for the low sensitivity of the large scale activation patterns to specific fiber organization.

\begin{table}[!ht]
    \centering
    \begin{tabular}{ c | c c c }
    \hline
        $r$ & 0.1 & 0.2 & 0.5 \\\hline
        LAT Error (ms) & 10.6 or 5.5\% & 7.8 or 4.3\% & 3.15 or 2.0\% \\ 
    \hline
    \end{tabular}
    \caption{Different anisotropy ratios. The percentage is with respect to the time it takes the activation to travel through the left atrium. For r = 0.1, 0.2, 0.5, that time is 192, 180, 155 ms respectively.}
    \label{tb:r}
\end{table}

\newpage

\section{The effects of other variables of the fiber organizations}
The reason for fiber organization having little effect on the activation patterns at the large scale could be partly due to significant differences in fiber orientation on the epicardial and endocardial layers (see Figure \ref{fig:delta theta map}) which cause reduction in apparent anisotropy. But the left atrial wall was not always two layers in all parts \cite{Ho2009, Ho2012}. To evaluate the contribution of this effect on the large scale activation we generated seven models in which epicardial layers were assigned the endocardial fiber orientations, thus producing identical fiber orientation in both layers, which is equivalent to completely uncouple the endocardial layer from the epicardium. 

We hypothesized that this modification should amplify the apparent anisotropy, and consequently the differences between activation patterns in different models. Using these modified models, we performed the same experiment as described in Section \ref{sec:42 simulations} and compared the results. It might be expected, the errors in  \ac{LAT} in the models with identical epicardial and endocardial fiber orientations was greater than in the original models, however, the difference was not very large (9.1 ms vs 7.8 ms). 

This suggests that the difference in the endocardium and epicardium fiber orientations is not the main factor responsible for low sensitivity of the large scale activation patterns to specific fiber organization. 

\newpage

\section{Fiber's cancellation effect}
\label{sec:cancellation effect}

Our data suggest that the effects of fiber organizations are cancelled because fiber organizations vary across the left atrium. This cancellation effect can be explained at two levels. 

\begin{figure*}[!ht]
\centering
\includegraphics[width = 1\textwidth]{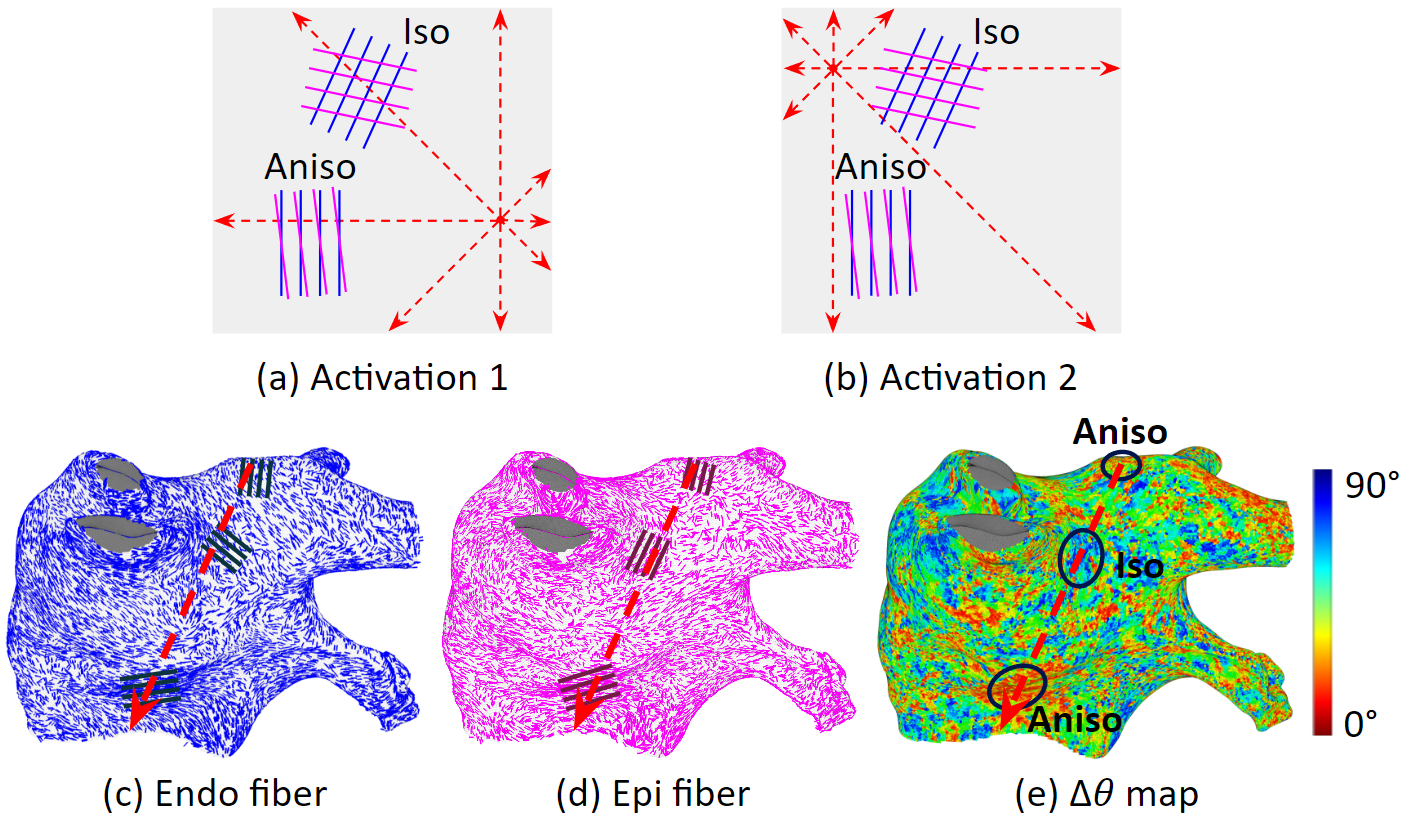}
\caption{The cancellation effect explained via the fibers: there is no macro level consistent fiber organization, therefore local fiber effects cancel each other. (a, b) show fiber’s local effect. Blue lines represent endocardium fiber, magenta lines represent epicardium fiber. Aniso region has a small $\Delta\theta$ (about 0\degree), and Iso region has a large $\Delta\theta$ (about 90\degree). Red arrows are activation wave traveling directions. For activation scenario (a), The Aniso region has a slow down effect on the activation; However, for activation scenario (b), The Aniso region has a speed up effect on the activation. And for both (a) and (b), the propagation speed in the Iso region are similar. (c, d, e) show that the local effects cancel each other at the macro level. (c) is endocardium fiber, (d) is epicardium fiber, and (e) is $\Delta\theta$ map. Along the activation path marked by the red dashed arrow, the propagation speed increases in the top Aniso region, then it goes trough an Iso region that does not change the speed much, finally it enters the bottom Aniso region and the speed decreases. In summary, activation speed first increases then decreases, therefore the effects cancel each other.}
\label{fig:cancellation effect}
\end{figure*}

At the micro level, depending on the location of the activation origin, wave propagation can be shaped by the local fiber organization. As shown in Figure \ref{fig:cancellation effect}(a, b), they represent the same tissue region but has different activations. Aniso region has small $\Delta\theta$ (about 0\degree), it is highly anisotropic, and have the strongest local effect on activation propagation; Iso region has large $\Delta\theta$ (about 90\degree), it is highly isotropic, and have the least effect on activation propagation. For Iso region, activation waves will travel through them in a similar manner regardless of the direction of travel. For Aniso region, depends on the direction of the activation wave, its travelling speed can be decreased (as shown in a) or increased (as shown in b). 

At the macro level, the local effects from the micro level will be cancelled by each other. As shown in Figure \ref{fig:cancellation effect}(c, d, e), along a global path, the activation wave can travel through areas that increase and areas that decrease propagation speed, resulting in an overall zero effect. 

\begin{figure}[!ht]
\centering
\includegraphics[width = 1\textwidth]{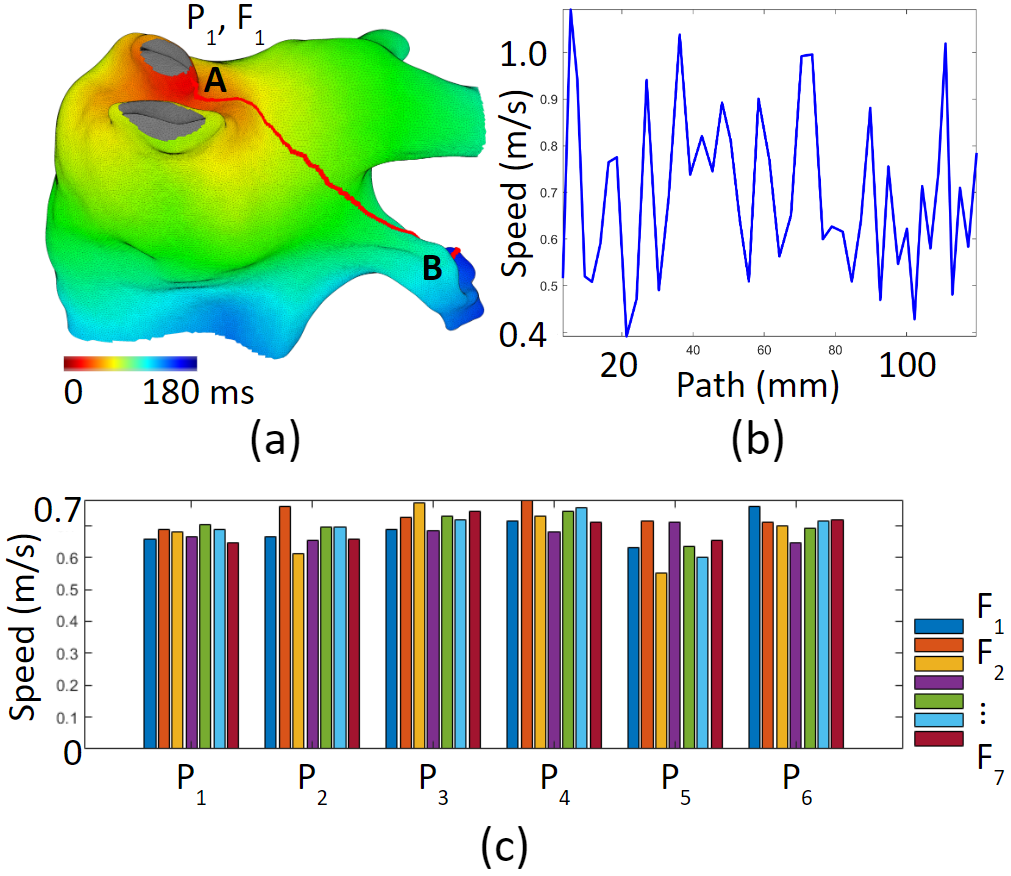}
\caption{An example of propagation speed along a path. (a) The path (red line) is the geodesic line from the earliest (point A) to the latest activation location (point B). (b) The path’s value is the geodesic length from the point along the path to point A. We can see that the propagation speed increases and decreases along the path. (c) The bar plots are the average propagation speed. We can see that for all the six different pacing scenarios, the average propagation speed values are similar regardless of which fiber organization was implemented.}
\label{fig:cv}
\end{figure}

Quantitatively, the propagation speed along a path is shown in Figure \ref{fig:cv}. The path is the geodesic line (minimum distance path) from the activation origin (point A) to the latest activation location (point B) as shown in (a), and it can be found via Dijkstra's algorithm. The path has a length of about 120 mm. The propagation speed of a point along the path is calculated as distance divided by time, where distance is the geodesic length from that point to point A, and the time is the \ac{LAT} values difference between that point and point A. (Note that propagation speed is not necessarily equal to conduction velocity, because conduction velocity is in the \ac{LAT} gradient direction while the path is not necessarily aligned with that. We did not compute conduction velocity along the path because it is not easy to get accurate results \cite{Cantwell2015TechniquesFA}.)

We can see that the propagation speeds vary around its average value along the path as shown in (b), and the average propagation speeds are similar regardless of fiber organizations as shown in (c). Such phenomenon happened for all 42 scenarios. Because of this phenomenon, the fiber organization’s local effect does not accumulate along the path, resulting in a small macro effect.

\newpage

\section{Limitations}
We found that fiber organization does not significantly affect activation patterns. However, there are important limitations to the scenarios we considered that informed this finding. 

1) We focused only on sustained sources, such as focal atrial tachycardia, which may occur following atrial fibrillation ablation. It is possible that non-stationary sources such as drifting rotors could be affected by fibers. 

2) We did not consider scars, which could play an important role in atrial fibrillation dynamics \cite{Gonzales2014}. 

3) The Mitchell-Schaeffer model we used is not a detailed ionic model, and it may not be a good choice to model complex rhythms such as atrial fibrillation. Still such a two-component model is good for modeling periodic propagation. We utilized the computationally more efficient mono-domain model. It is well established that more detailed bi-domain models are required for accurate simulation of electrical activity in the immediate vicinity of the stimulating electrodes and for modelling electrical defibrillation \cite{Roth2021}. With regard to accuracy, the bi-domain models have practically no advantages over mono-domain models for simulations of propagation patterns of external stimulus \cite{Potse2006}.

4) We simplify fiber organization into only two layers. The real left atrium has many more layers, and the number of layers also varies in different regions, as does the atrial thickness. If more layers were incorporated, we would need to study if the effects of fibers would become stronger as well as whether the fiber direction changes abruptly or gradually through the thickness. 

5) We examined fiber organizations of the left atrium, because the most common atrial fibrillation sources are in the left atrium; therefore, it has more available clinical electroanatomical mapping data than the right atrium. We have not examined whether our findings would hold in the right atrium.  

\newpage

\section{Conclusion}
We found that 1) Fiber organization varies significantly across different left atria and within a left atrium. 2) Fibers have local effects on activation propagation but such local effects cancel each other at the macro level. And 3) fibers do not significantly affect focal arrhythmia activation pattern. Fiber organization may not be essential for accurate heart modeling of arrhythmias in  the left atrium.

These results encourage us to develop a patient-specific heart model without fiber data. In the next chapter, we will explain in detail the math and computation of our left atrium model. Then in the chapter after, we will explain how we build such a fiber-independent heart model.

\chapter{\MakeUppercase{Computational model of the left atrium}}
\label{chapter:heart_modeling}

\section{Overview}
We have observed that, in the left atrium, fiber organizations do not have a large effect on activation patterns. This means it is possible for us to develop a patient-specific heart model without fiber data. We developed a method to tune the diffusion coefficients of the heart model to compensate for the lack of fiber data. Here in this chapter, we will explain the details of the heart model's math, and how to implement it in programming, and the next chapter will describe the model tuning. 

Figure \ref{fig:heart modeling} shows an overview of arrhythmia simulation. First, patient data are processed into heart model parameters, and an interactive optimization is implemented to make the model patient-specific. Next, give the simulation an initial state, which will initiate focal or rotor arrhythmias. Lastly, with the initial state, solve the heart model equations to simulate the arrhythmia.

\begin{figure}[!ht]
\centering
\includegraphics[width = 1\textwidth]{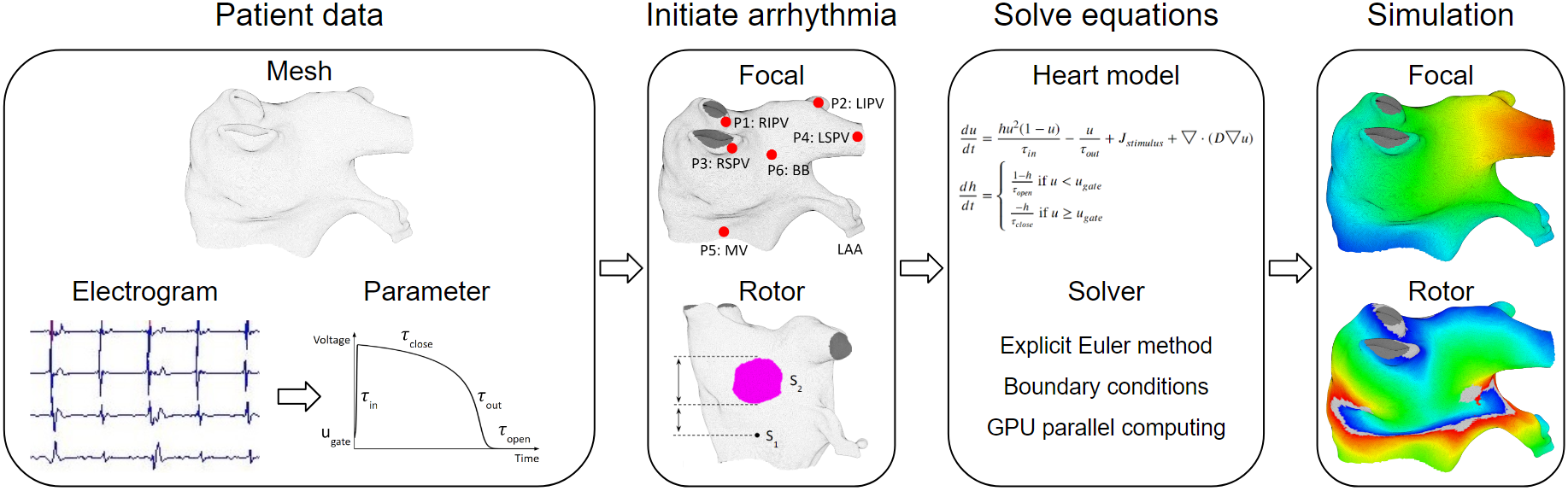}
\caption{Overview of arrhythmia simulation. First, patient data such as the left atrium 3D triangular mesh, endocardium electrogram recordings, and fiber orientations are processed and fed into the heart model. Then, initial conditions are given to simulate arrhythmia by solving the heart model differential equations.}
\label{fig:heart modeling}
\end{figure}

In the following sections, we will explain the heart model equations and their parameters, derive the discrete form the heart model equations for programming and GPU parallel computing, and show arrhythmia simulation examples. 

\newpage

\section{Heart model equations}
The heart model we use is called Mitchell-Schaeffer heart model \cite{Mitchell2003} as shown in \eqref{eq:mitchell schaeffer}. It models the inward current caused by sodium and calcium ion channels, outward current caused by potassium channels, and external stimulus current. This model is chosen because its simplicity makes it efficient in 3D numerical simulations, and the parameters provide direct insight into changes in electrophysiological behavior.

\begin{align}
\label{eq:mitchell schaeffer}
\begin{split}
\frac{du}{dt}&=\frac{hu^2(1-u)}{\tau _{in}}-\frac{u}{\tau _{out}}+J_{stimulus}+\bigtriangledown \cdot (D\bigtriangledown u)
\\
\frac{dh}{dt}&=\left\{\begin{matrix}\ \frac{1-h}{\tau_{open}} \ \text{if}\ u<u_{gate} \\ \ \frac{-h}{\tau_{close}} \ \text{if}\ u \geq u_{gate}\end{matrix}\right.
\end{split}
\end{align}

The variables are as follows:
\begin{itemize}
\item $u$ is the transmembrane voltage and $h$ is an inactivation gating variable for the inward current.
\item $\tau_{in}, \tau_{close}, \tau_{out}, \tau_{open}$ and $u_{gate}$ are parameters that control the action potential shape as shown in Figure \ref{fig:action potential shape}.
\item $J_{stimulus}$ is an external current applied locally as impulses to initiate action potential. We specify this impulse to have 10 ms duration and a magnitude of 20.
\item $\bigtriangledown \cdot (D\bigtriangledown u)$ is the diffusion term, responsible for action potential propagation. Here $\bigtriangledown \cdot$ is the divergence operator, and $\bigtriangledown$ is the gradient operator.
\end{itemize}

Fiber anisotropy is introduced via a $3 \times 3$ diffusion tensor $D$ according to \eqref{eq:D} \cite{Elaff2018}, 
\begin{align}
\label{eq:D}
\begin{split}
D &= d \left ( rI+\left ( 1-r \right )ff^\top \right )
\end{split}
\end{align}

The variables are as follows:
\begin{itemize}
\item $d$ is the diffusion coefficient that controls action potential propagation speed.
\item $r$ is the anisotropy ratio, a ratio of fiber's transverse to longitudinal diffusion coefficients, or the ratio of transverse to longitudinal conduction velocity squared.
\item $I$ is a $3 \times 3$ identity matrix.
\item $f$ is a $3 \times 1$ unit vector pointing along the fiber direction. $f^\top$ is the transpose of $f$ .
\end{itemize}

\begin{figure}[!ht]
\centering
\includegraphics[width = 0.5\textwidth]{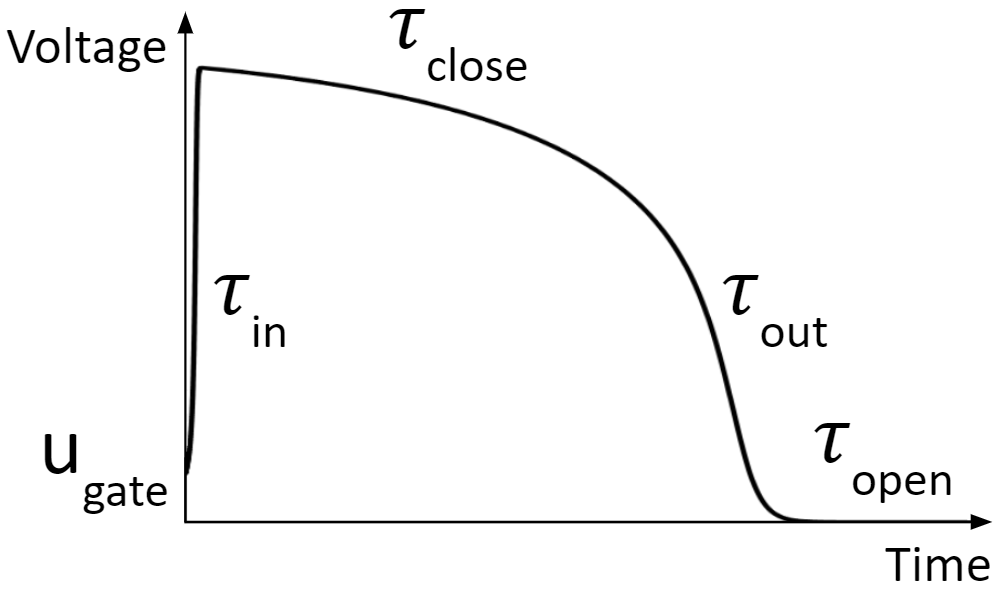}
\caption{The action potential ($u$) shape is governed by $\tau_{in}, \tau_{close}, \tau_{out}, \tau_{open}$ and $u_{gate}$.}
\label{fig:action potential shape}
\end{figure}

\begin{table}[!ht]
    \centering
    \begin{tabular}{ ccccccc }
    \hline
        $\tau_{in}$ & $\tau_{out}$ & $\tau_{open}$ & $\tau_{close}$ & $u_{gate}$ & $r$ & $d$ \\\hline
        0.3 ms & 6 ms & 120 ms & 80 ms & 0.13 mV & 0.2 & 1 \\ 
    \hline
    \end{tabular}
    \caption{Nominal parameter values.}
    \label{tb:parameter}
\end{table}

The parameters are given nominal values as shown in Table \ref{tb:parameter} \cite{Cabrera2017, Roney2019}. With this setting, the conduction velocity will be around 0.69 m/s, which is a typical value for the atrium. To make the heart model patient-specific, these parameters will need to be tuned according to patient data.

\newpage

\section{Transform left atrium 3D mesh to Cartesian nodes}
The left atrium geometry (3D triangular mesh) is from an electroanatomical mapping system (for example, Carto3 System of Biosense Webster, Inc. Irvine, California, United States). The meshes often needs to be refined, details please refer to Section \ref{sec:mesh_processing}. After mesh refinement, the mesh will have uniform triangles of edge length about 0.6 mm.

\begin{figure}[!ht]
\centering
\includegraphics[width = 1\textwidth]{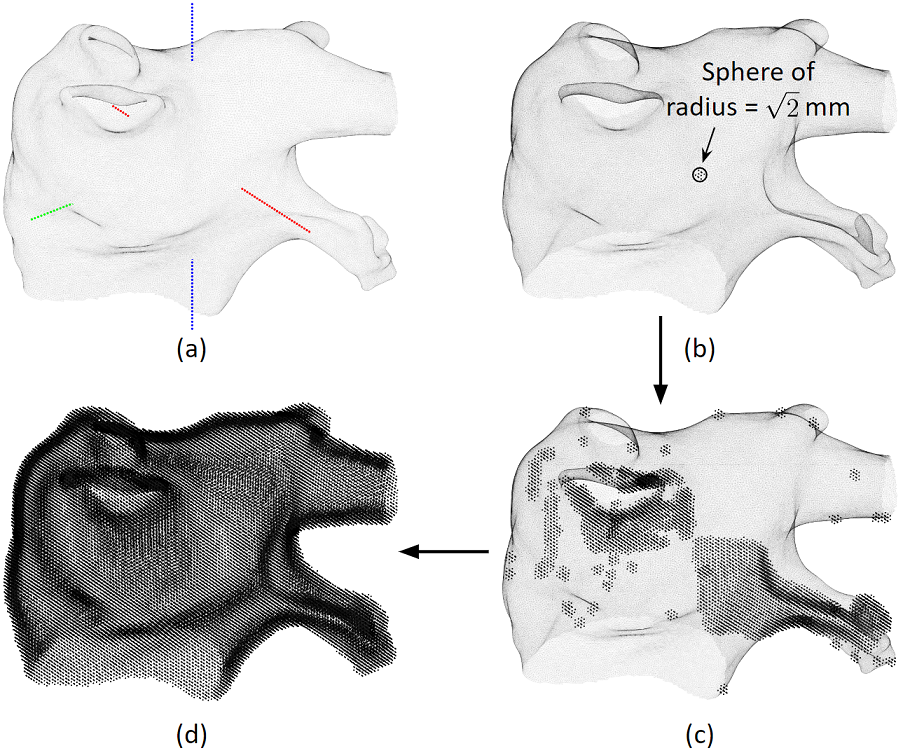}
\caption{Transform triangular mesh to Cartesian nodes. (a) The red, green and blue dots are 1 mm spacing Cartesian grids in x, y and z axis respectively. These grids defined the allowable x, y and z coordinates of the nodes. (b) For a mesh vertex, nodes that within a distance threshold ($d_{threshold}$) are created. (c) Loop through more vertices to create more nodes. (d) After looping through all vertices, all Cartesian nodes for this left atrium are created.}
\label{fig:create_nodes}
\end{figure}

To run simulation on the left atrium, we transform the triangular mesh to Cartesian nodes. We specify Cartesian node spacing to be $\Delta x= \Delta y = \Delta z = \Delta = 1$ mm. (This spatial resolution is chosen to be adequate for accurate simulation without being computationally demanding. Justification is in Appendix \ref{app:resolution}.) First, the x, y and z axes are discretized into 1 mm spacing grids (Figure \ref{fig:create_nodes}a), these grid values will be the allowable coordinates for the nodes. Then for a vertex, create nodes that are within $d_{threshold} = \sqrt{2}$ mm radius sphere (Figure \ref{fig:create_nodes}b). Lastly, loop through every vertex to create all the nodes (Figure \ref{fig:create_nodes}cd). The resulted left atrium tissue thickness is about 2 mm, which is in the range of clinically observed values \cite{Whitaker2016, Sun2018}.

\newpage

\section{Computation of the heart model equations}
To solve the differential equations \eqref{eq:mitchell schaeffer}, initial values at t = 0 are given ($u[0] = 0$ and $h[0] = 1$), then solutions of the next time step are computed using the explicit Euler method as shown in \eqref{eq:explicit Euler}. The time step is $\Delta t = 0.01$ ms (justification is in Appendix \ref{app:resolution}). 

\begin{align}
\label{eq:explicit Euler}
\begin{split}
u[t+1] &= \left( \frac{h[t](u[t])^2(1-u[t])}{\tau _{in}}-\frac{u[t]}{\tau _{out}}+J_{stimulus}+\bigtriangledown \cdot (D\bigtriangledown u[t]) \right) \Delta t + u[t]
\\
h[t+1] &= \left\{\begin{matrix}\ \frac{1-h[t]}{\tau_{open}} \Delta t + h[t] \ \text{if}\ u[t]<u_{gate} \\ \ \frac{-h[t]}{\tau_{close}} \Delta t + h[t] \ \text{if}\ u[t] \geq u_{gate}\end{matrix}\right.
\end{split}
\end{align}

To compute the diffusion term $\bigtriangledown \cdot (D\bigtriangledown u)$, we follow \cite{McFarlane2010} that assumed no-flux (Neumann) boundary conditions and used a 19-node stencil as shown in Figure \ref{fig:node neighbors}.

\begin{figure}[!ht]
\centering
\includegraphics[width = 0.5\textwidth]{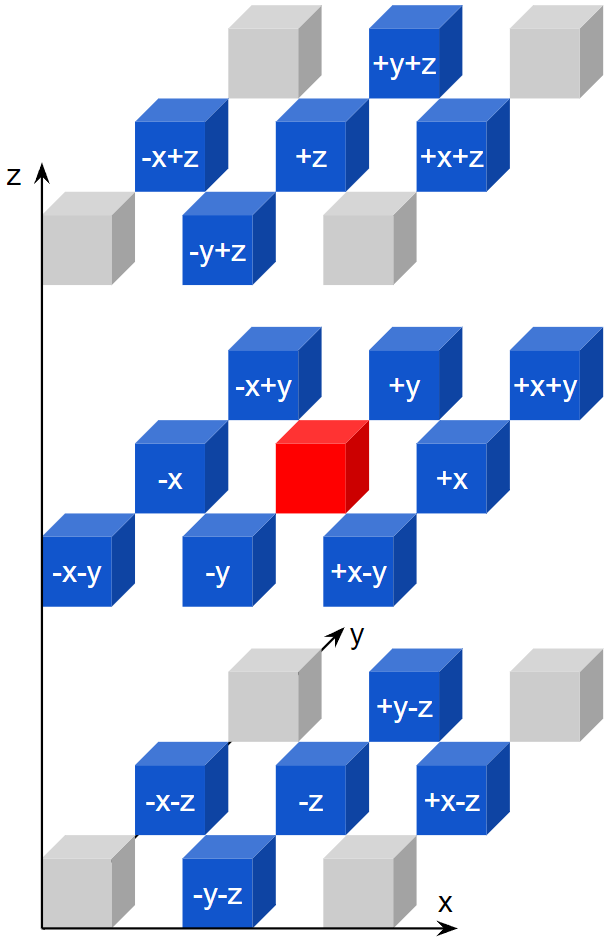}
\caption{Node neighbors. The red node has 26 neighbors, but only 18 of them are involved in solving the heart model equations, and they are colored blue. If the red node has coordinate $(x,y,z)$, then the neighbor blue node labelled "-x+y" has coordinate $(x-\Delta,y+\Delta,z)$.}
\label{fig:node neighbors}
\end{figure}

Give $D$ a matrix notation as \eqref{eq:D2}:

\begin{align}
\label{eq:D2}
D = d \left ( rI+\left ( 1-r \right )ff^\top \right ) = d \begin{bmatrix} D_{11} & D_{12} & D_{13} \\
D_{21} & D_{22} & D_{23} \\
D_{31} & D_{32} & D_{33} \end{bmatrix}
\end{align}

Then the diffusion term becomes:

\begin{align}
\label{eq:D3}
\bigtriangledown \cdot \left ( D\bigtriangledown u \right ) = \bigtriangledown \cdot \left ( d \begin{bmatrix} D_{11} & D_{12} & D_{13} \\
D_{21} & D_{22} & D_{23} \\
D_{31} & D_{32} & D_{33} \end{bmatrix} \begin{bmatrix} \frac{\partial u}{\partial x} \\
\frac{\partial u}{\partial y} \\
\frac{\partial u}{\partial z} \end{bmatrix}\right ) = d \left( \sum_{i=1}^{3}\sum_{j=1}^{3}\left ( \frac{\partial D_{ij}}{\partial x_i} \frac{\partial u}{\partial x_j} + D_{ij}\frac{\partial^2 u}{\partial x_i \partial x_j} \right ) \right)
\end{align}

Where $x_1 = x$, $x_2 = y$ and $x_3 = z$. In the following equations, we introduce a subscript notation: $\square_{...}$ represents the value for the node at coordinate $(x,y,z)$, which is the red node in Figure \ref{fig:node neighbors}; $\square_{.+-}$ represents the value for the node at coordinate $(x,y+\Delta,z-\Delta)$, which is the blue node labelled "+y-z" in Figure \ref{fig:node neighbors}. The partial derivatives can be approximated as \eqref{eq:derivatives}, and other terms in \eqref{eq:D3} can be approximated in similar manners. 

\begin{align}
\label{eq:derivatives}
\begin{split}
\frac{\partial u_{...}}{\partial x} &\approx \frac{u_{+..} - u_{-..}}{2 \Delta} \\
\frac{\partial^2 u_{...}}{\partial x^2} &\approx \frac{u_{+..}-2u_{...}+u_{-..}}{\Delta^2} \\
\frac{\partial^2 u_{...}}{\partial x \partial y} &\approx \frac{ u_{++.} - u_{-+.} - u_{+-.} + u_{--.} } {4\Delta^2}
\end{split}
\end{align}

\subsection{Boundary conditions}
As shown in Figure \ref{fig:node neighbors}, to compute the diffusion term for a node, it involves 18 neighbor nodes. However, some nodes may not have all those 18 neighbors because it is at the boundary. For those none-existing neighbors, we can multiply 0 to eliminate the associated terms, to satisfy the no-flux boundary conditions. Substitute \eqref{eq:derivatives} into \eqref{eq:D3}, rearrange the terms, and add in an indicating variable $\delta$, we then have an expression of the diffusion term with implicit no-flux boundary conditions as shown in \eqref{eq:diffusion term with boundary condition}. The indicator variable $\delta = 1$ if there is a Cartesian node at that coordinate specified in the subscript, and $\delta = 0$ if it is void at that coordinate.

\begin{align}
\label{eq:diffusion term with boundary condition}
\begin{split}
\bigtriangledown \cdot ( D\bigtriangledown u ) = \frac{d}{4\Delta^2} \{& 4\delta_{+..} D_{...}^{11}(u_{+..}-u_{...}) + 4\delta_{-..} D_{...}^{11}(u_{-..}-u_{...}) + 4\delta_{.+.} D_{...}^{22}(u_{.+.}-u_{...}) + \\
&4\delta_{.-.} D_{...}^{22}(u_{.-.}-u_{...}) + 4\delta_{..+} D_{...}^{33}(u_{..+}-u_{...}) + 4\delta_{..-} D_{...}^{33}(u_{..-}-u_{...}) + \\
&\delta_{+..}\delta_{-..} [\delta_{+..}\delta_{-..}(D_{+..}^{11}-D_{-..}^{11}) + \delta_{.+.}\delta_{.-.}(D_{.+.}^{21}-D_{.-.}^{21}) + \delta_{..+}\delta_{..-}(D_{..+}^{31}-D_{..-}^{31})] (u_{+..}-u_{-..}) + \\
&\delta_{.+.}\delta_{.-.} [\delta_{+..}\delta_{-..}(D_{+..}^{12}-D_{-..}^{12}) + \delta_{.+.}\delta_{.-.}(D_{.+.}^{22}-D_{.-.}^{22}) + \delta_{..+}\delta_{..-}(D_{..+}^{32}-D_{..-}^{32})] (u_{.+.}-u_{.-.}) + \\
&\delta_{..+}\delta_{..-} [\delta_{+..}\delta_{-..}(D_{+..}^{13}-D_{-..}^{13}) + \delta_{.+.}\delta_{.-.}(D_{.+.}^{23}-D_{.-.}^{23}) + \delta_{..+}\delta_{..-}(D_{..+}^{33}-D_{..-}^{33})] (u_{..+}-u_{..-}) + \\ 
&2\delta_{++.}\delta_{+-.} D_{...}^{12} (u_{++.}-u_{+-.}) + 2\delta_{--.}\delta_{-+.} D_{...}^{12} (u_{--.}-u_{-+.}) + \\
&2\delta_{+.+}\delta_{+.-} D_{...}^{13} (u_{+.+}-u_{+.-}) + 2\delta_{-.-}\delta_{-.+} D_{...}^{13} (u_{-.-}-u_{-.+}) + \\
&2\delta_{.++}\delta_{.+-} D_{...}^{23} (u_{.++}-u_{.+-}) + 2\delta_{.--}\delta_{.-+} D_{...}^{23} (u_{.--}-u_{.-+}) \}
\end{split}
\end{align}

\subsection{The indicator variable $\delta$}
Each node has an index $i$, typically, for a left atrium, the total number of nodes is about $N=100,000$. To find out $\delta$, we first find out the neighboring nodes indices. Create a $N \times 18$ matrix $neighbor\_id$, in which each row $i$ stores node $i$'s 18 neighbors (as shown in Figure \ref{fig:node neighbors}) in sequence: these neighbors are "+x", "-x", ..., "-x-z" as shown in Table \ref{tb:node neighbors}. For example, if we want to find the "-x+z" neighbor of a node at $(x,y,z)$, than check which node has coordinate $(x-\Delta,y,z+\Delta)$. 

For programming in Matlab, variable indices starts at 1, therefore $i$ starts at 1 and ends at N. If there is a node at that coordinate, then record that node's index; if there is not a node, then record 0. For example, the 120th row of $neighbor\_id$ may be as Table \ref{tb:node neighbors}. Take the sign of $neighbor\_id$ to get the value of $\delta$:

\begin{align}
\label{eq:delta}
\delta = sign(neighbor\_id)
\end{align}

\begin{table}[!ht]
    \centering
    \begin{tabular}{ cccccccccc }
    \hline
        $+x$ & $-x$ & $+y$ & $-y$ & $+z$ & $-z$ & $+x+y$ & $-x+y$ & $+x-y$ & $-x-y$ \\
        201 & 39 & 121 & 119 & 0 & 0 & 202 & 40 & 200 & 38 \\\hline\hline
        $+y+z$ & $-y+z$ & $+y-z$ & $-y-z$ & $+x+z$ & $-x+z$ & $+x-z$ & $-x-z$ & & \\
        0 & 0 & 0 & 0 & 0 & 0 & 0 & 0 & & \\\hline
    \hline
    \end{tabular}
    \caption{An example of a row of $neighbor\_id$. The neighbor nodes indices of node $i=120$ may look like this table. It means, for example, its "-x" neighbor is node $i=39$, and there is no neighbor at "-y+z". 
    }
    \label{tb:node neighbors}
\end{table}

\subsection{Simplify notations}
Many terms in \eqref{eq:diffusion term with boundary condition} are constants and can be computed and stored before computing simulation. We give these terms simplified notations as shown in \eqref{eq:parts}.

\begin{align}
\label{eq:parts}
\begin{split}
P_1 &= 4\delta_{+..} D_{...}^{11}, P_2 = 4\delta_{-..} D_{...}^{11}, P_3 = 4\delta_{.+.} D_{...}^{22} \\
P_4 &= 4\delta_{.-.} D_{...}^{22}, P_5 = 4\delta_{..+} D_{...}^{33}, P_6 = 4\delta_{..-} D_{...}^{33} \\
P_7 &= \delta_{+..}\delta_{-..} [\delta_{+..}\delta_{-..}(D_{+..}^{11}-D_{-..}^{11}) + \delta_{.+.}\delta_{.-.}(D_{.+.}^{21}-D_{.-.}^{21}) + \delta_{..+}\delta_{..-}(D_{..+}^{31}-D_{..-}^{31})] \\
P_8 &= \delta_{.+.}\delta_{.-.} [\delta_{+..}\delta_{-..}(D_{+..}^{12}-D_{-..}^{12}) + \delta_{.+.}\delta_{.-.}(D_{.+.}^{22}-D_{.-.}^{22}) + \delta_{..+}\delta_{..-}(D_{..+}^{32}-D_{..-}^{32})] \\
P_9 &= \delta_{..+}\delta_{..-} [\delta_{+..}\delta_{-..}(D_{+..}^{13}-D_{-..}^{13}) + \delta_{.+.}\delta_{.-.}(D_{.+.}^{23}-D_{.-.}^{23}) + \delta_{..+}\delta_{..-}(D_{..+}^{33}-D_{..-}^{33})] \\
P_{10} &= 2\delta_{++.}\delta_{+-.} D_{...}^{12}, P_{11} = 2\delta_{--.}\delta_{-+.} D_{...}^{12}, P_{12} = 2\delta_{+.+}\delta_{+.-} D_{...}^{13} \\
P_{13} &= 2\delta_{-.-}\delta_{-.+} D_{...}^{13}, P_{14} = 2\delta_{.++}\delta_{.+-} D_{...}^{23}, P_{15} = 2\delta_{.--}\delta_{.-+} D_{...}^{23}
\end{split}
\end{align}

Substitute \eqref{eq:parts} into \eqref{eq:diffusion term with boundary condition}, we have

\begin{align}
\label{eq:diffusion term with boundary condition simplified}
\begin{split}
\bigtriangledown \cdot ( D\bigtriangledown u ) = \frac{d}{4\Delta^2} \{& P_1 (u_{+..}-u_{...}) + P_2 (u_{-..}-u_{...}) + P_3 (u_{.+.}-u_{...}) + \\
&P_4 (u_{.-.}-u_{...}) + P_5 (u_{..+}-u_{...}) + P_6 (u_{..-}-u_{...}) + \\
&P_7 (u_{+..}-u_{-..}) + P_8 (u_{.+.}-u_{.-.}) + P_9 (u_{..+}-u_{..-}) + \\ 
&P_{10} (u_{++.}-u_{+-.}) + P_{11} (u_{--.}-u_{-+.}) + P_{12} (u_{+.+}-u_{+.-}) + \\
&P_{13} (u_{-.-}-u_{-.+}) + P_{14} (u_{.++}-u_{.+-}) + P_{15} (u_{.--}-u_{.-+}) \}
\end{split}
\end{align}

Substitute \eqref{eq:diffusion term with boundary condition simplified} into \eqref{eq:explicit Euler}, and we have the complete solution for solving the heart model equations.

\subsection{Heart simulation pseudo-code}
To compute the heart simulation, we need to compute the differential equations for each time step and for each node. Pseudo-code is shown in Algorithm \ref{al:heart_sim}. If implement in Matlab (MathWorks, Natick, Massachusetts, United States), the "for every node" loop can be made into vector form as one equation; if implement with GPU computing, the computation of each node can be paralleled. We implemented GPU computing using CUDA kernels (Nvidia, Santa Clara, California, United States). 

\begin{algorithm}
	\caption{Heart simulation pseudo-code}
	\begin{algorithmic}[]
        \State Assign initial value to $u[0]$ and $h[0]$
		\For {every time step}
    		\For {every node}
                \State Compute equation \eqref{eq:diffusion term with boundary condition simplified}
                \State Compute equation \eqref{eq:explicit Euler} to get $u[t+1]$ and $h[t+1]$
        		\State Update current value: $u[t]=u[t+1]$, and $h[t]=h[t+1]$
    		\EndFor
		\EndFor
	\end{algorithmic}
 \label{al:heart_sim}
\end{algorithm} 

\subsection{Compute unipolar electrogram}
The heart model equations \eqref{eq:mitchell schaeffer} compute the time sequence of action potentials $u$. Clinically during arrhythmia ablations, electrograms are recorded. To obtain the unipolar electrogram from an electrode, we need to integrate the action potentials of all nodes according to \eqref{eq:electrogram}, where $\overrightarrow{l}$ is the vector from the electrode to a node of the atrium (let $xyz_e$ be the coordinate of the electrode, $xyz_n$ be a node on the atrium, then vector $l=xyz_n-xyz_e$), and $l$ is the distance from a node to point A \cite{Shillieto2016,Tusscher2007,Bueno-Orovio2008}. Note that the $\cdot$ is the dot product. To prevent the "divide by 0" problem, set $l=1$ if $l<1$, this effectively means the electrode will always be at least 1 mm away from tissue. To generate patient-specific electrograms, will need to scale the magnitude according to patient data.

\begin{align}
\label{eq:electrogram}
\begin{split}
unipolar\_electrogram=&\sum_{l}^{ } D\bigtriangledown u \cdot \frac{\overrightarrow{l}}{l^{3}} \\
=&\sum_{l}^{ } d \begin{bmatrix} D_{11} & D_{12} & D_{13} \\ D_{21} & D_{22} & D_{23} \\ D_{31} & D_{32} & D_{33} \end{bmatrix} \begin{bmatrix} \frac{\partial u}{\partial x} \\ \frac{\partial u}{\partial y} \\ \frac{\partial u}{\partial z} \end{bmatrix} \cdot \frac{\begin{bmatrix}l_x \\ l_y \\ l_z\end{bmatrix}}{\left (\sqrt{l_x^2+l_y^2+l_z^2}  \right )^3} \\
=&\sum_{l}^{ } \frac{d}{\left (\sqrt{l_x^2+l_y^2+l_z^2}\right )^3} [ ( D_{11}\frac{\partial u}{\partial x} + D_{12}\frac{\partial u}{\partial y} + D_{13}\frac{\partial u}{\partial z} )l_x + \\
&( D_{21}\frac{\partial u}{\partial x} + D_{22}\frac{\partial u}{\partial y} + D_{23}\frac{\partial u}{\partial z} )l_y + ( D_{31}\frac{\partial u}{\partial x} + D_{32}\frac{\partial u}{\partial y} + D_{33}\frac{\partial u}{\partial z} )l_z ]
\end{split}
\end{align}

\subsection{GPU parallel computing}
We implemented GPU parallel computing to speed up the computation. The GPU computing flowchart is shown in Figure \ref{fig:gpu_computing}. First, variables and initial values of time 0 are allocated to the GPU. Next, for each time step, the computation of each node is done in parallel in the GPU. Lastly, results are analyzed and displayed. We achieved a simulation for 1 heartbeat in 5 seconds on a personal computer.

\begin{figure}[!ht]
\centering
\includegraphics[width = 1\textwidth]{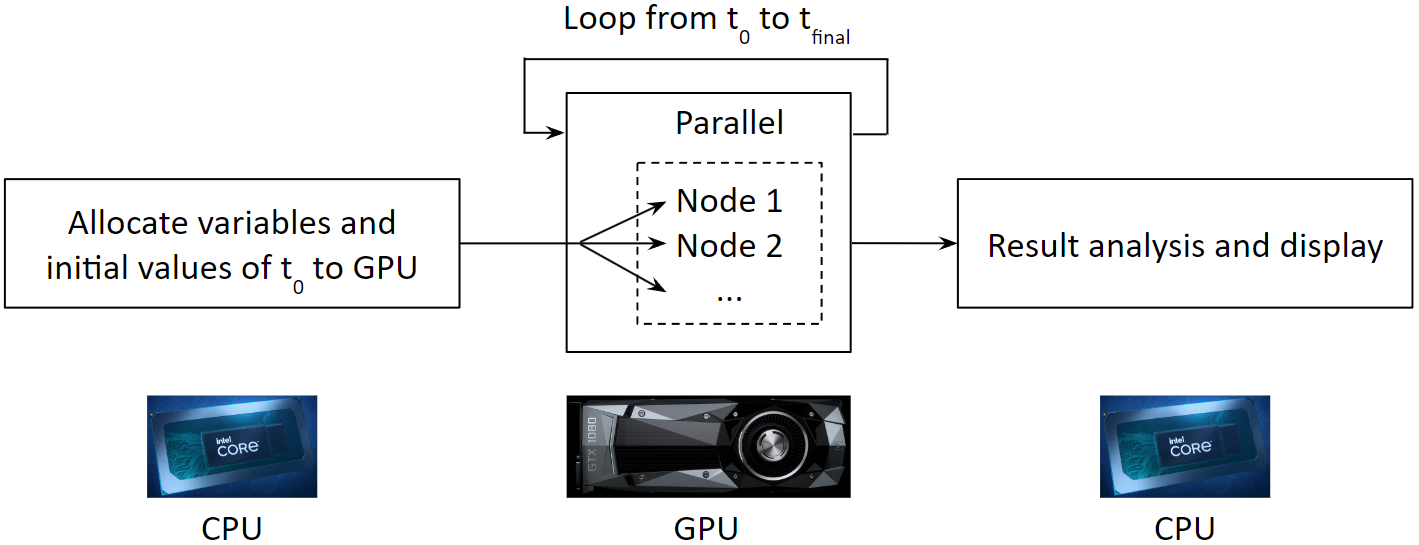}
\caption{A flowchart of GPU parallel computing for simulating arrhythmia.}
\label{fig:gpu_computing}
\end{figure}

\newpage

\section{Arrhythmia simulation examples}
We developed an user interface for assigning left atrium regional properties to initiate different arrhythmias. Here we present some of the most common arrhythmia our heart model can simulate. They are shown in Figure \ref{fig:arrhythmia_simulations}.

\begin{figure}[!ht]
\centering
\includegraphics[width = 1\textwidth]{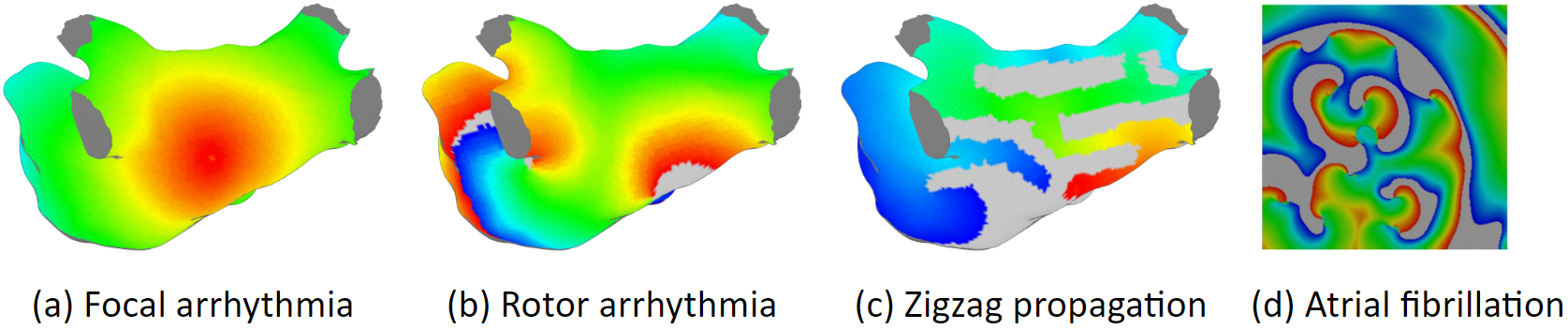}
\caption{Examples of arrhythmia simulations}
\label{fig:arrhythmia_simulations}
\end{figure}

Focal arrhythmia is shown in Figure \ref{fig:arrhythmia_simulations}(a). The fast heartbeat is activated from a point source. Rotor arrhythmia is shown in (b). It has spiral waves that cause irregular heartbeat. Zigzag propagation is shown in (c). The complex scar distributions create activation tunnels of various conduction velocities. Zigzag propagation is an explanation of slow conduction in the infarcted heart \cite{Bakker}. Atrial fibrillation is shown in (d). Its activation waves are chaotic: The activation patterns are not periodic, and there is no organizations.

\chapter{\MakeUppercase{Patient-specific heart model parameter tuning}}
\label{chapter:fiber_independent}

\section{Overview}
Over the past two decades there has been a steady trend towards the development of realistic models of cardiac conduction with increasing levels of detail \cite{o2011simulation, greene2022voltage, fenton2008models} of gross cardiac anatomy \cite{Ho1999, Ho2009, iaizzo2016visible}, myocardial fiber organization \cite{Ho2001, Fastl2018}, and ionic currents involved in the generation of cardiac action potentials \cite{Lopez-Perez2015, berman2021interactive, kaboudian2019real}. However, making models more realistic complicates their personalization and use in clinical practice due to limited availability of tissue and cellular scale data. 

The velocity of action potential propagation along fibers is typically 2-3 times faster than across fibers \cite{Valderrabano2007}, which makes fiber organization one of the key factors determining activation patterns in cardiac tissue \cite{franzone1993spread, fenton1998vortex}. Unlike atrial geometry, which can be readily obtained in a clinical setting via electroanatomical mapping \cite{Bhakta2008}, fiber organization cannot be obtained with a sufficient level of detail in live tissue \cite{Zhao2012}. To date, the best available resource for real patient fiber data is an ex-vivo \ac{DT-MRI} fiber database \cite{Pashakhanloo2016, Roney2021}, which required about 50 hours to scan each atrium \cite{Pashakhanloo2016}. 

One solution is to use synthetic fibers, which can be mathematically generated based on the heart's geometry \cite{Fastl2018, Krueger2011, Labarthe2021, Wachter2015, Saliani2021}. Alternatively, fibers from existing databases can be registered, as it has been found that some patients' fibers can be generalized to many different patients \cite{Roney2021}. However, fibers obtained through these methods do not represent the true fibers. 

In previous two chapters we concluded that fiber organization do not have a large effect on activation patterns, therefore, it is possible to develop a patient-specific heart model without fiber data. The major advantage of a fiber-independent model is that it only requires an atrial geometry, which can be readily obtained in clinical setting from the electroanatomical mapping \cite{Bhakta2008,He2019}. We derive a method for tuning diffusion coefficients of the fiber-independent model, enabling it to compensate for the absence of fibers and become patient-specific. 

These are two terminologies we use throughout this chapter: 1) Fiber-inclusive model: A heart model with intrinsic endocardial and epicardial fibers. 2) Fiber-independent model: A heart model without fibers, instead, have tuned diffusion coefficients. We treat the simulation of the fiber-inclusive model as the ground truth, and use it to tune the fiber-independent model. We conduct comprehensive evaluations of our fiber-independent model for atrial tachycardia, demonstrating its ability to produce highly accurate activation patterns in focal and rotor arrhythmias. The evaluations are based on comparisons of our model to a fiber-inclusive model as shown in Figure \ref{fig:model_performance_evaluation}. Fibers are from a publicly available \ac{DT-MRI} data of 7 ex-vivo hearts \cite{Roney2021}. The comparison is carried out in 51 cases of focal and rotor arrhythmias located in different regions of the atria. 

\begin{figure}[!ht]
\centering
\includegraphics[width = 1\textwidth]{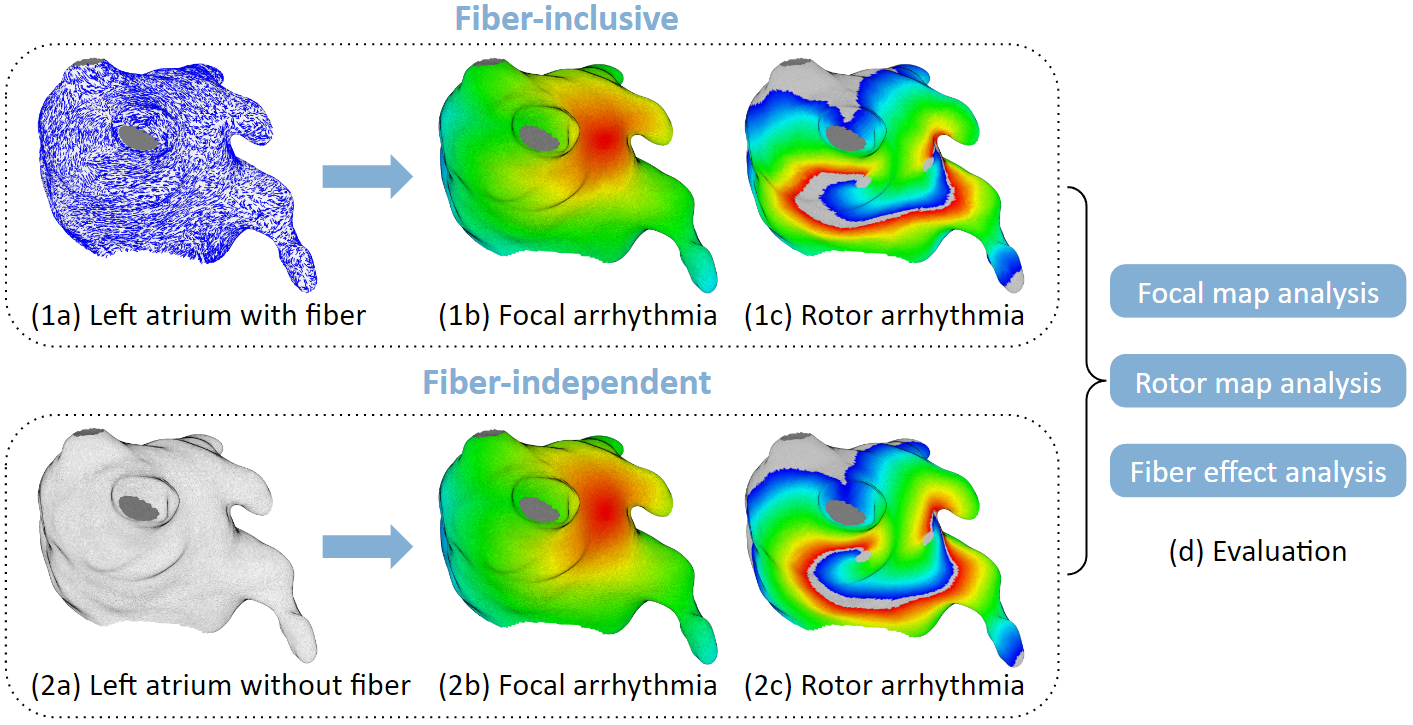}
\caption{Model performance evaluation. (1a,1b,1c) In-silico experiments on the left atria with a fiber-inclusive model. This model includes endocardial and epicardial fibers, but for artistic reason, only one fiber layer is shown in (1a). (2a,2b,2c) In-silico experiments with a fiber-independent model. This model does not have fiber data, instead, it has tuned diffusion coefficients. (d) Evaluate the fiber-independent model by comparing activation patterns of the focal and rotor arrhythmias to the fiber-inclusive model.}
\label{fig:model_performance_evaluation}
\end{figure}

\newpage

\section{Arrhythmia simulations setup}
We specify 6 different focal arrhythmia locations on each atrium. The pacing sites are shown in Fig. \ref{fig:focal_setup}. P$_1$: \ac{RIPV}, P$_2$: \ac{LIPV}, P$_3$: \ac{RSPV}, P$_4$: \ac{LSPV}, P$_5$: \ac{MV}, and P$_6$: \ac{BB}. These locations are chosen because they are clinically identifiable, also they cover a wide variety of scenarios.

\begin{figure}[!ht]
\centering
\includegraphics[width = 1\textwidth]{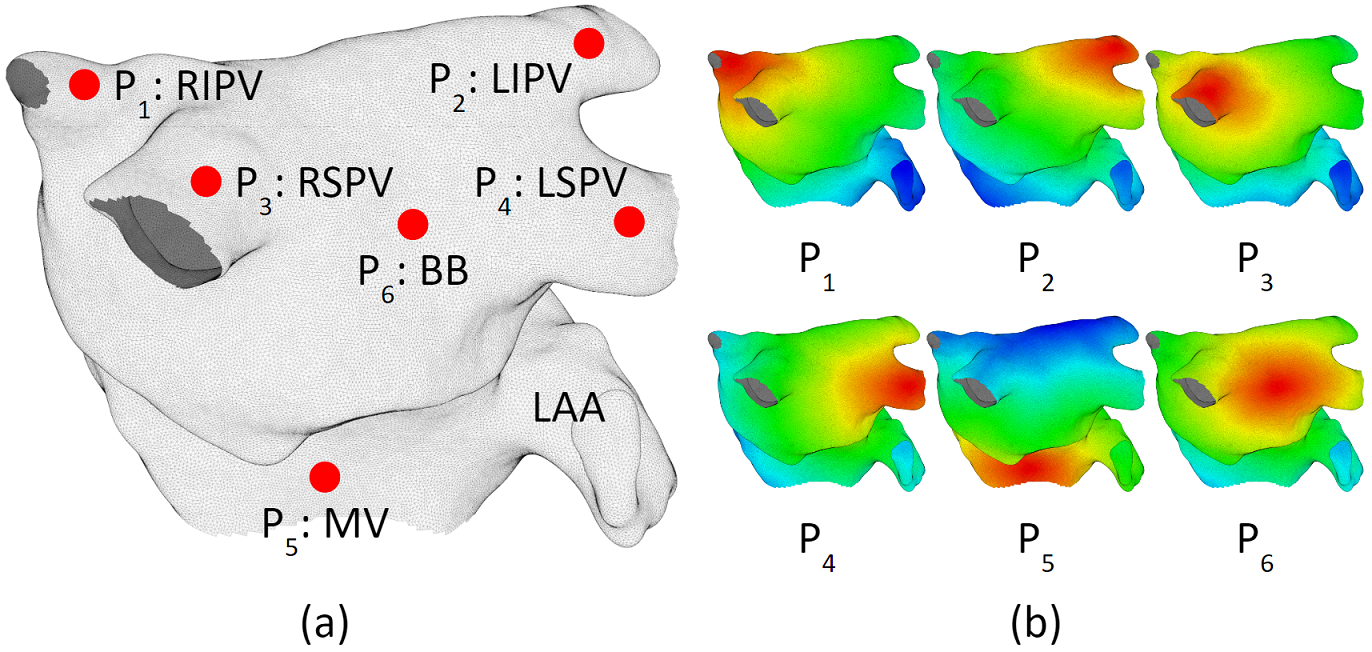}
\caption{Focal arrhythmia simulations setup. (LA$_3$ is shown here.) (a) Pacing sites for focal arrhythmia simulation. (b) Examples of activation maps obtained for each of the pacing locations. Red represents early activation and blue represents late activation. The focal source is located at the center of the red region. LAA: left atrial appendage.}
\label{fig:focal_setup}
\end{figure}

For rotor arrhythmia simulations, we use a method illustrated in Fig. \ref{fig:rotor_setup}, which produces a pair of rotors in a desired location. The method utilizes two electrical stimuli S$_1$ and S$_2$: S$_1$ stimulus is applied focally at the \ac{MV} region, initiating a propagating wave, S$_2$ stimulus depolarizes a larger area (magenta) and is applied after S$_1$ at the tail of the propagating action potential. The size and the position of S$_2$ defines the rotor location. To make the rotor stable in some of the simulations, we place small (3-5 mm) non-conducting fibrotic patches in the rotation centers as anchors \cite{Zemlin2012}.

\begin{figure}[!ht]
\centering
\includegraphics[width = 1\textwidth]{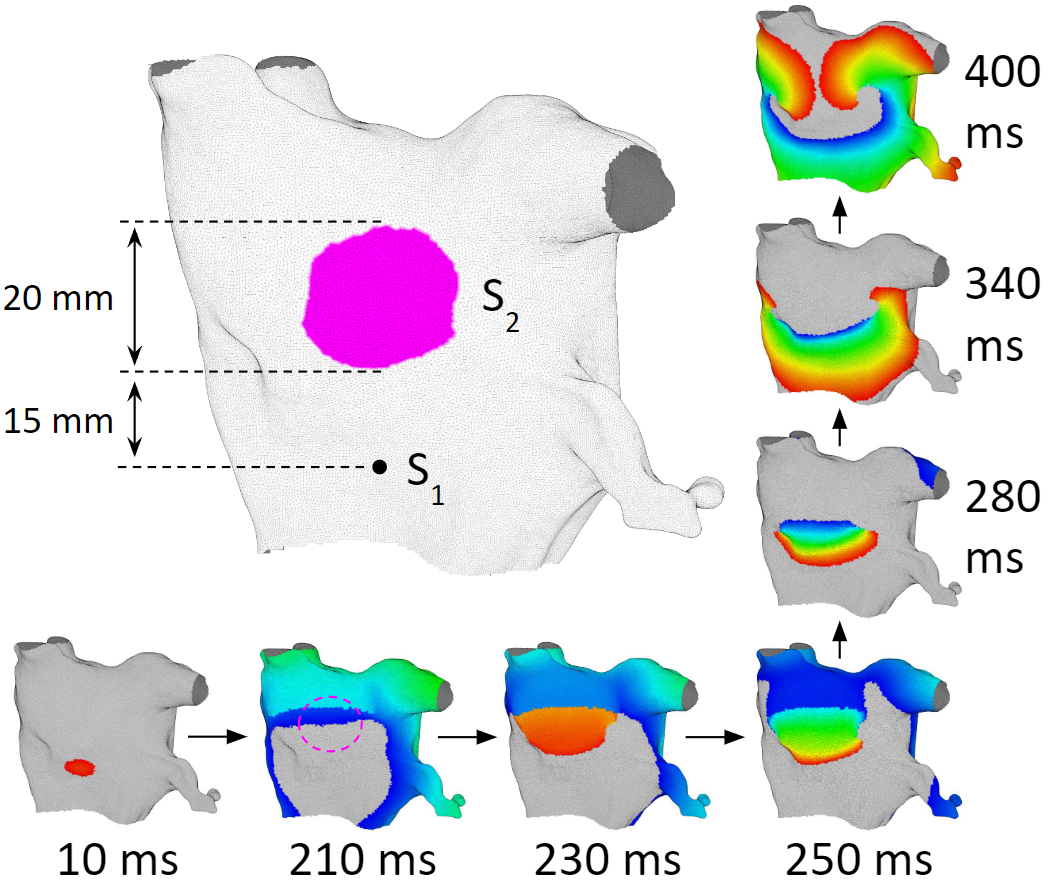}
\caption{Rotor arrhythmia simulations setup. (LA$_1$ is shown here.) S$_1$ and S$_2$ are the locations of stimuli. S$_1$ stimulus is applied focally at the \ac{MV} region, initiating a propagating wave (see the 10 ms snapshot). S$_2$ stimulus depolarizes a larger area (the magenta  region). It is applied after S$_1$ at the tail of the propagating action potential. 210 and 230 ms snapshots show the activation right before and after the S$_2$ stimulus, respectively. Magenta dashed circle in the 210 ms snapshot indicates the S$_2$ area. Snapshots 250-400 ms show the formation of a pair of rotors at the intersection of S$_2$ with the tail of S$_1$ wave.}
\label{fig:rotor_setup}
\end{figure}

\newpage

\section{Patient-specific model parameter tuning without myocardial fiber data}
\label{sec:tune diffusion}

The fiber-independent model uses an isotropic spatially uniform diffusion. The diffusion coefficient is tuned to approximate the effect of fiber organization on tissue conductivity. For tuning we used the P$_5$ activation map, generated using the respective fiber-inclusive model as the ground truth. The selection of the P$_5$, as opposed to other pacing locations, is intended to simulate the tuning procedure in clinical setting. P$_5$ is a point in the \ac{MV} region, where the coronary sinus catheter resides during ablation procedure. An electroanatomical map \cite{He2019} with pacing from the coronary sinus can be acquired at the beginning of the ablation procedure using Pentaray or Lasso catheters. 

We developed an optimization process to tune the diffusion coefficients $d$, the flowchart is illustrated in Figure \ref{fig:diffusion_tuning}. We process the patient data to obtain the true local activation times. Then we run a simulation with some initial guess diffusion values $d_0$, and process the simulation result to obtain the simulated local activation times.
These two sets of activation times are compared and the diffusion values are updated according to Equation \ref{eq:empirical_risk_minimization}.

\begin{equation}
\label{eq:empirical_risk_minimization}
d^*=\underset{d}{argmin}\ mean\left( \left| LAT_{simulation}(d)-LAT_{patient} \right| \right)
\end{equation}

New simulations are run and diffusion values are adjusted in each iteration. The tuning is done when the difference between patient activation time and simulation activation time is minimized. For faster tuning, we exploit the relation between conduction velocity and diffusion coefficient according to Equation \ref{eq:d_tuned_all_points}, which reduces the number of iterations needed to reach the optimum. In the equation, $N$ is the total number of nodes. $T_{0,n}$ and $T_{1,n}$ are the traveling times for the activation to get to this $n$-th node from the pacing site (P$_5$) in the fiber-independent and fiber-inclusive models, respectively.

\begin{figure}[!ht]
\centering
\includegraphics[width = 1\textwidth]{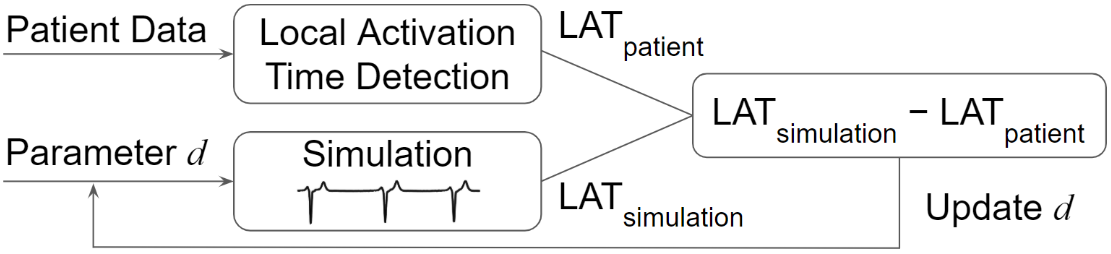}
\caption{Flowchart of diffusion tuning.}
\label{fig:diffusion_tuning}
\end{figure}

\begin{align}
\label{eq:d_tuned_all_points}
d_{tuned} = d_{0} \frac{1}{N} \sum_{n=1}^{N} \left(\frac{T_{0,n}}{T_{1,n}}\right)^2 
\end{align}

The formula \eqref{eq:d_tuned_all_points} is based on the notion that in any location the activation time is inversely proportional to conduction velocity ($ T\propto1/CV $) and that in the reaction diffusion systems the ratio of conduction velocities ${CV_0}$ and ${CV_1}$ evaluated at different diffusion coefficients $d_0$ and $d_1$ is equal to square root of the ratio of diffusion coefficients $CV_0/CV_1$ = $(d_0/d_1)^{1/2}$. Accordingly, the ratio of conduction times at a given location at different diffusion coefficients can be expressed as $T_0/T_1$ = $CV_1/CV_0$ = $(d_1/d_0)^{1/2}$. After a simple transformation one can obtain a formula linking the values of diffusion coefficients to activation times $d_1/d_0$ = $(T_0/T_1)^2$ that we utilize for tuning the fiber-independent model. 

The model tuning process takes approximately 18 seconds on a personal computer with an Intel Core i7-8700 CPU (3.20GHz) and an Nvidia GeForce GTX 1080 GPU. The required patient data (the electroanatomical map) is typically acquired at the beginning of an ablation procedure and can be done in 3 minutes. Therefore, the time to tune our model does not add significant burden in clinical practice, considering that an arrhythmia ablation procedure typically runs for 3 to 6 hours.

\newpage

\section{Performance evaluation metrics}
We calculate the absolute \ac{LAT} error and accuracy of every vertices between the two models. The error is defined according to \eqref{eq:error},
\begin{align}
\label{eq:error}
\text{Error} = \frac{1}{M} \sum_{m=1}^{M} \left| LAT_{1,m} - LAT_{2,m} \right|
\end{align}
where $M$ is the total number of vertices, and $LAT_{1,m}$ and $LAT_{2,m}$ are the \ac{LAT} values of the $m$-th vertex from fiber-inclusive and fiber-independent model respectively.

Accuracy of a vertex is defined according to \eqref{eq:accuracy},
\begin{align}
\label{eq:accuracy}
\text{Accuracy} = \left(1 - \frac{\left| LAT_{1} - LAT_{2} \right|}{T} \right) \times 100\%
\end{align}
where $LAT_{1}$ and $LAT_{2}$ are the \ac{LAT} values of a same vertex of the fiber-inclusive and fiber-independent model respectively. $T$ is calculated from fiber-inclusive model simulation data. For focal arrhythmia, $T=Range(LAT)$, it represents the time for the activation wave to travel across the entire left atrium. For rotor arrhythmia, $T=rotor\ cycle\ length$, it represents the time for the rotor to rotate 360 degrees.

\newpage

\section{Model performance on focal arrhythmias}
\begin{figure}[!ht]
\centering
\includegraphics[width = 1\textwidth]{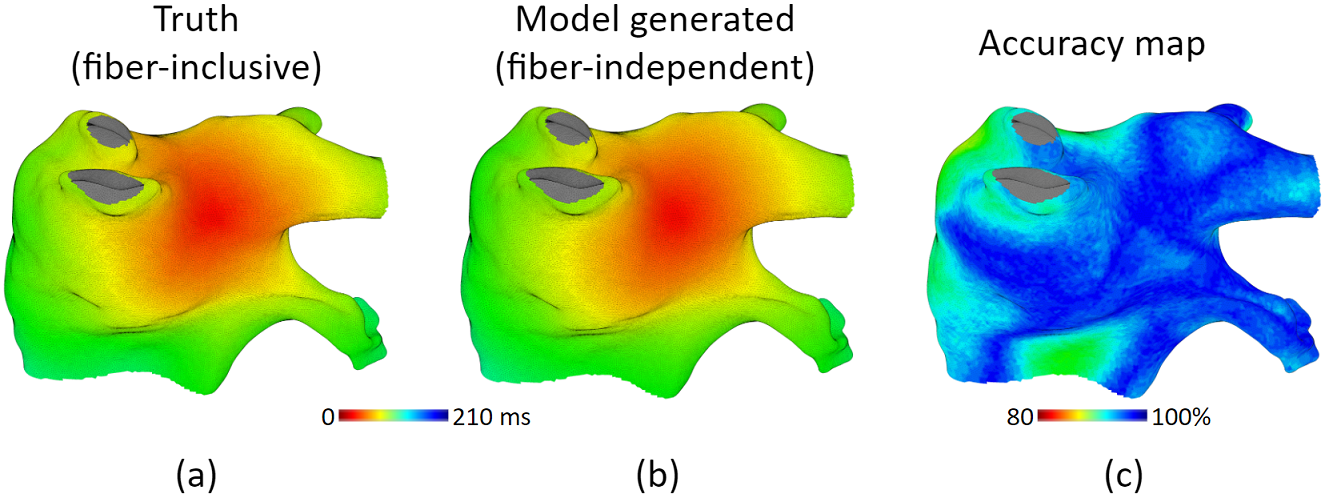}
\caption{Focal arrhythmia comparison. (a) is the ground truth, which includes fiber data in simulation. (b) is generated by our fiber-independent model, which no fiber data was given to the model, and the diffusion coefficients were tuned to compensate for the lack of fiber data. (c) is the accuracy map calculated according to Equation \eqref{eq:accuracy}, with 100\% meaning no difference between the two models.}
\label{fig:focal_lat_map_example}
\end{figure}

We generate 42 focal arrhythmias: 7 left atria each with 6 pacing scenarios. As illustrated in Figure \ref{fig:focal_lat_map_example}, our model can generate focal arrhythmia with high accuracy. Figure \ref{fig:focal lat map} in the appendix shows more detailed results. Quantitative results are in the following sections.

\subsection{Local activation time error}
Results are summarized in Table \ref{tb:lat err focal arrhythmia}. The overall error is 9$\pm$7 ms, or 96\% accuracy. Details of the accuracy is shown in Figure \ref{fig:focal accuracy}.  

\begin{table}[!ht]
    \centering
    \begin{tabular}{ c | c c c c c c c | c }
    \hline
 & LA$_1$ & LA$_2$ & LA$_3$ & LA$_4$ & LA$_5$ & LA$_6$ & LA$_7$ & \textbf{Avg} \\ \hline
P$_1$ & 4$\pm$3 & 6$\pm$6 & 6$\pm$5 & 11$\pm$7 & 5$\pm$4 & 10$\pm$8 & 14$\pm$10 & \textbf{8$\pm$7} \\ 
P$_2$ & 6$\pm$4 & 7$\pm$6 & 14$\pm$9 & 9$\pm$8 & 14$\pm$8 & 9$\pm$6 & 9$\pm$6 & \textbf{10$\pm$8} \\ 
P$_3$ & 8$\pm$6 & 4$\pm$3 & 7$\pm$5 & 9$\pm$8 & 13$\pm$7 & 10$\pm$7 & 9$\pm$9 & \textbf{9$\pm$7} \\ 
P$_4$ & 9$\pm$6 & 7$\pm$6 & 9$\pm$7 & 6$\pm$5 & 15$\pm$8 & 11$\pm$9 & 9$\pm$6 & \textbf{10$\pm$7} \\ 
P$_5$ & 5$\pm$3 & 5$\pm$4 & 10$\pm$8 & 5$\pm$4 & 6$\pm$4 & 8$\pm$5 & 7$\pm$5 & \textbf{7$\pm$5} \\ 
P$_6$ & 7$\pm$6 & 7$\pm$5 & 9$\pm$6 & 8$\pm$6 & 8$\pm$6 & 9$\pm$6 & 9$\pm$6 & \textbf{8$\pm$6} \\ \hline 
\textbf{Avg} & \textbf{6$\pm$5} & \textbf{6$\pm$5} & \textbf{9$\pm$7} & \textbf{8$\pm$7} & \textbf{10$\pm$8} & \textbf{9$\pm$7} & \textbf{10$\pm$8} & \textbf{9$\pm$7} \\ \hline 
    \end{tabular}
    \caption{Local activation time error (ms). Focal arrhythmia LAT errors between fiber-inclusive and fiber-independent models. Average error for all scenarios is 9$\pm$7 ms.}
    \label{tb:lat err focal arrhythmia}
\end{table}

\begin{figure}[!ht]
\centering
\includegraphics[width = 0.8\textwidth]{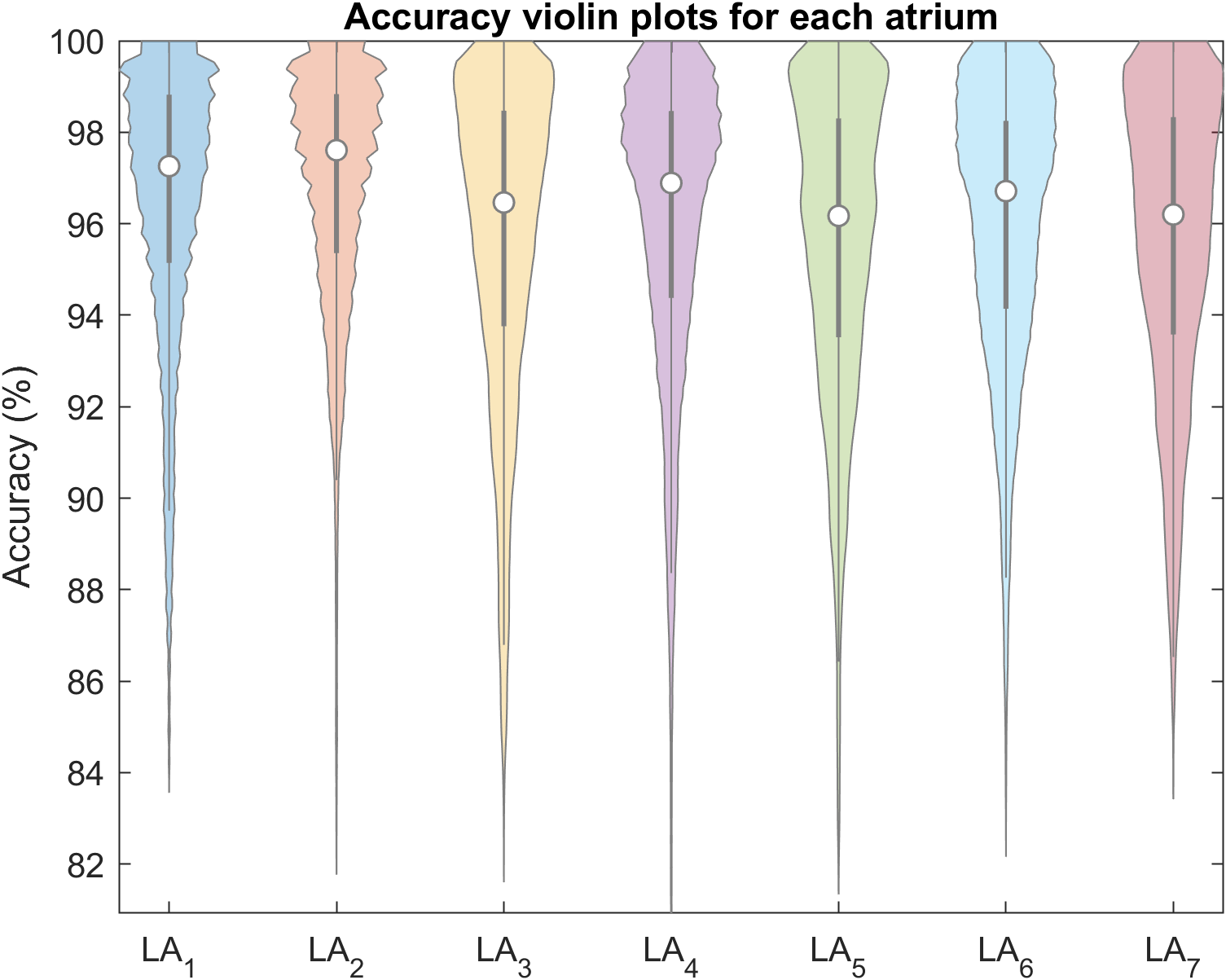}
\caption{Violin plots of focal arrhythmia local activation time accuracy of each left atrium. The overall average accuracy is 96\%.}
\label{fig:focal accuracy}
\end{figure}

\subsection{Latest activation location error}
One of the clinically relevant characteristics  that we use to evaluate the predictive capabilities of fiber-independent model is the accuracy of prediction of the regions with the latest activation (within 10 ms of the maximum \ac{LAT}.)

Figure \ref{fig:latest activation location} shows the location of these regions in $LA_6$ and $LA_2$ in fiber inclusive and fiber-independent models for focal arrhythmias $P_1$ and $LA_6$, respectively.   The Euclidean distance between fiber-inclusive and fiber-independent model's latest activation locations are summarized in Table \ref{tb:latest activation location}. The average distance is 7.5 mm (the size of the left atrium is 106 mm $\times$118 mm$\times$116 mm, 7.5 is 6.4\% of 118). However, we observe in some cases that the fiber-independent model fail to predict all of the latest activation locations, see Figure \ref{fig:latest activation location}(b). This happened 4 times, or 9.5\% of all 42 scenarios. 

The Euclidean distance between fiber-inclusive and fiber-independent model's latest activation locations are summarized in Table \ref{tb:latest activation location}. The average distance is 7.5 mm (the size of the left atrium is 106 mm $\times$118 mm$\times$116 mm, 7.5 is 6.4\% of 118). However, we observe in some cases that the fiber-independent model fail to predict all of the latest activation locations, see Figure \ref{fig:latest activation location}(b). This happened 4 times, or 9.5\% of all 42 scenarios. 

\begin{table}[!ht]
    \centering
    \begin{tabular}{ c | c c c c c c c | c }
    \hline
 & LA$_1$ & LA$_2$ & LA$_3$ & LA$_4$ & LA$_5$ & LA$_6$ & LA$_7$ & \textbf{Avg} \\ \hline
P$_1$ & 0.5 & 5.8 & 0.5 & 4.8 & 7.9 & 4.8 & 1.1 & \textbf{3.6} \\ 
P$_2$ & 7.1 & 12.3 & 4.7 & 5.8 & 7.4 & 3.9 & 5.7 & \textbf{6.7} \\ 
P$_3$ & 6.1 & 7.7 & 3.7 & 3.2 & 3.4 & 5.3 & 16.3 & \textbf{6.5} \\ 
P$_4$ & 2.6 & 8.8 & 3.7 & 31.9 & 5.9 & 5.2 & 14.1 & \textbf{10.3} \\ 
P$_5$ & 18.2 & 8.0 & 21.4 & 33.3 & 9.1 & 10.6 & 1.7 & \textbf{14.6} \\ 
P$_6$ & 2.5 & 2.1 & 3.2 & 7.0 & 0.4 & 7.4 & 1.8 & \textbf{3.5} \\ \hline 
\textbf{Avg} & \textbf{6.2} & \textbf{7.4} & \textbf{6.2} & \textbf{14.3} & \textbf{5.7} & \textbf{6.2} & \textbf{6.8} & \textbf{7.5} \\ \hline 
    \end{tabular}
    \caption{Latest activation location difference (mm). Focal arrhythmia latest activation location difference between fiber-inclusive and fiber-independent models. The overall average is 7.5 mm.}
    \label{tb:latest activation location}
\end{table}

\begin{figure}[!ht]
\centering
\includegraphics[width = 0.8\textwidth]{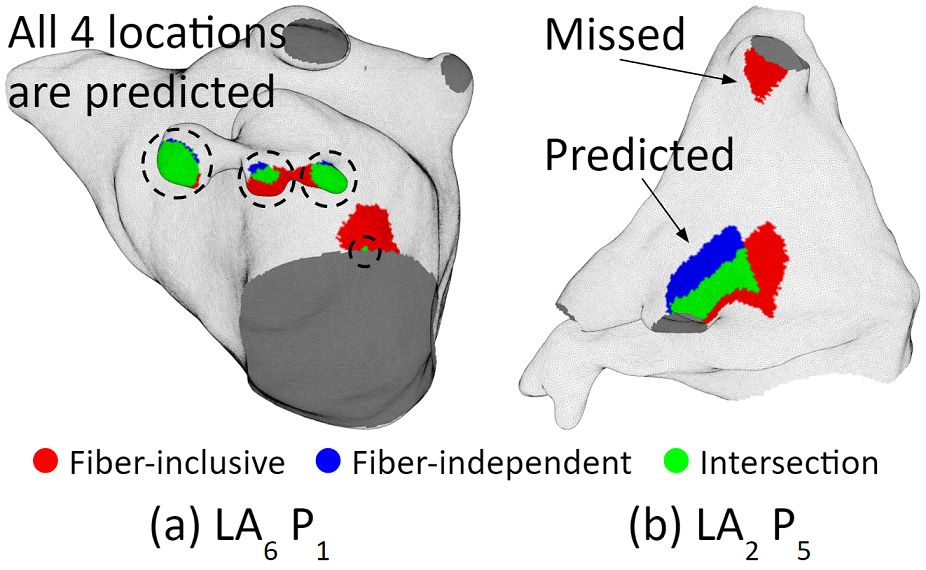}
\caption{Focal arrhythmia latest activation location plots. Red and blue represent results of fiber-inclusive and fiber-independent model respectively, and green is the overlapping area. (a) Latest activation locations of P$_1$ focal arrhythmia on left atrium 6. We can see that the regions of the two models overlap well, resulting a small error. (b) The case of P$_5$ focal arrhythmia on left atrium 2. We can see that the error is small in the red-blue-green overlapping region, however, the fiber-independent model did not predict the upper red region. This happened 4 times, accounted for 9.5\% of all 42 scenarios.}
\label{fig:latest activation location}
\end{figure}

\newpage

\section{Model performance on rotor arrhythmias}
\begin{figure}[!ht]
\centering
\includegraphics[width = 1\textwidth]{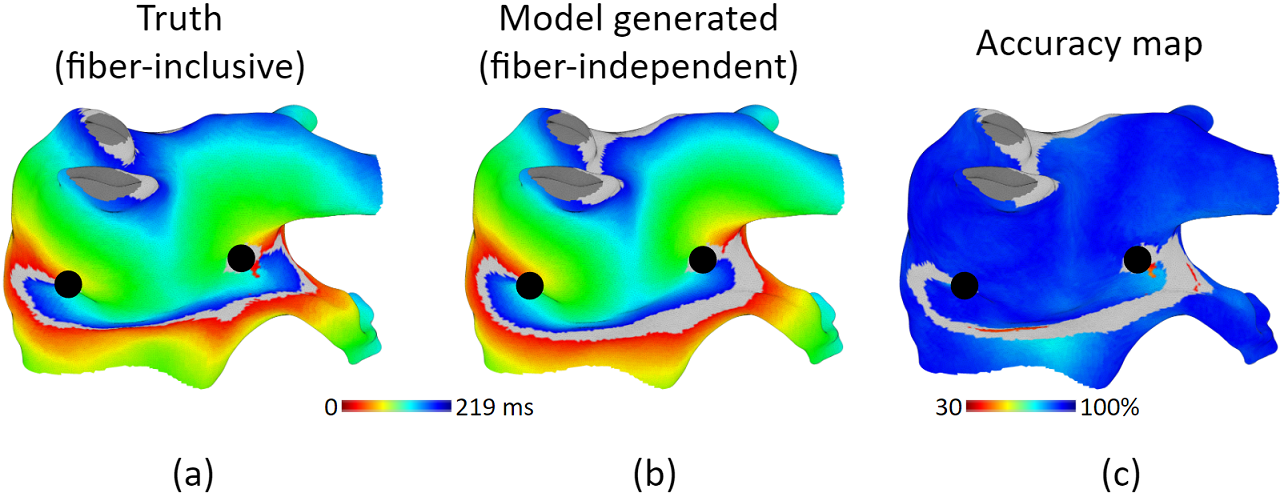}
\caption{Rotor arrhythmia comparison. A pair of rotors are generated on each left atrium. These rotors were made stable by placing small non-conducting patches as anchors (marked with black dots) at the rotation center, which allowed us to compare the activation patterns of the rotor arrhythmias between models. (a) and (b) show local activation time maps of fiber-inclusive and fiber-independent models respectively. (c) shows the accuracy maps calculated according to Equation \eqref{eq:accuracy}, with 100\% meaning no difference between the two models.}
\label{fig:rotor_lat_map_example}
\end{figure}

Atrial flutter, macro re-entry and atrial fibrillation can have rotating activation waves, which can be represented as rotors. We create stable rotors on the 7 left atria. Figure \ref{fig:rotor_lat_map_example} shows the comparisons (more details in Figure \ref{fig:rotor lat map} in the appendix). We can see the activation wavefronts are slightly different between the two models, however, the over all activation patterns are similar. Quantitative analysis are shown in Table \ref{tb:lat err rotor arrhythmia}. The overall \ac{LAT} error is 14$\pm$16 ms, or 93\% accuracy.

\begin{table}[!ht]
    \centering
    \begin{tabular}{ c c c c c c c | c }
    \hline
LA$_1$ & LA$_2$ & LA$_3$ & LA$_4$ & LA$_5$ & LA$_6$ & LA$_7$ & \textbf{Avg} \\ \hline
8$\pm$10 & 13$\pm$16 & 14$\pm$14 & 12$\pm$11 & 14$\pm$18 & 15$\pm$15 & 23$\pm$20 & \textbf{14$\pm$16} \\ \hline 
    \end{tabular}
    \caption{Local activation time error (ms). Rotor arrhythmia LAT errors between fiber-inclusive and fiber-independent models. Average error for all scenarios is 14$\pm$16 ms.}
    \label{tb:lat err rotor arrhythmia}
\end{table}

\begin{figure}[!ht]
\centering
\includegraphics[width = 0.8\textwidth]{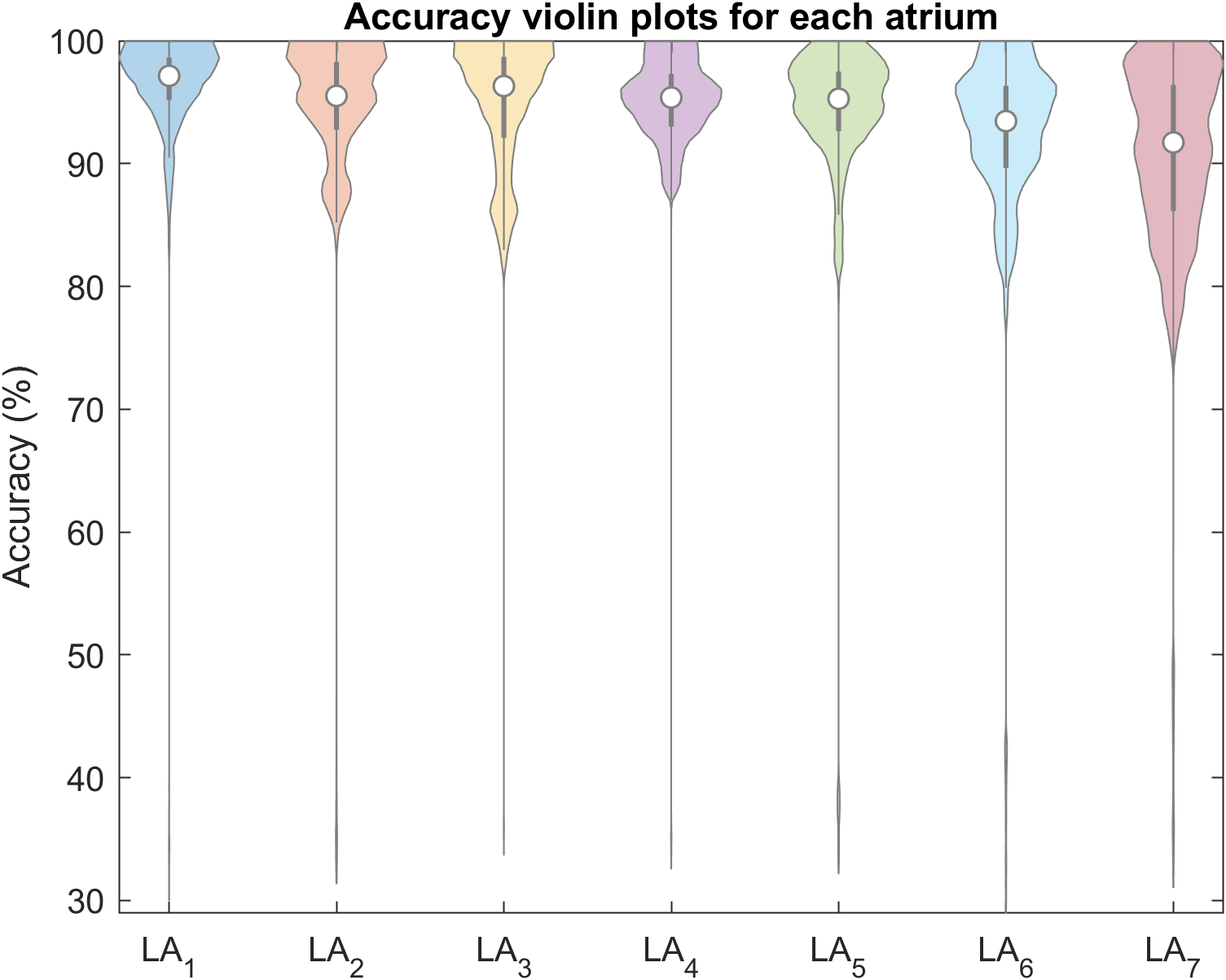}
\caption{Violin plots of rotor arrhythmia local activation time accuracy of each left atrium. The overall average accuracy is 93\%.}
\label{fig:rotor accuracy}
\end{figure}

Details of the accuracy is shown in Figure \ref{fig:rotor accuracy}. We can see that the majority accuracy are higher than 90\%. However, we notice that there are a small amount of accuracy in the 30\%-40\% range. These low accuracy regions are the red regions in the accuracy maps of Figure \ref{fig:rotor lat map} row 3 (this figure is in the appendix). For example, on LA$_6$, the low accuracy region is marked by an arrow, it is located in between the wave front (red, marked by an arrow) and wave tail (blue, marked by an arrow). Wave front has \ac{LAT} value of 0 ms, wave tail has \ac{LAT} value of 219 ms, these two values are far apart, thus small deviation in the wave front / wave tail location can result in large error, or low accuracy.

To observe if model performance is different for different rotor locations, we create two more rotor arrhythmias on LA$_{6}$, make a total of three different rotor arrhythmias for this atrium. LA$_{6}$ is chosen, because in Table \ref{tb:lat err rotor arrhythmia}, it has an average performance among the seven atria. As shown in Figure \ref{fig:different_rotor_location}, the \ac{LAT} errors are similar: Left column \ac{LAT} error is 15$\pm$15 ms (93\% accuracy), middle column is 15$\pm$12 ms (94\% accuracy), right column is 12$\pm$13 ms (94\% accuracy).

\begin{figure}[!ht]
\centering
\includegraphics[width = 1\textwidth]{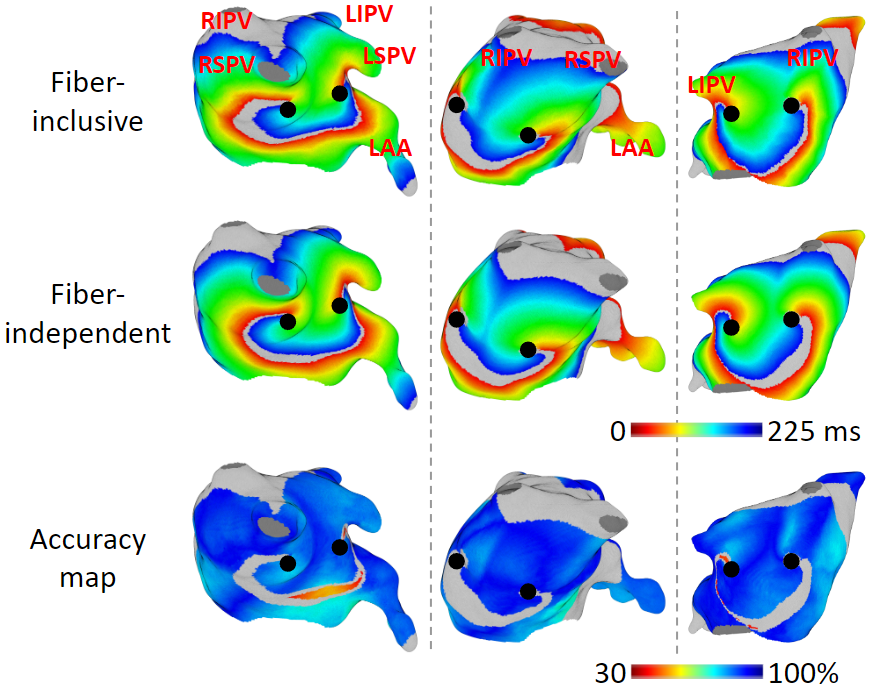}
\caption{Local activation time maps of different rotor arrhythmias on LA$_{6}$. All atria here are the same one but with different view angles. The activation patterns between the two models are similar.}
\label{fig:different_rotor_location}
\end{figure}

We also find there are slight differences in rotor cycle lengths between the two models. Details are summarized in Table \ref{tb:rotor cycle length}.

\begin{table}[!ht]
    \centering
    \begin{tabular}{ c | c c c c c c c | c }
    \hline
         & LA$_1$ & LA$_2$ & LA$_3$ & LA$_4$ & LA$_5$ & LA$_6$ & LA$_7$ & \textbf{Avg} \\ \hline
CL$_1$ & 210 & 223 & 217 & 216 & 233 & 214 & 217 & \textbf{219} \\ 
CL$_2$ & 212 & 215 & 219 & 213 & 227 & 219 & 217 & \textbf{217} \\ \hline 
    \end{tabular}
    \caption{Rotor average cycle length (ms). Rotor average cycle length. CL$_1$: fiber-inclusive rotor cycle length. CL$_2$: fiber-independent rotor cycle length.}
    \label{tb:rotor cycle length}
\end{table}

\subsection{Stable vs meandering rotors}
We conduct experiments on a 2D plane with different fiber orientations as depicted in Figure \ref{fig:rotor slab experiment}. The \ac{PS} location (or rotor center location) plots indicate that rotors are relatively stable in cases (a)-(d), while they exhibit more meandering in cases (e) and (f). (Please refer to Appendix \ref{app:ps_detection} for details of \ac{PS} detection.) Studies have shown that fiber gradients can cause rotor to meander \cite{Rogers1994}, which explains why rotors move more in case (f) where the curved fibers have larger gradients.

\begin{figure}[!ht]
\centering
\includegraphics[width = 1\textwidth]{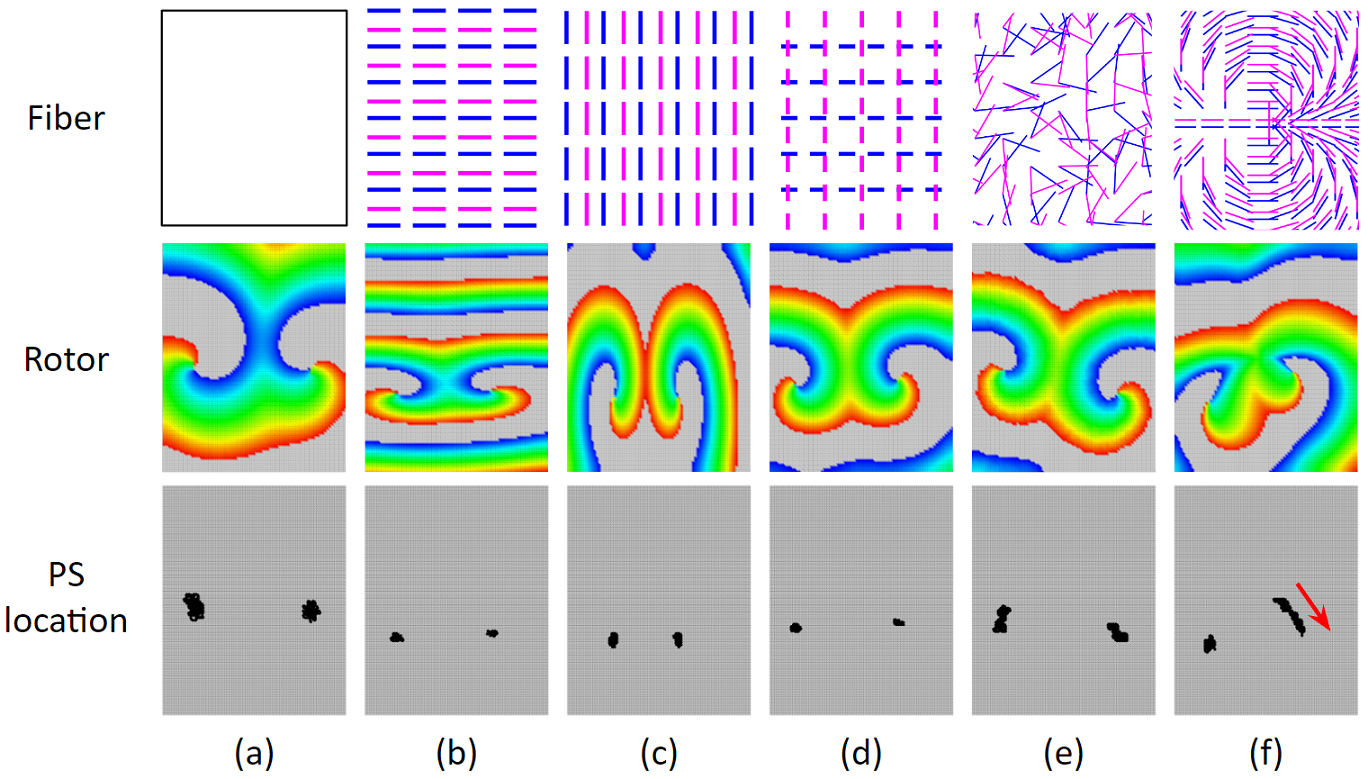}
\caption{Rotor arrhythmia experiments on a slab with different fiber organizations. Slab size is 80 mm $\times$ 100 mm $\times$ 2 mm. Each simulation produced 4 seconds of rotor arrhythmia. Row 1 are fibers: Blue represents endocardium fiber and magenta represents epicardium fiber. Column (a) has no fiber; column (b) fibers are in x direction; column (c) fibers are all in y direction; For column (d), endocardium fiber are in the x direction, epicardium fibers are in the y direction; column (e) has random fibers; and column (f) has curved fibers. Row 2 shows activation movie screenshots at time of 2600 ms. Row 3 shows the trajectories of the phase singularities (PS). Location tracking of the PS begins after the rotors finished their first rotation, ends 3 seconds or about 17 rotations later. We can see that PS locations moves the most in (f) because of the continuously changing fiber gradients.}
\label{fig:rotor slab experiment}
\end{figure}

In the left atrium, the combination of fiber orientations and atrial geometry curvature may create large fiber gradients that can have a significant impact on rotor meandering. We run un-anchored rotor arrhythmias on all 7 left atria. The average distances of the nearest fiber-inclusive and fiber-independent rotor pairs range from 5 to 19 mm, and rotors often meanders further away as time progresses. Some of the \ac{PS} trajectories of a 3-second interval are shown in Figure \ref{fig:unanchored rotor}. We can see that un-anchored rotors can meander long distances in fiber-inclusive models, but un-anchored rotors are still quite stable in fiber-independent models. This shows that fiber-independent models have a weak predictability on un-anchored rotors. However, clinically observed rotors rarely last more than a few rotations \cite{Fedorov2018}, and research found that stable rotors are ablation targets \cite{Narayan2013, Krummen2015, Kawata2013, Calvo2017}. 

\begin{figure}[!ht]
\centering
\includegraphics[width = 1\textwidth]{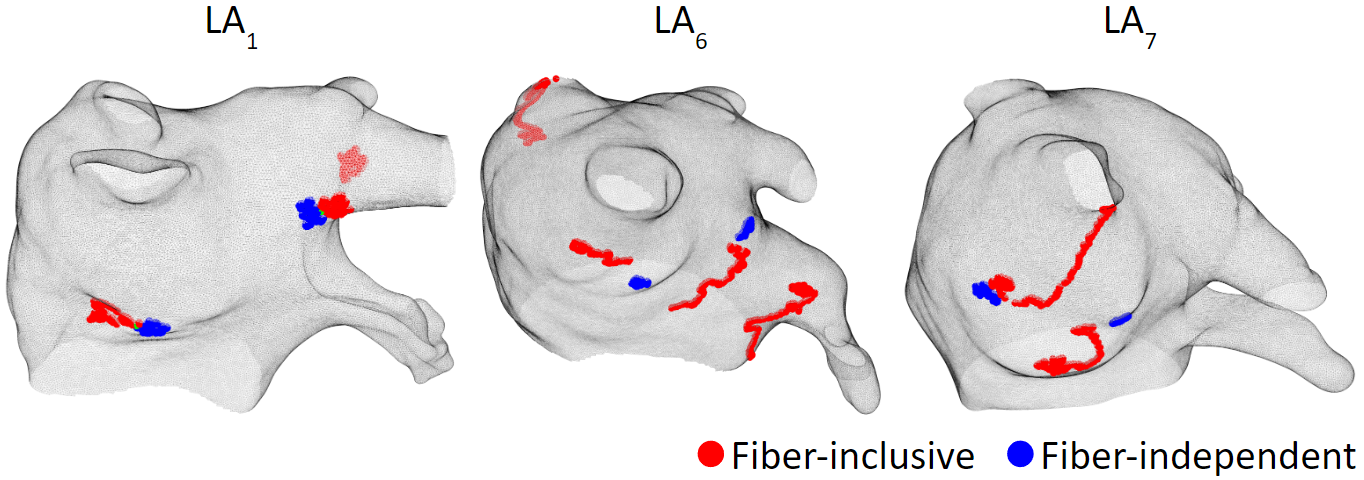}
\caption{Phase singularity locations for un-anchored rotors. The initial location of the rotors of the two models are similar. But as time progresses, rotors can meander away in fiber-inclusive models. However, un-anchored rotors in fiber-independent models are still stable.}
\label{fig:unanchored rotor}
\end{figure}

\newpage

\section{Limitation}
We find that our fiber-independent left atrium model can accurately reproduce patient-specific focal and stable rotor arrhythmias. However, there are limitations to this finding. 

We do not consider scars in this study, which could play an important role in atrial fibrillation dynamics \cite{Gonzales2014}. However, since 79\% of first time ablation patients do not have scars \cite{Verma2005}, our model can still be helpful for clinical practice. 

The model is not equipped to predict the dynamics of meandering rotors. However, this limitation may not be as critical in clinical setting, given that only stable rotors are the primary ablation targets.

Among the several parameters, we only tune the diffusion coefficient. For most of the tachycardia, flutter, and macro re-entry cases, the action potential parameters we choose will work well. However, for more complex rhythms such as atrial fibrillation, due to its short cycle length, an accurate tuning of the action potential parameters may be crucial. 

The Mitchell-Schaeffer model we use is not a detailed ionic model, and it may not be a good choice to model complex rhythms such as atrial fibrillation. Still such a two-component model is good for modeling periodic arrhythmias. 

It is well established that more detailed bi-domain models are required for accurate simulation of electrical activity in the immediate vicinity of the stimulating electrodes and for modelling electrical defibrillation \cite{Roth2021}. However, we utilized the computationally more efficient mono-domain model. With regard to accuracy, the bi-domain models have no advantages over mono-domain models for simulating action potential propagations \cite{Potse2006}.

We simplify fiber organization into only two layers. The real left atrium has many more layers, and the number of layers also varies in different regions, as does the atrial thickness. If more layers were incorporated, we would need to study if the effects of fibers would become stronger.

\newpage

\section{Conclusion}
We show that 1) fiber-independent left atrium model with tuned conduction can produce accurate activation maps of focal and stable rotor arrhythmias. 2) The model can be tuned to be patient-specific in a less than 4 minutes and can run in real-time to identify potential ablation targets during ablation procedure.

So far, we have explained the details of heart modeling, and showed that our model can accurately generate patient-specific arrhythmia simulations. To put it to clinical use, we need to work on clinical data. That will be covered in the following chapters.

\chapter{\MakeUppercase{Clinical data processing: atrium mesh and electrogram}}
\label{chapter:patient_data_processing}

\section{Overview}
In this chapter, we will describe the tools, user interfaces, and algorithms we developed for processing clinical data. As shown in Figure \ref{fig:patient_data_processing_overview}, during left atrium arrhythmia ablation procedure, the physician maneuvers mapping catheters inside the left atrium to collect endocardium electrograms, which are processed and displayed as electroanatomical maps on the left atrium mesh. With the help of these electroanatomical maps, physician identifies arrhythmia sources and apply radiofrequency ablations to stop abnormal activations and restore healthy sinus rhythm. 

\begin{figure}[!ht]
\centerline{\includegraphics[width = 1\textwidth]{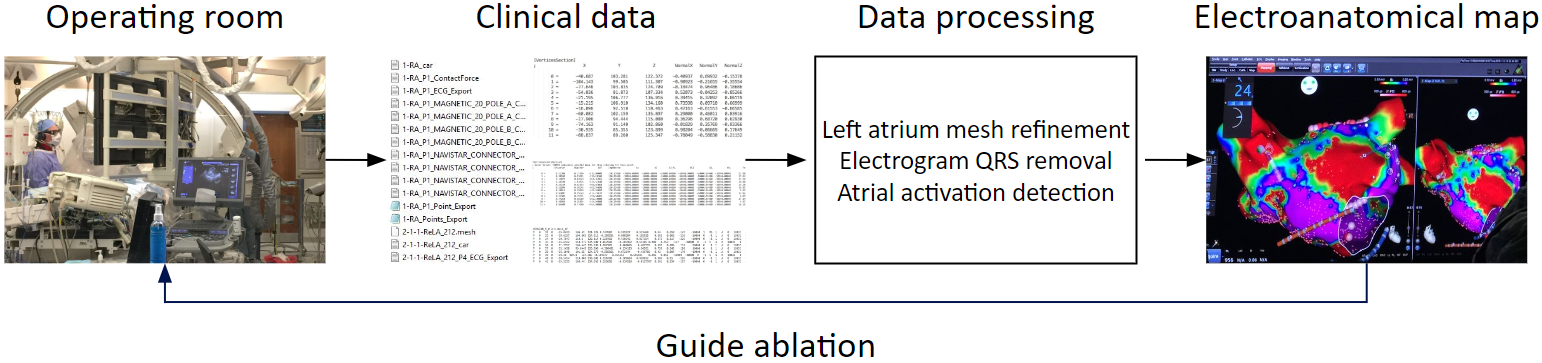}}
\caption{From clinical data to electroanatomical maps for guiding left atrium arrhythmia ablation procedure.}
\label{fig:patient_data_processing_overview}
\end{figure}

Clinical data are exported from the mapping system which include 3D triangular mesh, catheter locations, electrograms, and more. The data are in proprietary format. We need to decipher them. In addition, atrium mesh and electrograms need further processing. We will explain the details in the rest of this chapter.

\newpage

\section{Clinical data contents}
The clinical arrhythmia ablation system, Carto3 System, stores clinical data during the operation. Data are recorded via the electrodes on the catheters, which are threaded through a vein to the heart. There are mainly two types of catheters: mapping catheters and ablation catheters. Figure \ref{fig:catheter} shows examples of some commonly used catheters: Lasso, Pentaray and Decapolar catheters are mapping catheters that used for collecting electrograms at different locations of the left atrium endocardium; Ablation catheter is used to create tissue lesions that block abnormal electrical conduction.

\begin{figure}[!ht]
\centerline{\includegraphics[width = 1\textwidth]{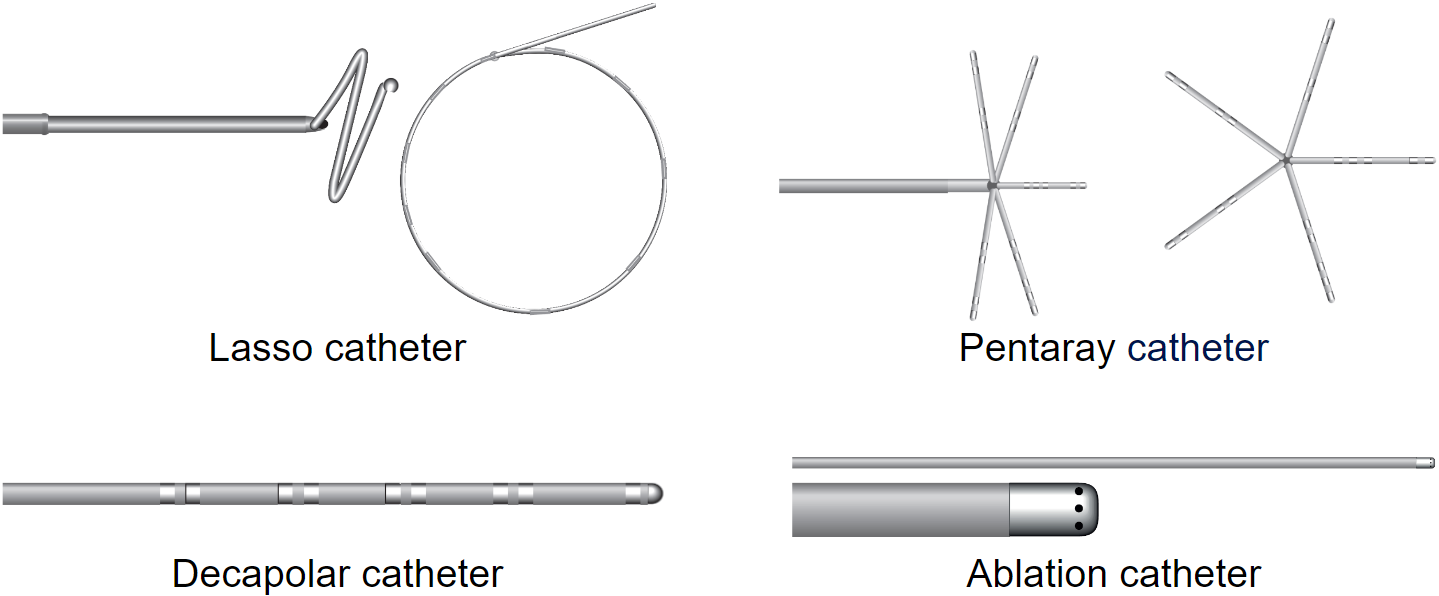}}
\caption{Commonly used catheters. Lasso, Pentaray and Decapolar catheters are mapping catheters that record endocardium electrograms. Ablation catheters deliver radiofrequency energy to kill abnormal myocardial cells, this results in tissue lesions that block abnormal electrical conduction. The bright dots on the catheter are electrodes: the Lasso catheter has 20 unipolar and 19 bipolar electrodes; the Pentaray catheter has 20 unipolar and 15 bipolar electrodes; the Decapolar catheter has 10 unipolar and 5 bipolar electrodes.}
\label{fig:catheter}
\end{figure}

Catheters Data are grouped into electroanatomical maps. One map contains a left atrium 3D triangular mesh and multiple segments of catheter electrogram recordings. Figure \ref{fig:create_mesh} shows the mesh creation process: At the beginning of the arrhythmia ablation procedure, the physician maneuvers the mapping catheter around the inside of the left atrium. The outer bound of the catheter trajectories becomes the atrium geometry. 

\begin{figure}[!ht]
\centerline{\includegraphics[width = 1\textwidth]{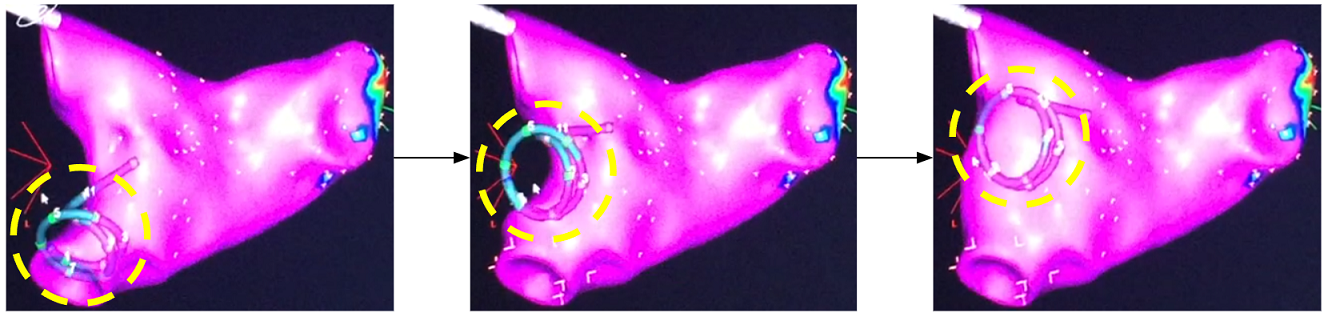}}
\caption{Creation of the left atrium 3D triangular mesh. The physician maneuvers the mapping catheter (marked inside the yellow dashed circle) around the inside of the left atrium while the catheter's outer locations becomes the atrium geometry. (Note: A Lasso catheter is shown in this example. Pentaray catheter can also be the mapping catheter for atrium mesh creation.)}
\label{fig:create_mesh}
\end{figure}

These data can be exported as one compressed file (.zip format) that contains a variety of sub-files as summarized in Figure \ref{fig:available_clinical_data}. These sub-files are the left atrium 3D triangular mesh, catheter locations, electrograms, and other files as shown in Figure \ref{fig:data_file}. For one patient, the common size of the compressed file is 2 GB, which will extract to 20 GB. Details of the sub-files we utilize are explained in the following list:

\begin{figure}[!ht]
\centerline{\includegraphics[width = 0.8\textwidth]{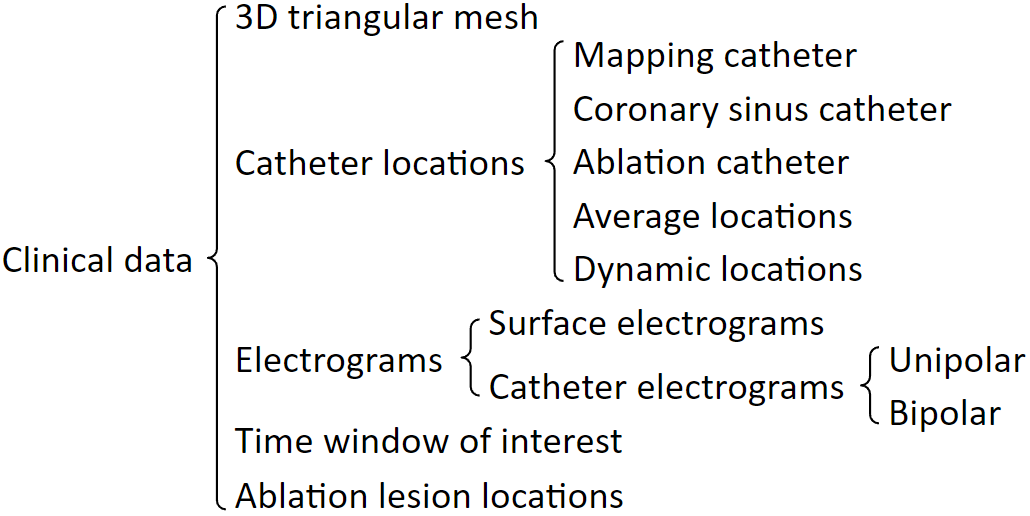}}
\caption{Available clinical data. }
\label{fig:available_clinical_data}
\end{figure}

\begin{figure}[!ht]
\centerline{\includegraphics[width = 1\textwidth]{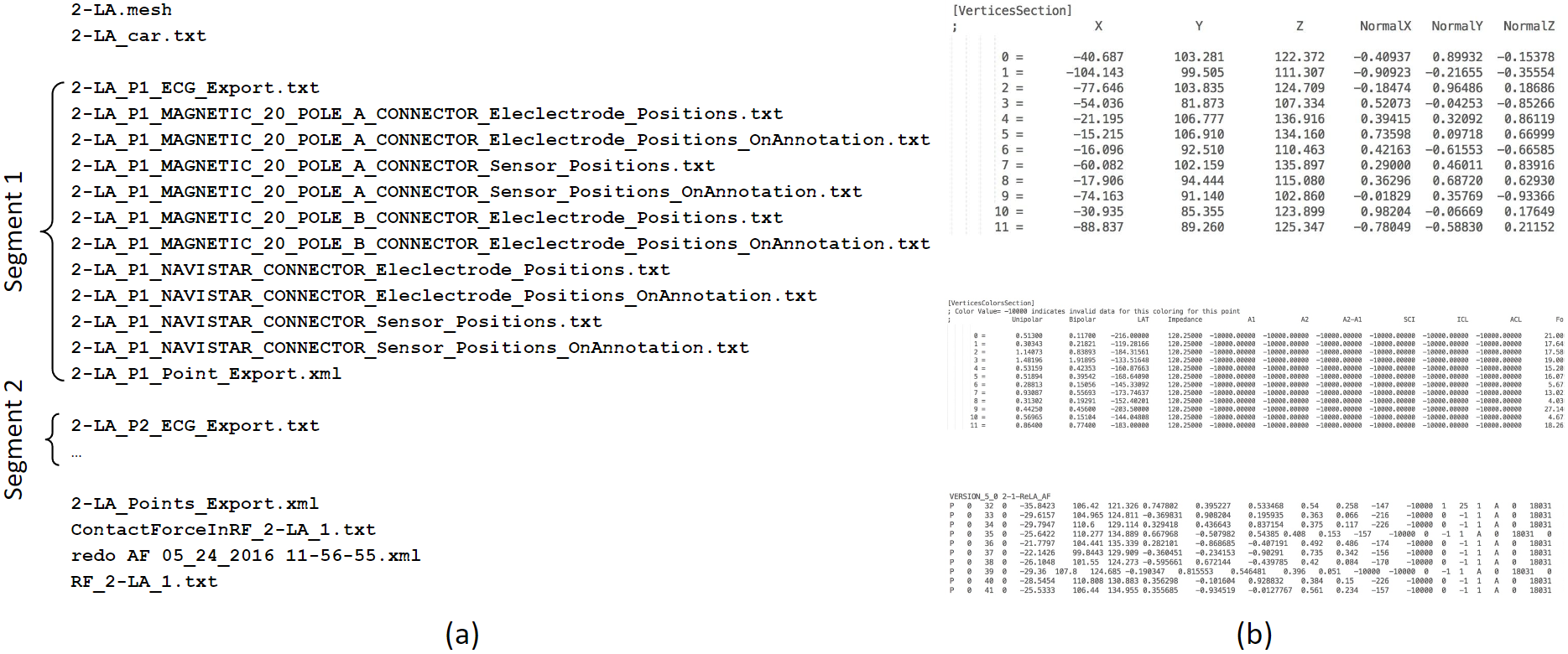}}
\caption{Clinical data files. (a) The contents of the compressed file. All associated files of the same electroanatomical map are starting with the same name prefix, and the same recording segments are having the same name prefix. (b) Examples of the contents of some sub-files.}
\label{fig:data_file}
\end{figure}

\vspace{-15pt}
\begin{itemize} [itemsep=-10pt]
\item "x.mesh" contains atrium geometry mesh information: vertices xyz coordinates, triangle lists, vertices normal, triangles normal, etc.
\item "x\_car.txt" contains all electrode sample points' xyz coordinates from all the catheter recording segments.
\item "x\_Px\_ECG\_Export.txt" contains electrogram files of a recording segment: surface electrogram, unipolar and bipolar mapping catheter electrograms, unipolar and bipolar coronary sinus catheter electrograms.
\item "x\_Px\_MAGNETIC\_20\_POLE\_A\_CONNECTOR\_Eleclectrode\_Positions.txt" or \\"x\_Px\_MAGNETIC\_20\_POLE\_B\_CONNECTOR\_Eleclectrode\_Positions.txt" contains the mapping catheter or the coronary sinus catheter electrodes' dynamic locations of a recording segment. 
\item "x\_Px\_MAGNETIC\_20\_POLE\_A\_CONNECTOR\_Eleclectrode\_Positions\_OnAnnotation.txt" or \\"x\_Px\_MAGNETIC\_20\_POLE\_B\_CONNECTOR\_Eleclectrode\_Positions\_OnAnnotation.txt" contains the mapping catheter or the coronary sinus catheter electrodes' average location of a recording segment.
\item "x\_Px\_NAVISTAR\_CONNECTOR\_Eleclectrode\_Positions.txt" contains the ablation catheter's dynamic locations of a recording segment.
\item "x\_Px\_NAVISTAR\_CONNECTOR\_Eleclectrode\_Positions\_OnAnnotation.txt" contains the ablation catheter's average location of a recording segment.
\item "x\_Px\_Point\_Export.xml" contains the time window of interest for the electrograms of a recording segment.
\item Inside folder "VisiTagExport", file "Sites.txt" contains the xyz coordinates of the ablation lesions.
\end{itemize}

The "MAGNETIC\_20\_POLE\_A" and "MAGNETIC\_20\_POLE\_B" mentioned above are catheter wiring connection boxes. Either connection box can accept mapping catheter or coronary sinus catheter, and different physicians can have their own preferences in these hardware connections. 

\newpage

\section{Left atrium mesh processing}
\label{sec:mesh_processing}

The clinical left atrium 3D triangular mesh can have a lot of problems such as: holes, floating faces, bridges, deep concave regions, etc. We developed an automated program to fix these problems and refine the geometry. The main idea is illustrated in Figure \ref{fig:refine_geometry}: First, transform the mesh into volumetric image; Next, slice this volumetric image in the xy, xz, and yz planes and fill the holes in these slices; Then delete the isolated voxels in this "slices hole-filled" volumetric image; Last, convert the surface of this volumetric image to a 3D triangular mesh. This automated program will result in a refined mesh which has smooth surface with the preserved left atrium shape constructed with uniform triangles.

\begin{figure}[!ht]
\centering
\includegraphics[width=1\textwidth]{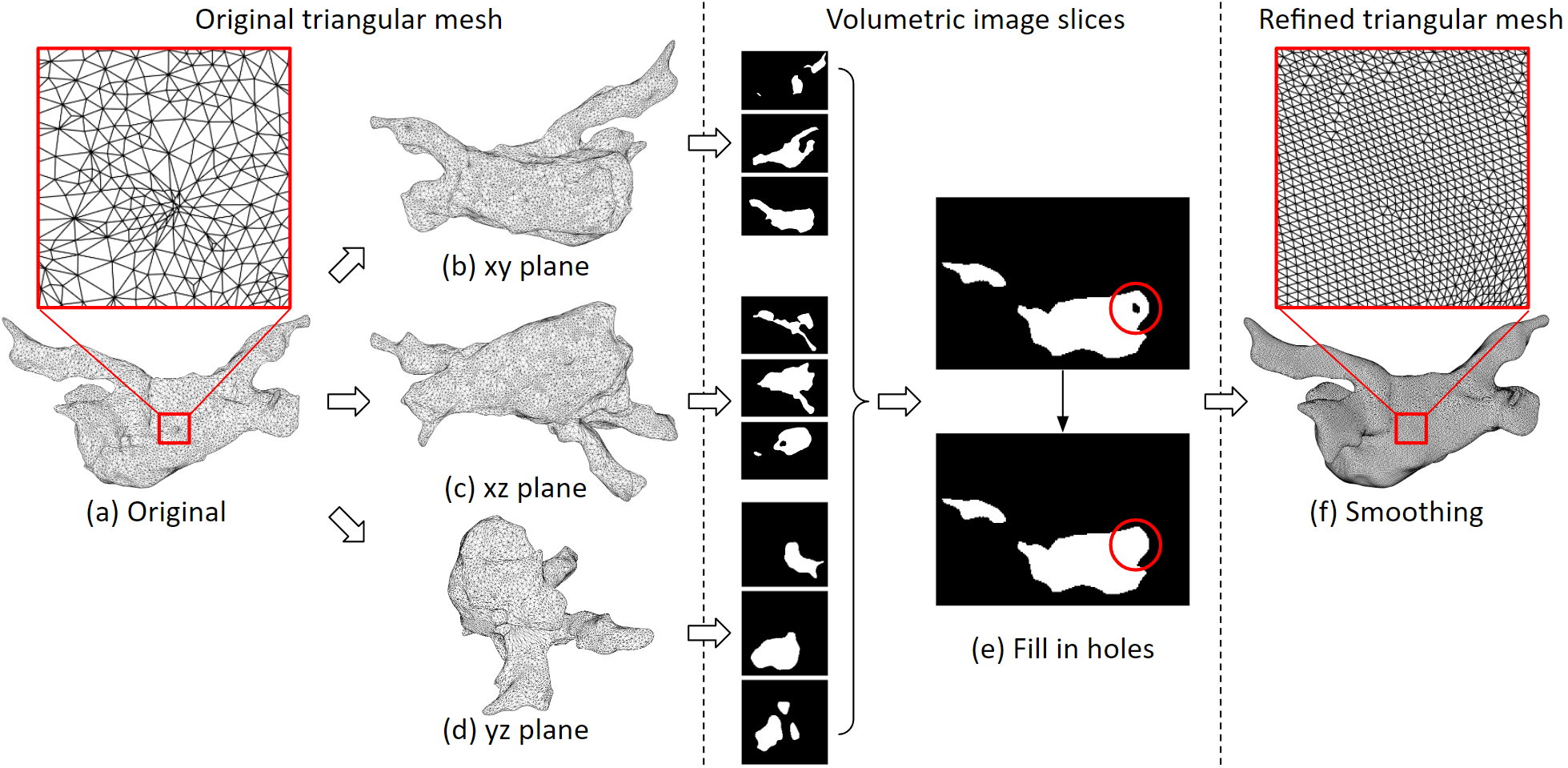}	
\caption{Fix the mesh problems and refine the geometry. (a) Clinical left atrium 3D triangular mesh. It can have many defects and the triangles are not uniform. (b), (c), (d) The top down views of the xy, xz, and yz planes. (e) The mesh is converted into a volumetric image, and the holes of the slices are filled. Here, a hole is marked inside the red circle. (f) After converting the surface of the volumetric image to triangular mesh and apply a Laplace smoothing, this refined triangular mesh has smooth surface with uniform triangles.}
\label{fig:refine_geometry}
\end{figure}

A user interface is also built to further edit the mesh. To cut out the pulmonary veins and the mitral valve, user first select the region to be removed, then the program will delete the selected vertices and triangles. Figure \ref{fig:cut_holes} shows an example of cutting holes in the mesh.

\begin{figure}[!ht]
\centering
\includegraphics[width=1\textwidth]{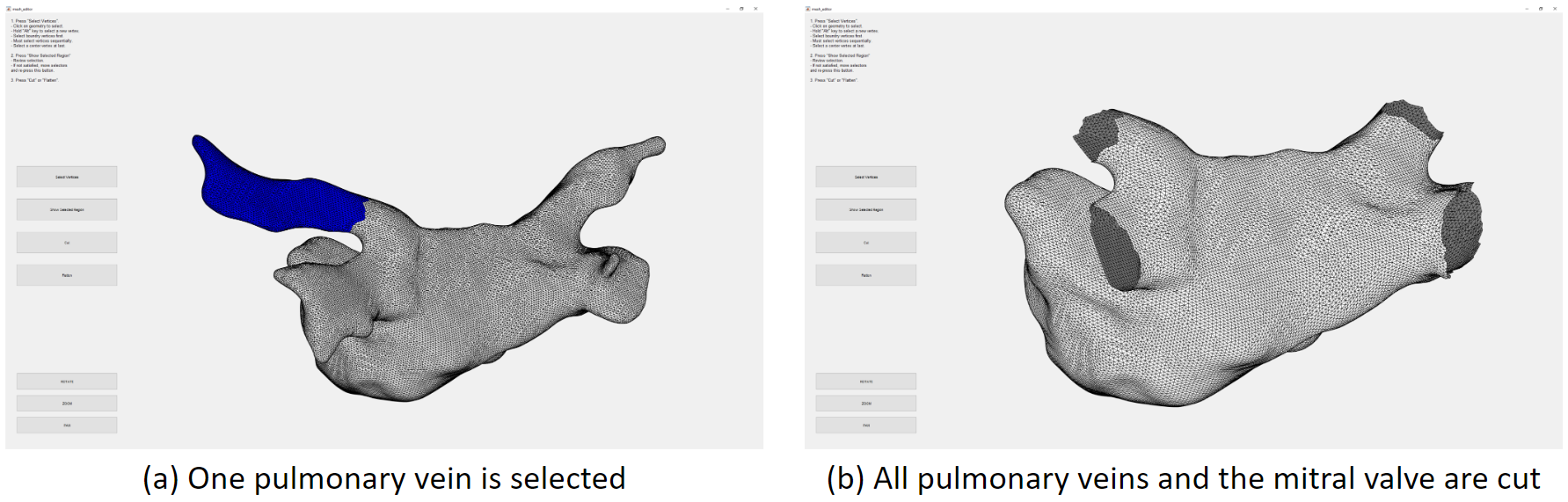}	
\caption{An example of cutting holes using the mesh editing user interface. (a) One pulmonary vein is selected and marked blue, which will be cut out. (b) The left atrium after the 4 pulmonary veins and the mitral valve are cut out.}
\label{fig:cut_holes}
\end{figure} 

The region selection algorithm is illustrated in Figure \ref{fig:select_region}. First, boundary vertices of the desired region and a vertex anywhere in the center is selected. Next, vertices along the geodesic lines in between the boundary vertices are found via the Dijkstra's algorithm. Then iteratively find the outer neighboring vertices of the center vertex, which results in an expanding region until such region hit the boundary vertices. In such a way, all vertices within the boundary are selected.

\begin{figure}[!ht]
\centering
\includegraphics[width=1\textwidth]{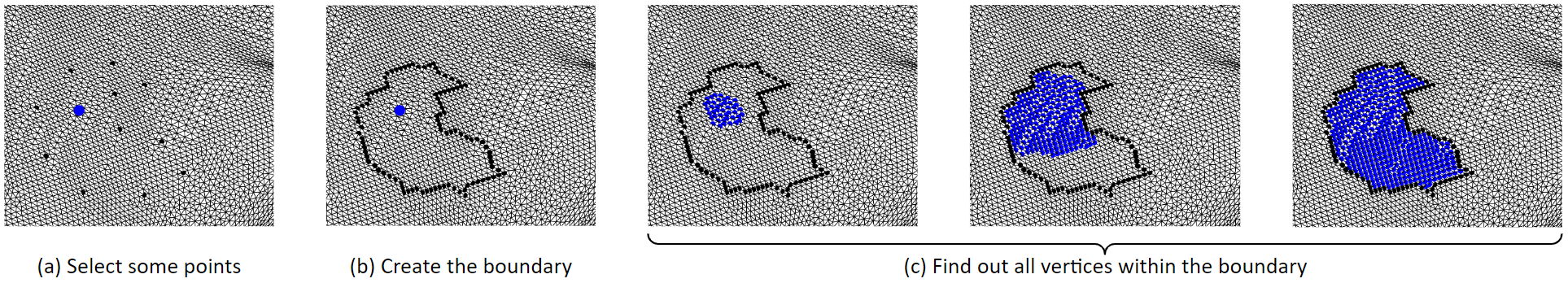}	
\caption{An illustration of the region selection algorithm. (a) Boundary vertices of the desired region and a vertices anywhere inside the region is selected. (b) Vertices along the geodesic lines in between the boundary vertices are found. (c) Starting from the center vertex, the outer neighboring vertices are found "layer-by-layer" as finding more and more outer rings of a cut onion. All vertices within the desired region will be selected after this "onion ring expansion" procedure.}
\label{fig:select_region}
\end{figure}

If there is a sharp spike or a deep concave that is not desirable, user can select that region and flatten it as shown Figure \ref{fig:fill_hole}. First, the undesirable region is cut out, then the resulting hole is filled in with triangles. 

\begin{figure}[!ht]
\centering
\includegraphics[width=1\textwidth]{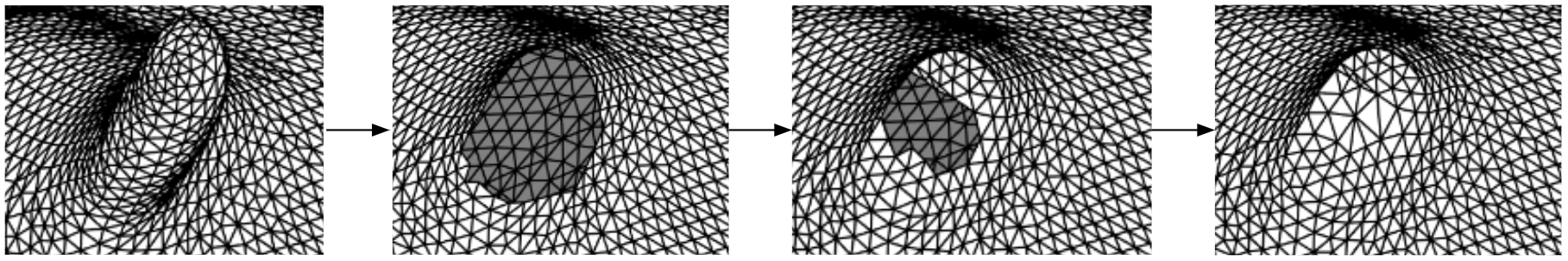}	
\caption{An example of flatten a region. Here an undesirable sharp spiky region is being flattened. First, the spiky region is selected and cut out, then the hole is filled with triangles from the outer perimeter till the center.}
\label{fig:fill_hole}
\end{figure}

\newpage

\section{Atrium electrogram QRS interference removal}
QRS is the sections of surface electrograms that has strong signal magnitude. It represents the activation of the ventricle. The ventricle activation is much stronger than the atrium, therefore the atrium endocardium electrograms often recorded this ventricle signal, which will lead to false detection of the atrium activations. Figure \ref{fig:far_field_interference} shows examples of QRS interference. We developed an algorithm to remove this noise. Figure \ref{fig:QRS_removal_algorithm} summarized the QRS removal algorithm's main steps. 

\begin{figure}[!ht]
\centering
\includegraphics[width=1\textwidth]{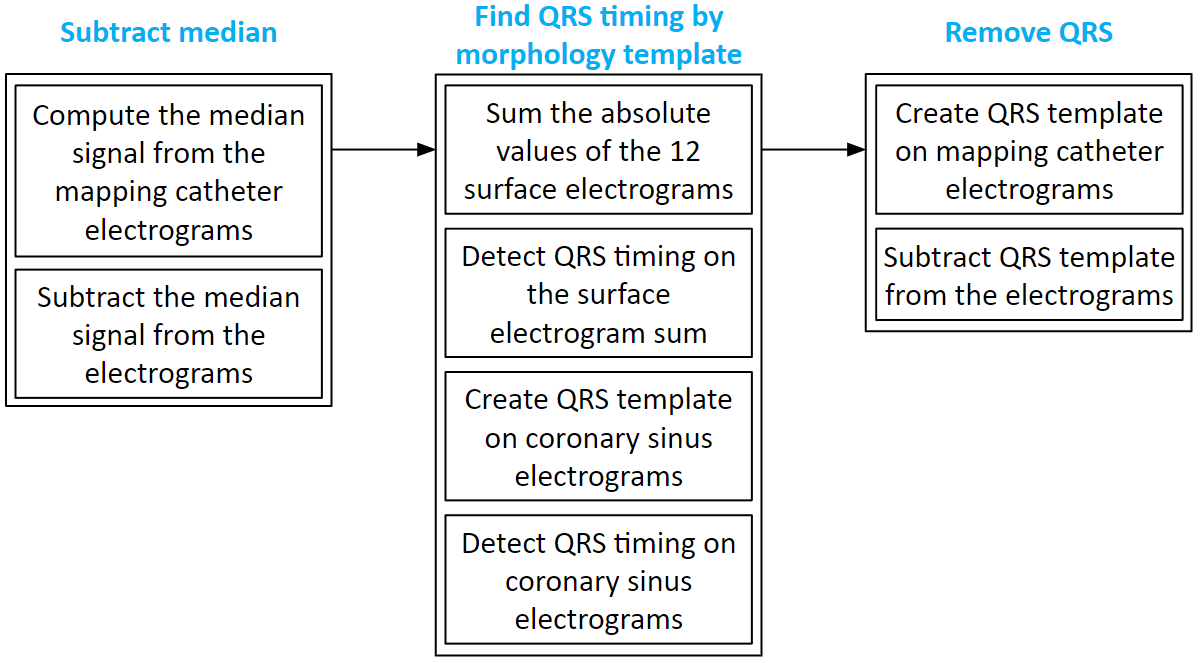}
\caption{QRS removal algorithm. }
\label{fig:QRS_removal_algorithm}
\end{figure}

First, the median signal is subtracted from each electrogram. Figure \ref{fig:1_original_median} shows an example of before and after subtracting the median signal. In this example, the Pentaray catheter is the mapping catheter. It has 20 electrodes (named "20B\_1", ..., "20B\_20") which record 20 electrograms at the same time. The bottom "median" signal in red is the median of these 20 electrograms. The magenta signals are the original electrograms, and the blue signals are median subtracted. We can see that at around 750, 1400 and 2000 ms, there are QRS interference on the original signal (magenta), but after the median subtraction (blue), the QRS are mostly removed.

\begin{figure}[!ht]
\centering
\includegraphics[width=1\textwidth]{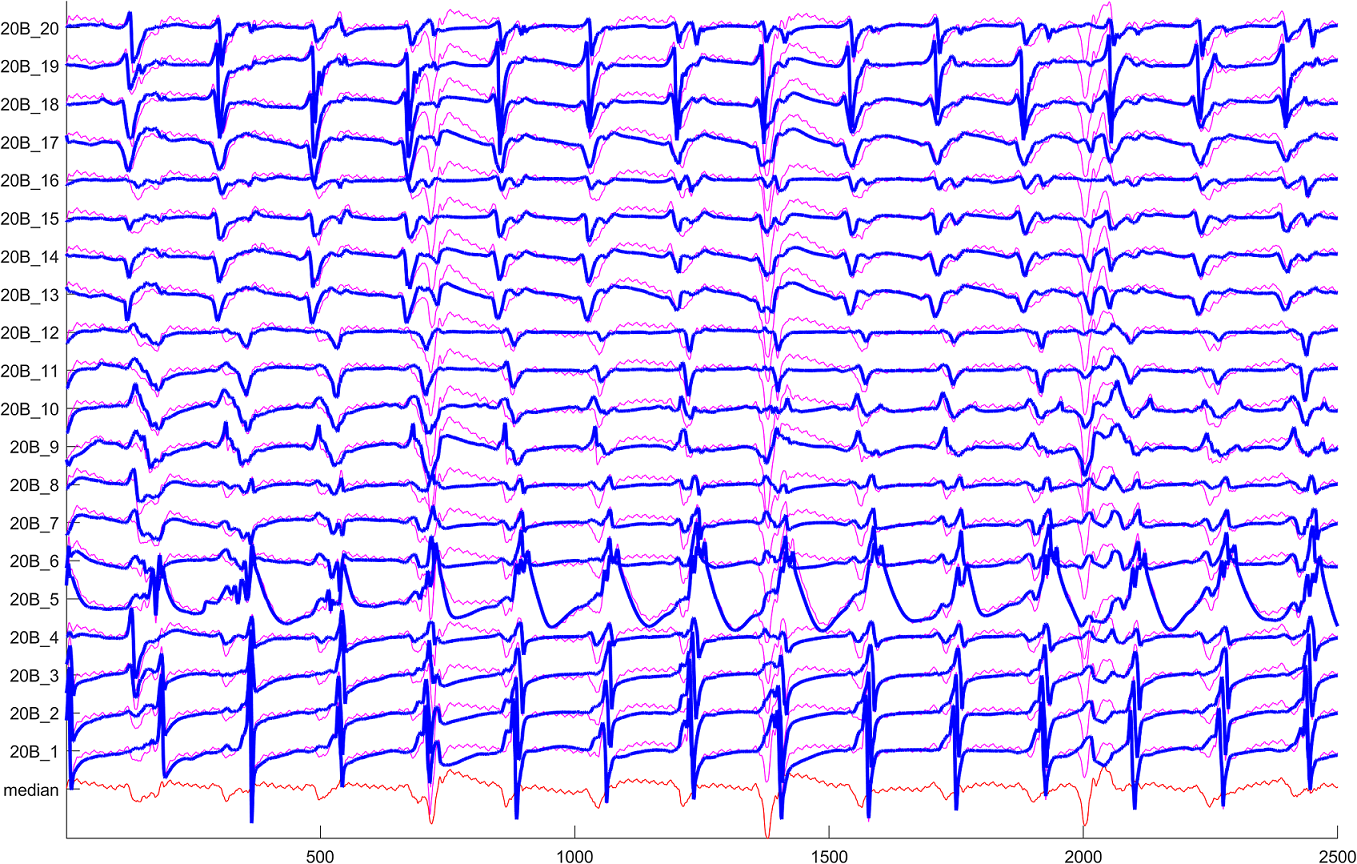}	
\caption{Subtract the median signal.}
\label{fig:1_original_median}
\end{figure}

Then, we find the initial QRS timing on the 12 surface electrograms. As shown in Figure \ref{fig:2_QRS_timing_on_surface_egm}, a sum of all absolute values of the 12 electrograms are plotted at the bottom marked magenta. The peaks of this magenta signal is the QRS timing, marked with black vertical lines. 

\begin{figure}[!ht]
\centering
\includegraphics[width=1\textwidth]{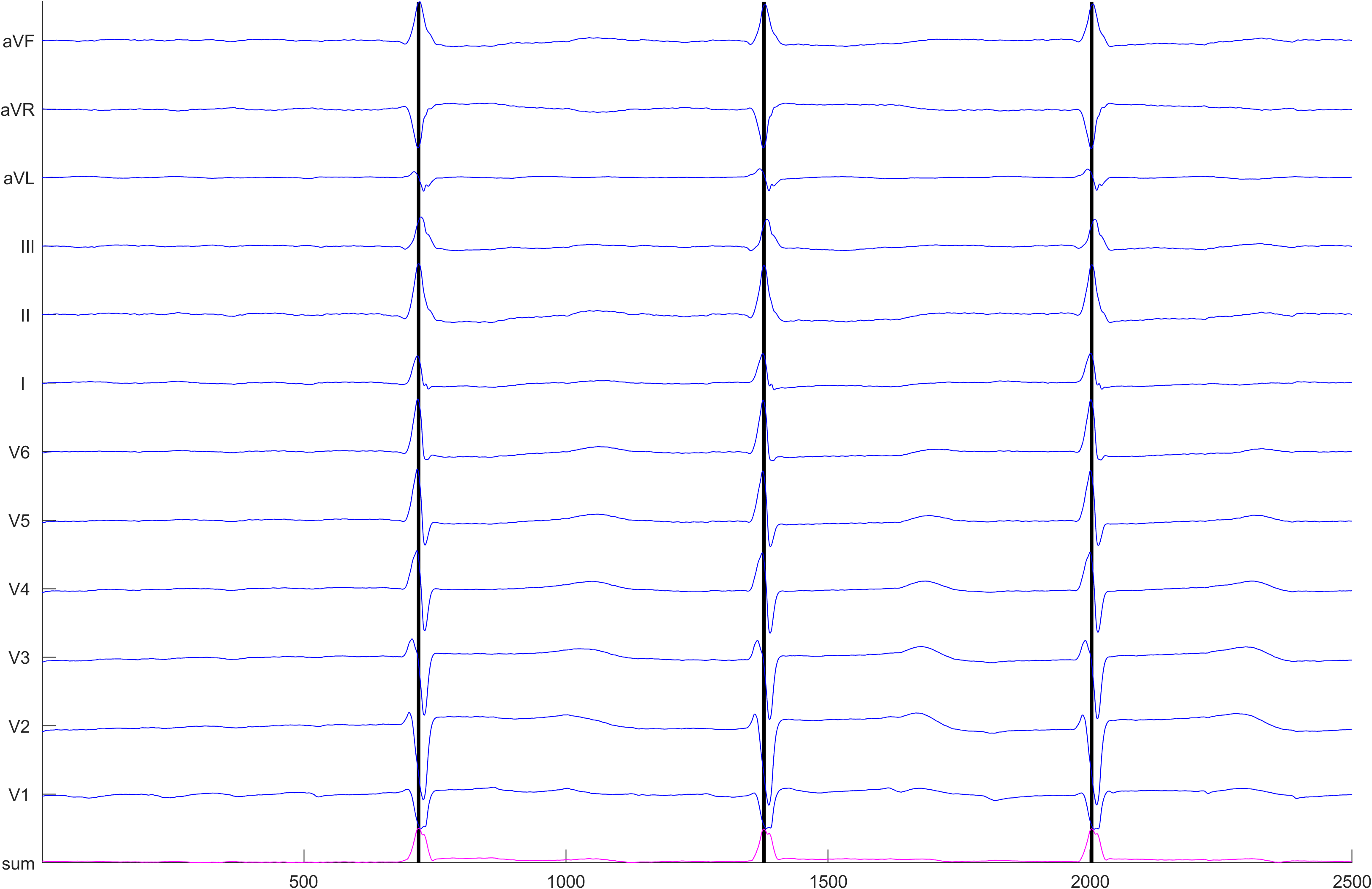}	
\caption{Find QRS timing on the 12 surface electrograms.}
\label{fig:2_QRS_timing_on_surface_egm}
\end{figure}

Next, create QRS templates on the coronary sinus electrograms around the surface electrogram QRS timings as shown in Figure \ref{fig:3_cs_QRS}(a). Using those coronary sinus QRS templates to cross correlate with the coronary sinus electrograms, and sum the correlation coefficients results in a signal as shown in the bottom red one of Figure \ref{fig:3_cs_QRS}(b). Find the peaks of this signal results in a more accurate QRS timing.

\begin{figure}[!ht]
\centering
\begin{tabular}{cc}
\includegraphics[height=5cm]{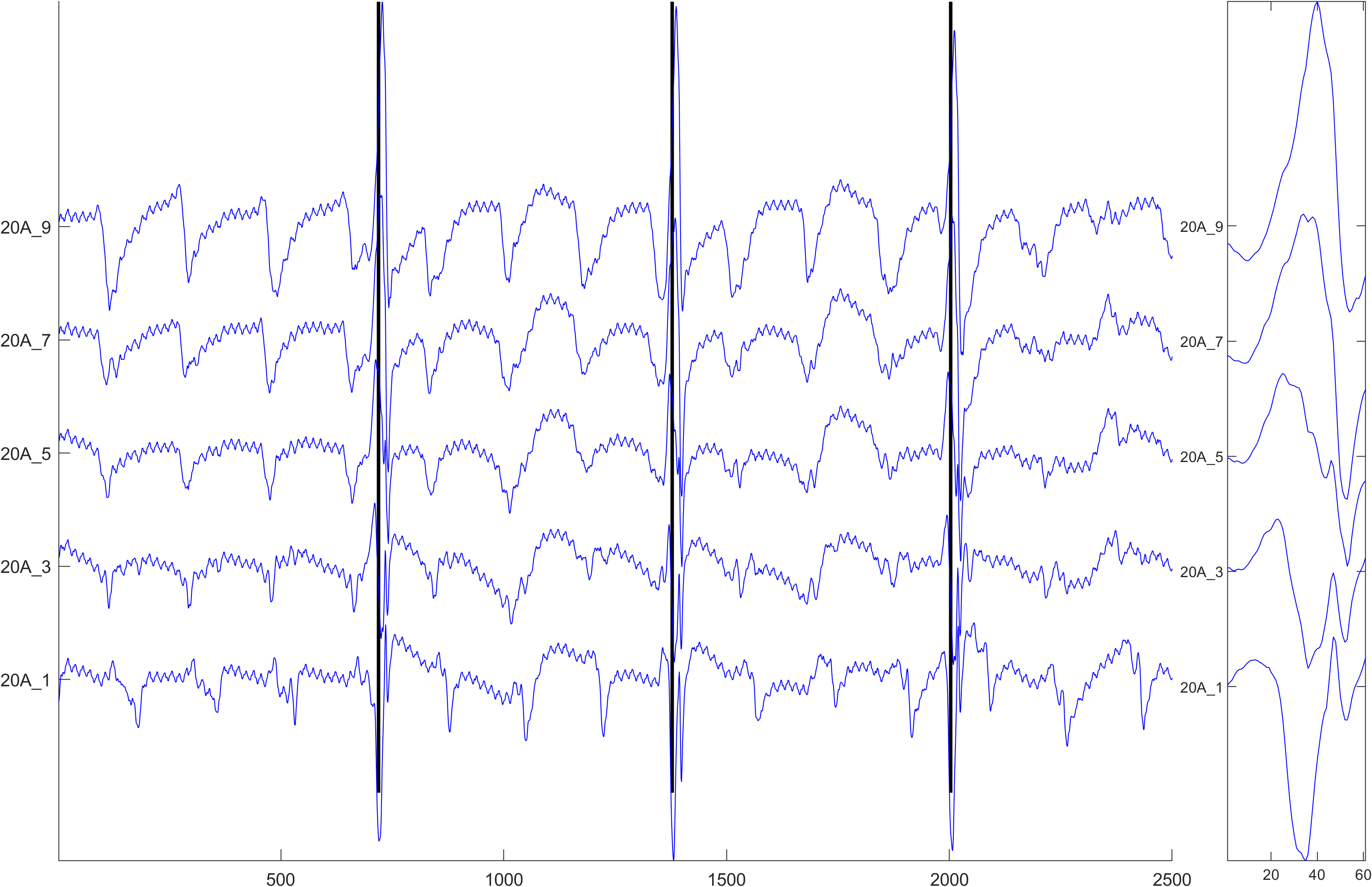} & \includegraphics[height=5cm]{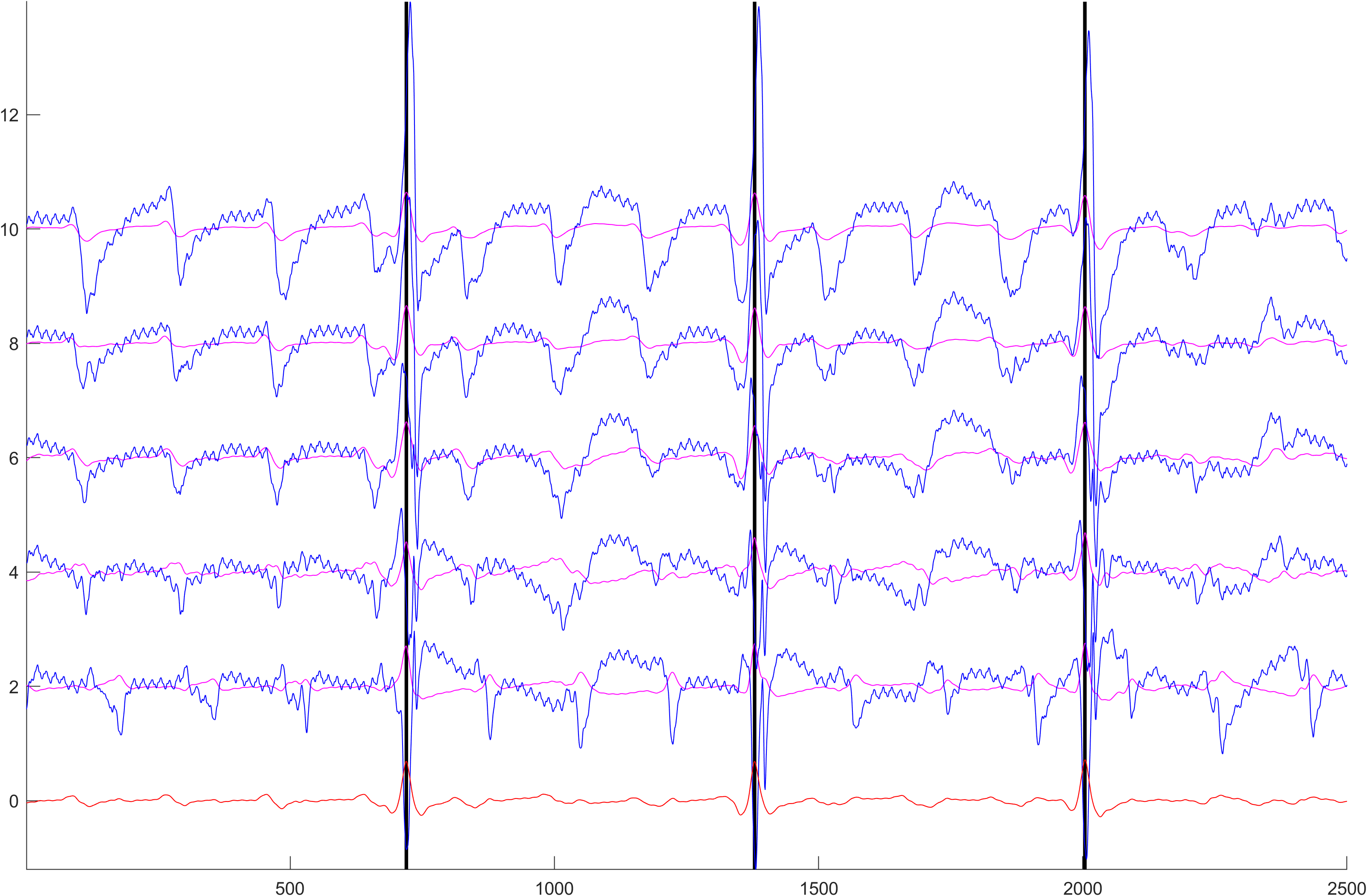} \\
(a) & (b) \\
\end{tabular}
\label{tb:performance_summary_lat_plot}
\caption{(a) Create QRS templates on the coronary sinus electrograms. (b) Detect the QRS timings on the coronary sinus electrograms. These timings are more accurate than the surface electrogram QRS timings, because the coronary sinus location is closer to the left atrium.}
\label{fig:3_cs_QRS}
\end{figure}

Then, for each mapping catheter electrogram, create its own QRS template around the QRS timings as shown in Figure \ref{fig:5_egm_QRS_template}. And lastly, subtract the QRS template as shown in Figure \ref{fig:6_QRS_subtracted}.

\begin{figure}[!ht]
\centering
\includegraphics[width=1\textwidth]{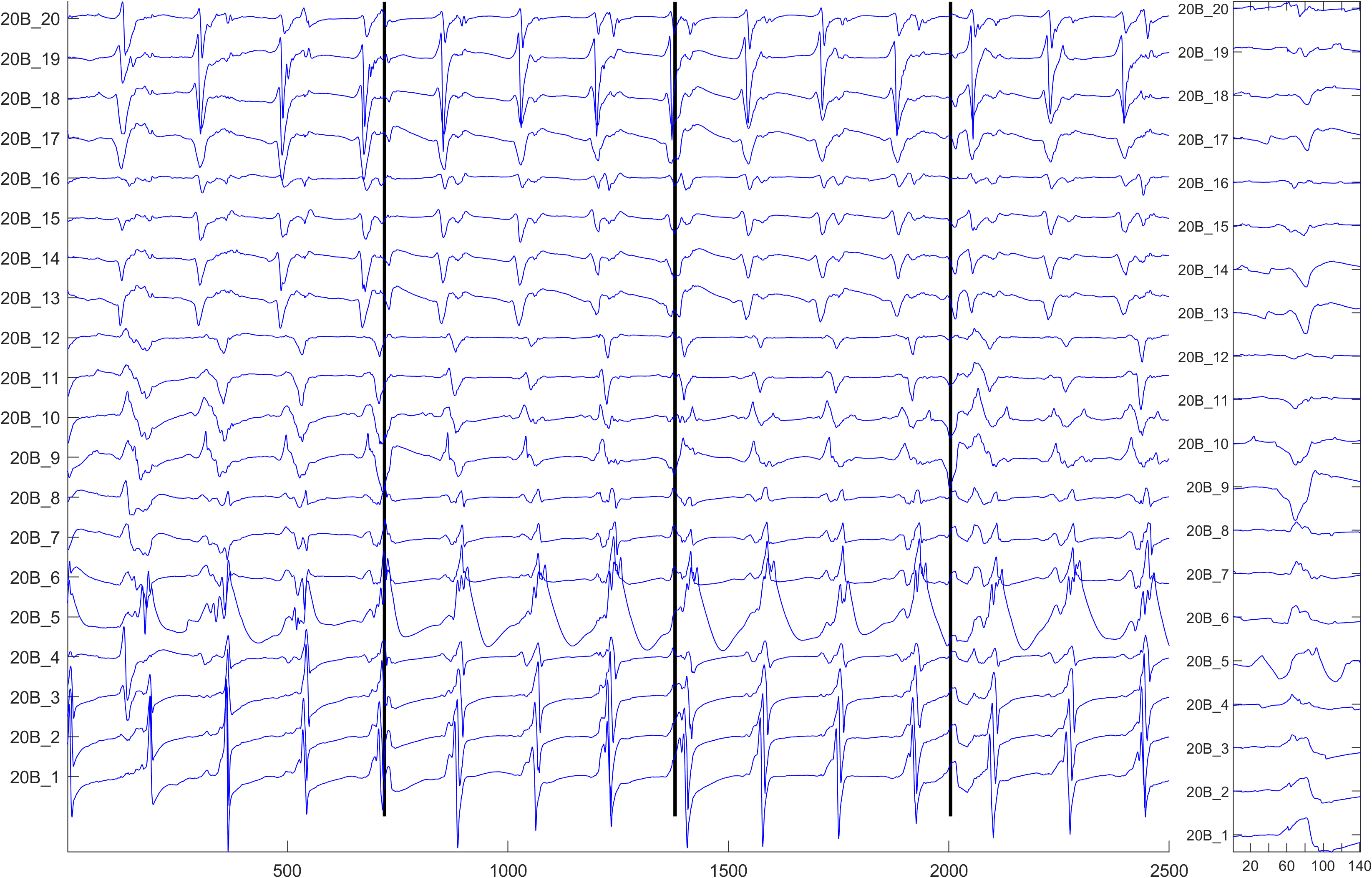}	
\caption{Create QRS template on each mapping catheter electrogram.}
\label{fig:5_egm_QRS_template}
\end{figure}

\begin{figure}[!ht]
\centering
\includegraphics[width=1\textwidth]{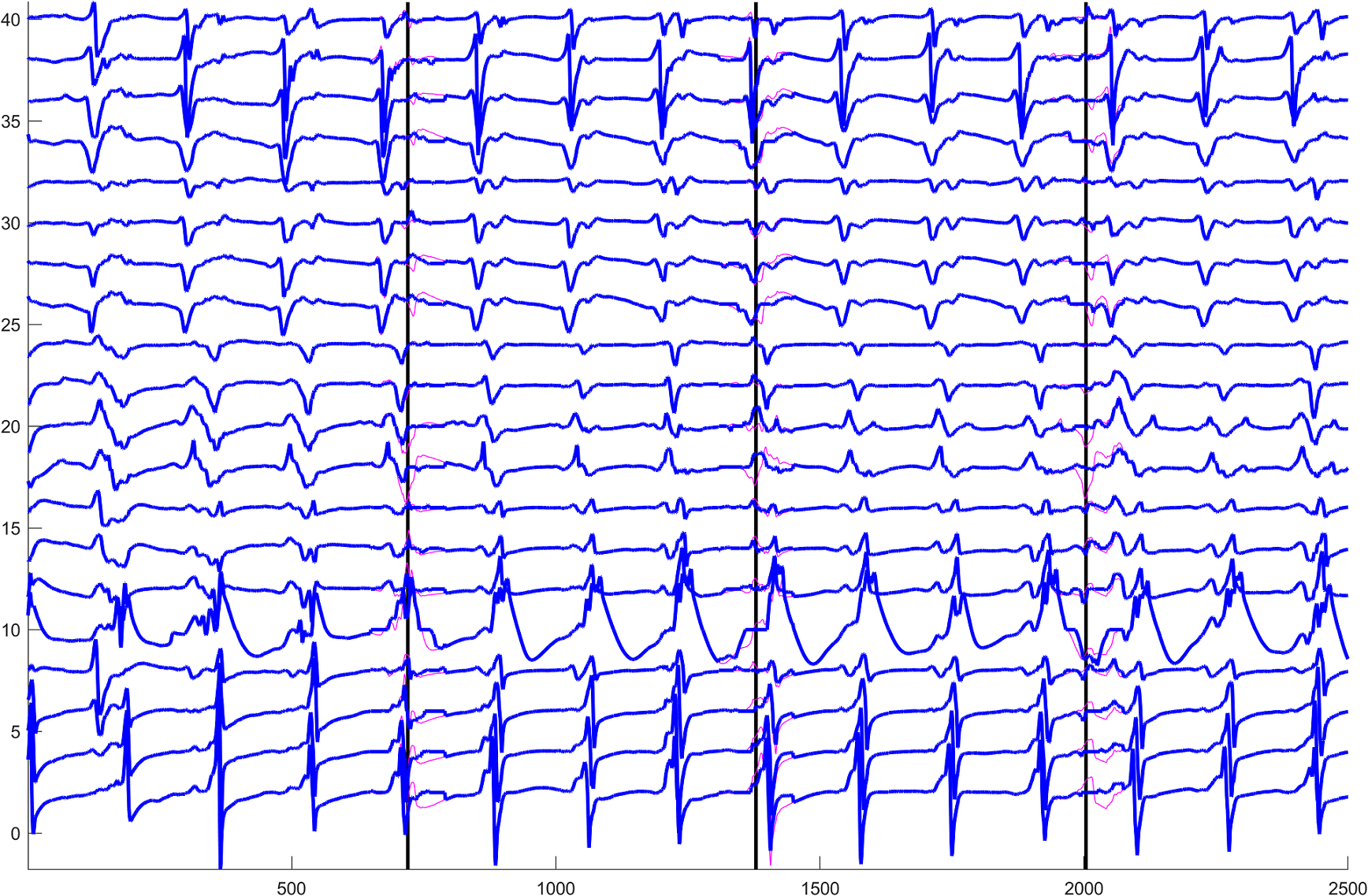}	
\caption{Remove residual QRS interference by subtracting the QRS template. Magenta are the original electrograms, and blue are the QRS subtracted ones.}
\label{fig:6_QRS_subtracted}
\end{figure}

Figure \ref{fig:recording_2_QRS_subtration} shows some examples of before and after QRS removal. We can see that regardless of the signal strength, the QRS interference are removed and left with clean and clear atrial electrogram.

\begin{figure}[!ht]
\centering
\includegraphics[width=1\textwidth]{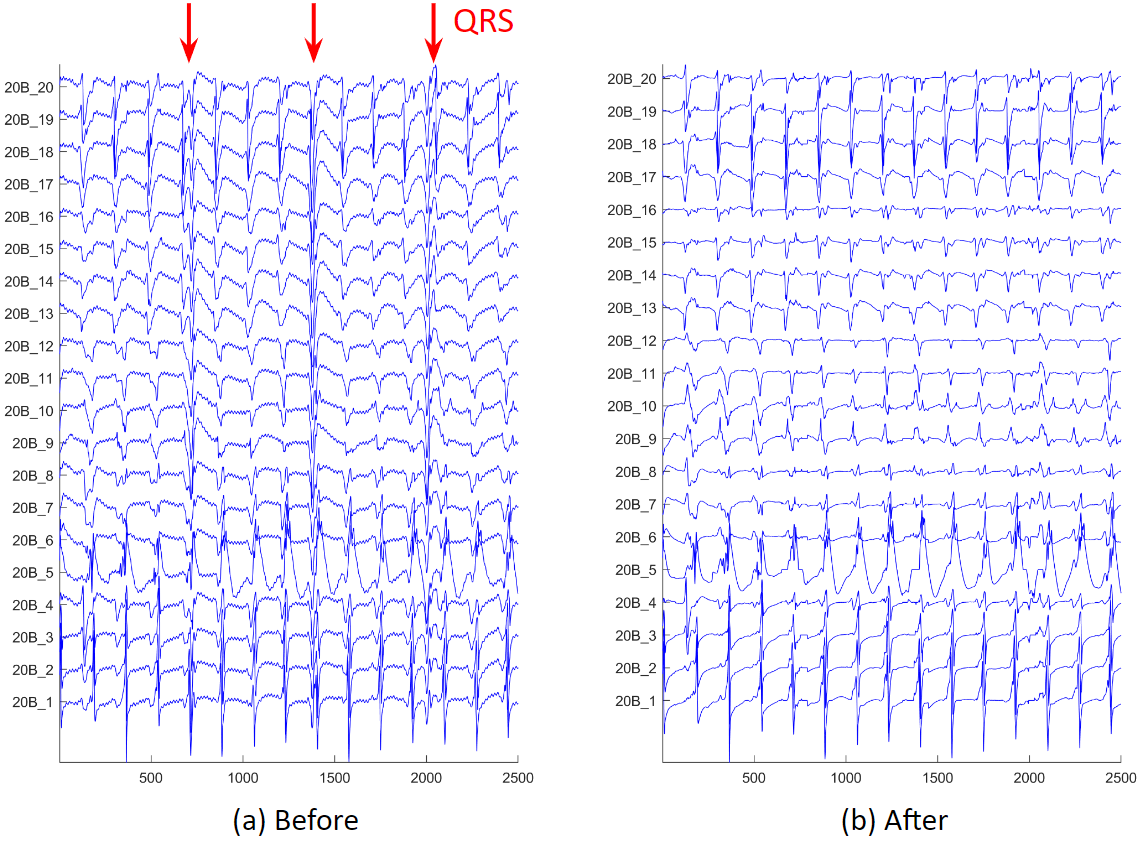}	
\caption{Before and after QRS removal. We can see that in (b), the ventricle activations are removed, and all the atrium activations preserved.}
\label{fig:recording_2_QRS_subtration}
\end{figure}


\newpage

\section{Atrial activation timing detection}
After QRS removal, we have clean and clear atrial electrograms. To detect atrial activation timings, we developed an algorithm as shown in Figure \ref{fig:activation_detection_algorithm}. The main steps are illustrated in Figure \ref{fig:activation_detection}. (a) plots a ventricle QRS interference removed electrogram. It is taken its absolute values and smoothed as shown in (b). Then find out peaks on this signal. The morphology template (c) is created as the median values of the red dashed box marked signals in (a). Then cross correlate the morphology template to (a), the peaks will be the activation timings as shown in (d). Lastly, as shown in (e), Check long intervals between consecutive peaks, if the interval is too long, detect additional peaks that above a signal strength threshold. As marked with a red dashed circle, a lower magnitude peak is added.

\begin{figure}[!ht]
\centering
\includegraphics[width=1\textwidth]{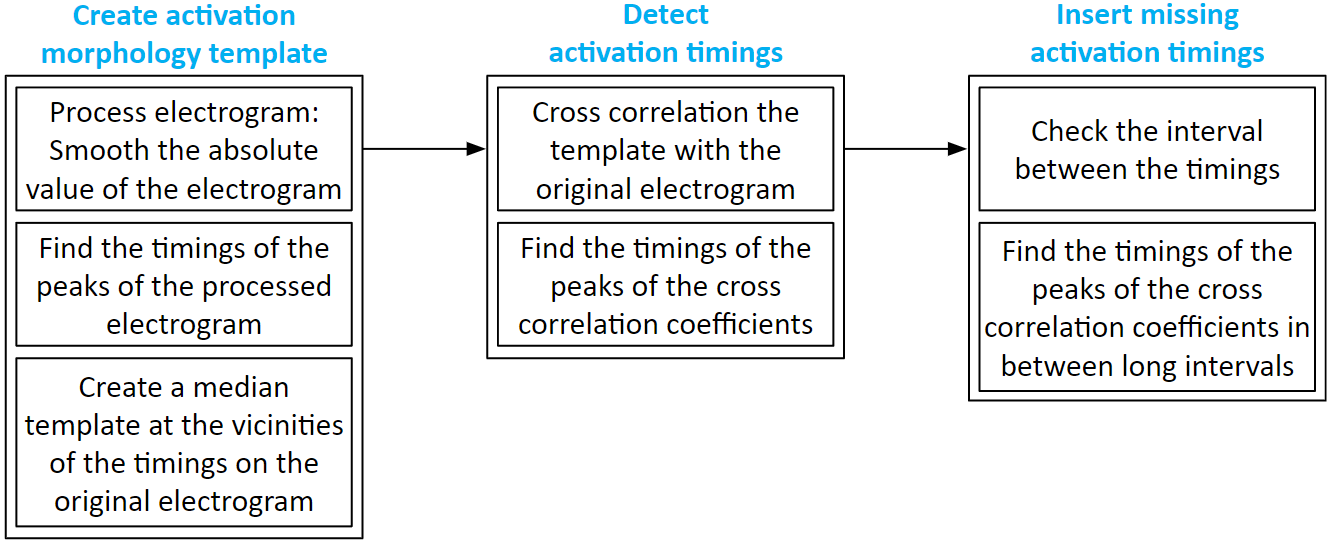}	
\caption{Activation detection algorithm.}
\label{fig:activation_detection_algorithm}
\end{figure}

\begin{figure}[!ht]
\centering
\includegraphics[width=1\textwidth]{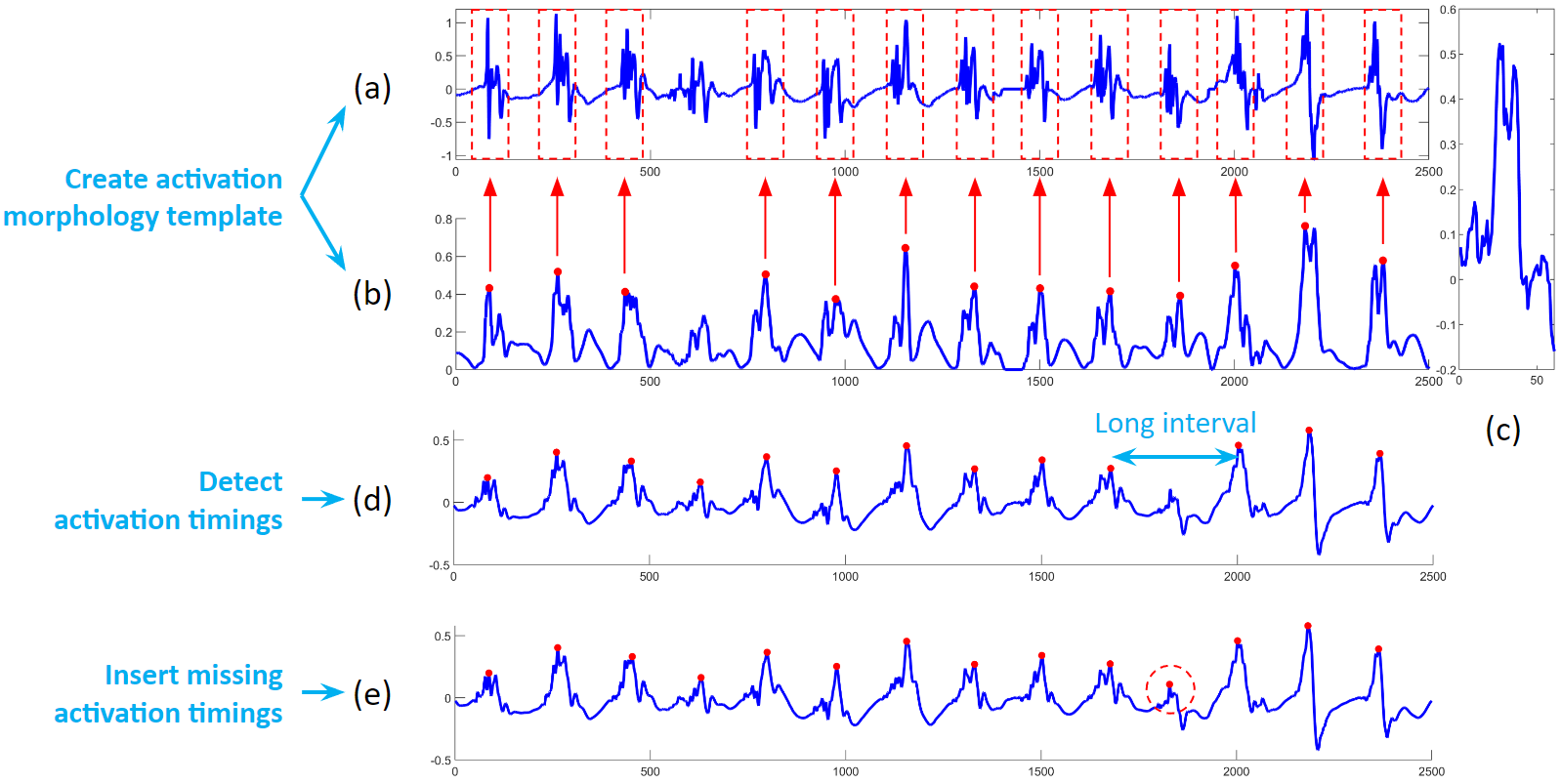}	
\caption{Atrial activation timing detection. }
\label{fig:activation_detection}
\end{figure}

Figure \ref{fig:2_activation_timing} shows atrial activation detection results marked in red dots with activation morphology template of each electrogram shown in the right panel. If the electrogram signal is too weak that results in questionable activation timing, it should be excluded. 

\begin{figure}[!ht]
\centering
\includegraphics[width=1\textwidth]{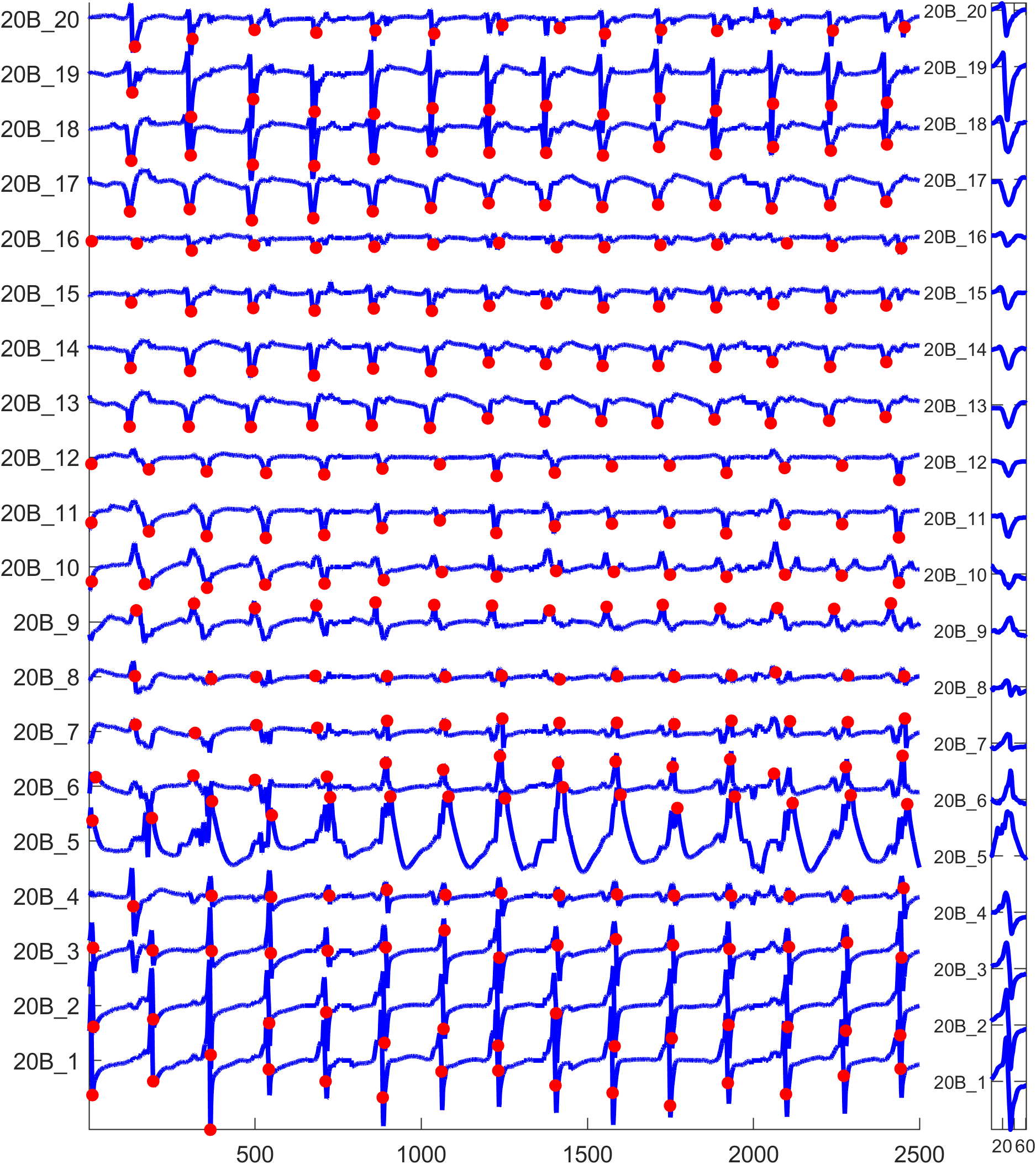}	
\caption{Atrial activation timing detection results of clear electrograms.}
\label{fig:2_activation_timing}
\end{figure}

\newpage

\section{Improve atrial fibrillation voltage map accuracy}
A voltage map is often used to demarcate low voltage areas and preserved voltage areas during catheter ablation therapy to treat atrial fibrillation. Low voltage areas correspond to areas of diseased atrium (fibrosis) or dense scars from prior ablations. Identifying low voltage area can help in planning ablation strategies especially in patients requiring repeat ablation procedures for arrhythmia recurrences, for example, atrial fibrillation and atypical atrial flutter.

The cutoff threshold voltage for a voltage map that determines low voltage areas has been established for maps obtained while the patient is in sinus rhythm \cite{Kapa_2014}\cite{Squara_2014}. However, this same threshold is not applicable to maps collected during atrial fibrillation.

For example, Figure \ref{fig:threshold_problem}(a) depicts the sinus rhythm map with a cutoff threshold of 0.45 mV. The regions below the threshold are clearly delineated from the healthy regions (magenta). (b), (c), (d) are examples of voltage maps in atrial fibrillation when applying different thresholds. (b) The threshold is too small, resulting in too few scar regions. (c) The threshold is too large, resulting in too much scar regions. (d) When an appropriate threshold is applied, the scar regions are most accurately displayed.

\begin{figure}[!ht]
\centering
\includegraphics[width=1\textwidth]{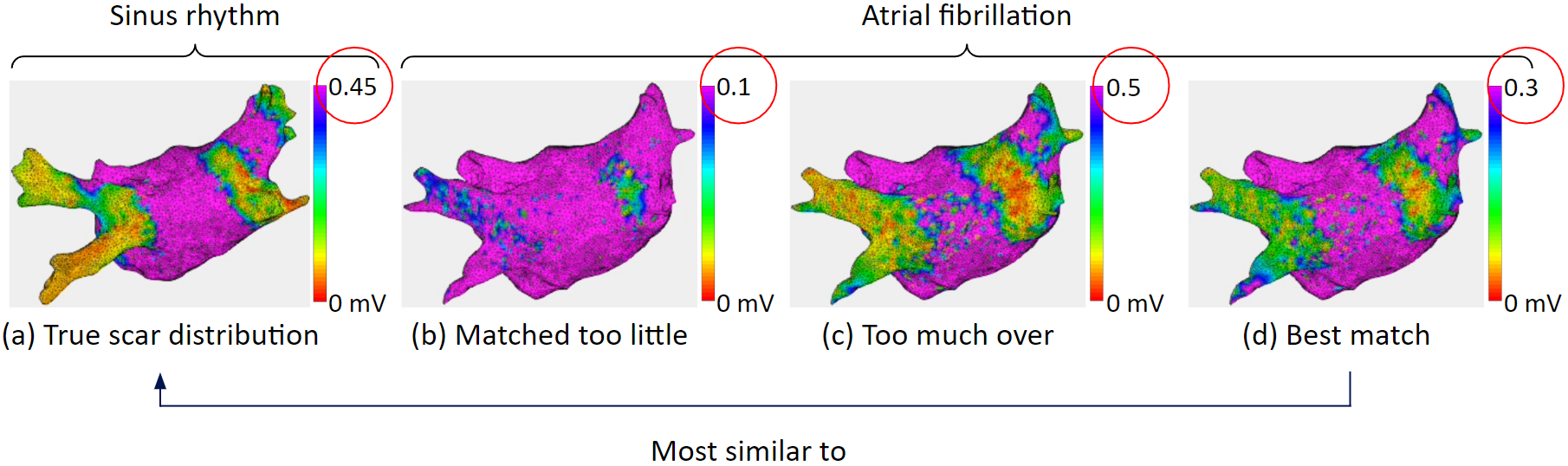}
\caption{Example of need for a threshold to be used on voltage maps obtained during atrial fibrillation. Magenta area is healthy tissue, other colors are scar tissue.}
\label{fig:threshold_problem}
\end{figure}

Several studies have shown that the local atrial signal acquired during atrial fibrillation is lower than in sinus rhythm. Thus, identifying low voltage area during electroanatomical mapping in atrial fibrillation should require a lower cutoff voltage \cite{Rodriguez_2018}. However, determining a consistent threshold that can be applied to all patients remains challenging. A prerequisite is determining the best match of low voltage areas that can be obtained between the sinus rhythm map and the atrial fibrillation map and thereby finding the best threshold to be applied on a patient-by-patient basis.

Our goal here is: Given a set of measurements during sinus rhythm and atrial fibrillation for a patient, maximize the topologically matched low voltage areas between the derived sinus rhythm and atrial fibrillation map.

We processed a cohort of 7 male patients data. These patient underwent repeat catheter ablation for atrial fibrillation at the Hospital of the University of Pennsylvania. Patient demographics are: age $66 \pm 0.7$ years, height $5'10'' \pm 4''$, weight $238.9 \pm 50$ lb and ejection fraction $49.4\%$. Details are in Table \ref{table:patient_population}. For each patient, voltage maps were sequentially obtained using Carto3 (Biosense Webster) during sinus rhythm and atrial fibrillation. Carto3 fill threshold was 5 mm, and filters were set at 2 to 240 Hz for unipolar electrograms, 16-500 Hz for bipolar electrograms, and 0.5-200 Hz for surface electrogram recordings.

A total of 46,589 data points were included in analysis. The sinus rhythm voltage map with 0.45 mV threshold depicted the scar regions. We applied various thresholds on the atrial fibrillation voltage map to find the one that resulted scar regions best matched the scars in sinus rhythm map. 

\begin{table}[!ht]
\centering
\begin{tabular}{ c|ccccccc } 
    \hline
    Patient & 1 & 2 & 3 & 4 & 5 & 6 & 7 \\ \hline
    Age & 66 & 58 & 70 & 70 & 69 & 58 & 76 \\
    Height (inch) & 71 & 75 & 72 & 62 & 70 & 72 & 68 \\
    Weight (pound) & 290 & 260 & 266 & 243  & 267 & 200 & 146 \\
    Ejection Function (\%) & 65 & 50 & 25 & 50 & 55 & 60 & 40 \\
    \hline
\end{tabular}
\caption{Patient population}
\label{table:patient_population}	
\end{table}

We implemented this scar region matching process in 2 different methods. One is the baseline method, which the atrium mesh defects were removed, electrograms noise were removed, and electrode voltage data is linearly interpolated onto the mesh. The other one is the "Omni.+GP" method, which the atrium mesh and electrograms were processed the same way, in addition, the bipolar voltages were corrected using the omni-bipolar algorithm, and a Gaussian process regression based interpolation was implemented. 

To reduce such dependency, we derive the omni-directional bipolar voltage. For each sample point, we select the unipolar electrodes in its vicinity and compute all possible bipolar electrodes. We approximate the omnipolar voltage as the largest bipolar amplitude in this set. For example, in Figure \ref{fig:omnipolar_algorithm}, 1-5 are unipolar electrodes that near the unipolar electrode 0. Each edge represent one possible bipolar. We can create that via the subtraction of the two unipolars. Here we have a total of 15 (6 choose 2) possible bipolars. We pick the one that has the largest amplitude and obtain its peak-to-peak voltage to represent the peak-to-peak voltage of electrode 0.

\begin{figure}[!ht]
\centering
\includegraphics[width = 0.4\textwidth]{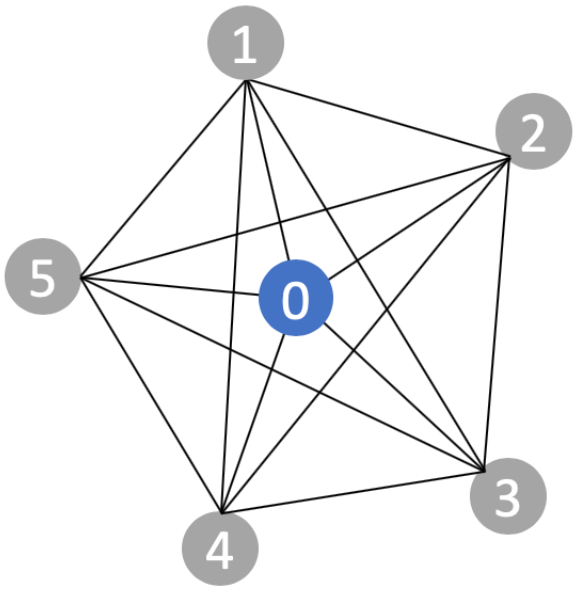}
\caption{For unipolar electrode 0 and its neighboring unipolar electrodes 1 through 5, the lines represents all 15 possible bipolar electrodes we can create from these 6 unipolar electrodes.}
\label{fig:omnipolar_algorithm}
\end{figure}

Figure \ref{fig:omni_overview} illustrates that in a bipolar voltage map, as shown in (a), healthy tissue can sometimes misclassified as scar. After applying the omni-directional bipolar algorithm, the voltage map is corrected as shown in (b).

\begin{figure}[!ht]
\centering
\includegraphics[width=1\textwidth]{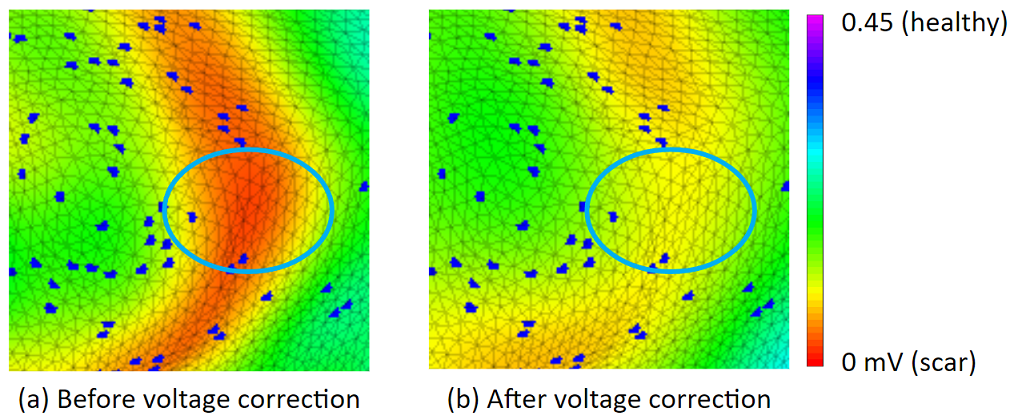}
\caption{Benefits of using omni-directional bipolar voltages. (a), (b) Regions of previously low voltage (red in a) are increased after computing omni-directional bipolar voltages (yellow in b). (Note: The blue dots are electrode sample sites.)}
\label{fig:omni_overview}
\end{figure}

Gaussian process regression (\ac{GPR}) based interpolation is a smoothed interpolation. If we model the endocardium as a surface and define the entire set of samples as, $\mathcal{D} = \{\mathbf{x}_n, y_n\}^N_{n=1}$, where inputs $\mathbf{X} = \{\mathbf{x}_n\}^N_{n=1}$ correspond to the locations on the mesh and  $\mathbf{y} = \{y_n\}^N_{n=1}$ are the voltage value at that location. Interpolating from these measured samples to the remainder of the mesh can be thought of as determining the estimates of the voltages at locations $\mathbf{X^*}$. The two major sources of interpolation error are the low measurement density and measurement noise. Both of these can be accounted for by modeling them using a Gaussian process $GP(m(\mathbf{x},k(\mathbf{x},\mathbf{x}')))$ \cite{Rasmussen_2006}, which is characterized by the mean $m(\mathbf{x})$ and the covariance $k(\mathbf{x},\mathbf{x}')$ kernel functions. We assume the common zero mean function and use the squared exponential function $k(\mathbf{x},\mathbf{x}')= \exp(-|| \mathbf{x}-\mathbf{x}' || / (2 \cdot l^2))$. 

Figure \ref{fig:interpolation_comparison}(a) shows how the linear interpolation can result in regions that are classified as low voltage areas, due to interpolation error. (b) shows \ac{GPR}-based interpolation can improve the boundaries of low voltage areas by considering the surrounding measurements.

\begin{figure}[!ht]
\centering
\includegraphics[width=1\textwidth]{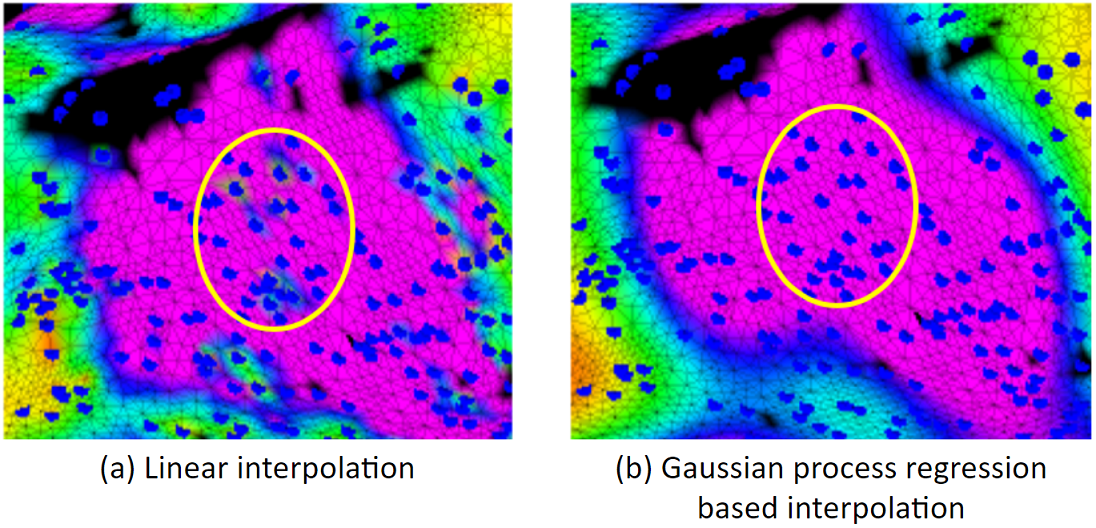}
\caption{(a) Linear interpolation. Such a simple interpolation can introduce data outliers. The greenish spots within the yellow circle can be interpolation noise. (b) \ac{GPR}-based interpolation removes the greenish spots within the yellow circle, resulted in a more accurate voltage map.}
\label{fig:interpolation_comparison}
\end{figure}

For determining the optimum threshold, a search in the range of 0-0.45 mV is performed to maximize the product of sensitivity and specificity. Here, sensitivity $= \frac{\text{TP}}{TP+FN}$ and specificity $= \frac{\text{TN}}{TN+FP}$, where a true positive (TP) indicates that the corresponding face on the anatomical mesh is detected as a low voltage area and the true label is a low voltage area.

Table \ref{table:patient_specific_result} summarizes the results. On average, there was a 3.00\% improvement in the geometric mean of the "Omni.+GP" method compared to the baseline method. Moreover, the "Omni.+GP" method exhibited a 7.88\% improvement in sensitivity and a 0.30\% improvement in specificity. ROC curves were obtained for each of the methods and the \acl{AUC} was computed as shown in Figure \ref{fig:ROC_baseline_gaussian_omni_all_patients}. The "Omni.+GP" method showed an average of 3.91\% improvement in terms of the \ac{AUC}. 

\definecolor{Gray}{gray}{0.85}

\begin{table*}[!ht]
	\centering
	\resizebox{1\textwidth}{!}{
		\begin{tabular}{c|c|c|c|c|c|c|c|c|c|c|c|c|c|c} \hline
		\multicolumn{1}{c|}{Patient} &\multicolumn{5}{c|}{Baseline} &\multicolumn{5}{c|}{Omni.+GP} &\multicolumn{4}{c}{Percentage of Improvement}\\ \hline
		&  Sens. & Spec. & GM & AUC & Voltage Threshold &Sens. & Spec. & GM. & AUC & Voltage Threshold & $\Delta$Sens. & $\Delta$Spec. & $\Delta$GM & $\Delta$AUC\\ \hline
1 & 76.18 & 63.56 & 69.58 & 0.77  & 0.26  & 84.92 & 70.74 & 77.51 & 0.85  & 0.23  & 11.47 & 11.30 & 11.40 & 10.39 \\
2 & 54.87 & 57.64 & 56.24 & 0.57  & 0.09  & 72.72 & 47.49 & 58.77 & 0.61  & 0.11  & 32.53 & -17.61 & 4.50  & 7.02 \\
3 & 77.38 & 71.48 & 74.37 & 0.80  & 0.16  & 91.22 & 69.59 & 79.67 & 0.86  & 0.15  & 17.89 & -2.64 & 7.13  & 7.50 \\
4 & 79.52 & 78.08 & 78.80 & 0.85  & 0.26  & 85.42 & 77.07 & 81.14 & 0.89  & 0.30  & 7.42  & -1.29 & 2.97  & 4.71 \\
5 & 58.08 & 66.88 & 62.32 & 0.67  & 0.36  & 69.24 & 56.71 & 62.66 & 0.66  & 0.38  & 19.21 & -15.21 & 0.55  & -1.49 \\
6 & 80.61 & 74.64 & 77.57 & 0.83  & 0.32  & 72.87 & 83.41 & 77.96 & 0.85  & 0.21  & -9.60 & 11.75 & 0.50  & 2.41 \\
7 & 70.24 & 52.57 & 60.77 & 0.63  & 0.10  & 53.54 & 60.87 & 57.09 & 0.61  & 0.17  & -23.78 & 15.79 & -6.06 & -3.17 \\ \hline
\rowcolor{Gray} Average & 70.98 & 66.41 & 68.52 & 0.73  & 0.22  & 75.70 & 66.55 & 70.69 & 0.76  & 0.22  & 7.88  & 0.30  & 3.00  & 3.91 \\
Std  & 10.49 & 9.19  & 8.87  & 0.11  & 0.11  & 12.72 & 12.34 & 10.65 & 0.13  & 0.09  & 18.96 & 13.32 & 5.54  & 4.95 \\ \hline
		\end{tabular}
	}
	\caption{Results for patient-specific performance for all patients}
    \label{table:patient_specific_result}
\end{table*}

\begin{figure}[!ht]
\centering
\includegraphics[width=1\textwidth]{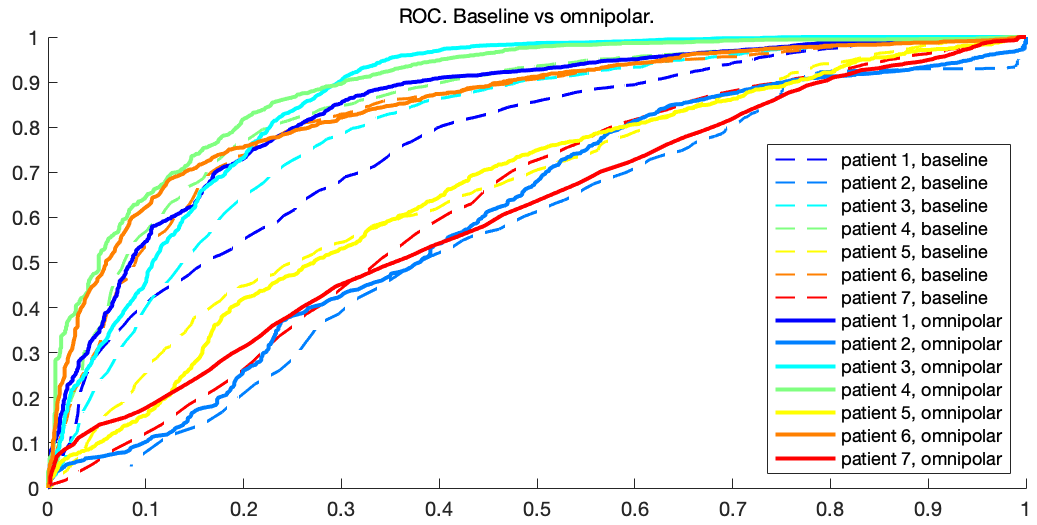}
\caption{ROC curve comparing, baseline method vs proposed method using omni-directional bipolar voltages and \ac{GPR}-based interpolation. The proposed method shows improved or similar performance across various thresholds.}
\label{fig:ROC_baseline_gaussian_omni_all_patients}
\end{figure}

\chapter{\MakeUppercase{Electroanatomical maps and arrhythmia source detection}}
\section{Overview}
In the previous chapter, we show the clinical data and how to process them. In this chapter, we will transform these data into colored maps on the left atrium that can help find out arrhythmia source locations. Such a colored map is called an electroanatomical map, where the color represents a metric processed from the endocardium electrograms. 

Each electroanatomical map contains about 1,500 electrogram recordings spread across the endocardium. As these recordings are taken at different times, they do not provide synchronous view of the whole atrium. However, sinus rhythm and stable tachycardia are periodic rhythms, thus the time asynchronous problem can be solved by aligning different recording segments to a reference channel. The reference channel can be one of the surface electrodes, an absolute sum of all of the surface electrodes, or one of the coronary sinus catheter electrodes. 

The two most frequently used electroanatomical maps during clinical arrhythmia ablation procedure are voltage map and local activation time map. We develop several additional maps that are helpful for physicians to analyze arrhythmia source locations. In the following sections, we will explain these maps: Voltage map, local activation time map, cycle length map, dominant frequency map, fractionation map, synchrony map, conduction velocity vector map, phase singularity intensity map.

The process of creating these electroanatomical maps are laid out in Figure \ref{fig:create_map}. We first process these patient data. From the electrogram, we can derive voltage, dominant frequency, fractionation, and synchrony. Then we detect the activation timings, which we can derive local activation time and cycle length. From the activation timing, we can create an activation movie, which we can derive conduction velocity and phase singularity. Each of these derived values can be interpolated to the atrium mesh. Then we can convert the values to color, and it becomes a map.

\begin{figure}[!ht]
\centering
\includegraphics[width = 1\textwidth]{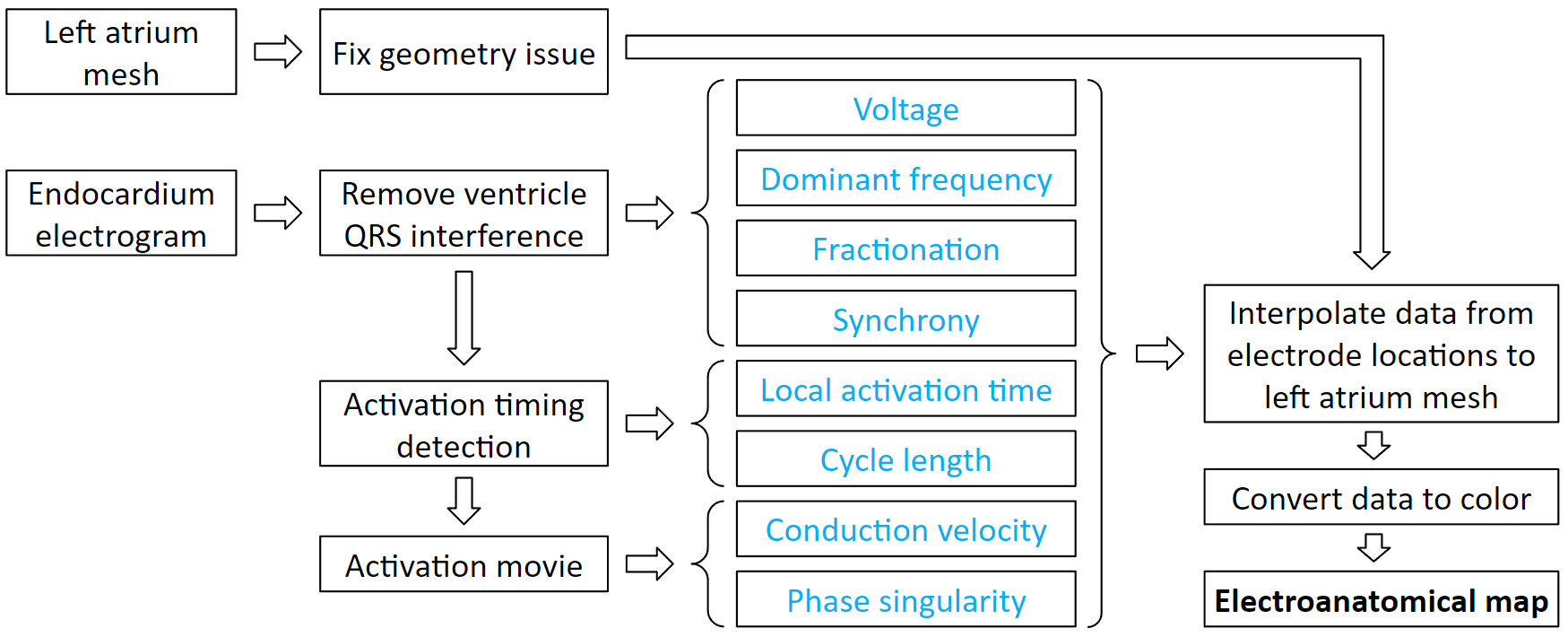}
\caption{Flowchart of creating various electroanatomical maps.}
\label{fig:create_map}
\end{figure}

Furthermore, we develop an user interface for screening patient data as shown in Figure \ref{fig:patient_data_observer}. User can manually marked electrograms good or bad, so that bad electrograms will be excluded.

\begin{figure}[!ht]
\centering
\includegraphics[width = 1\textwidth]{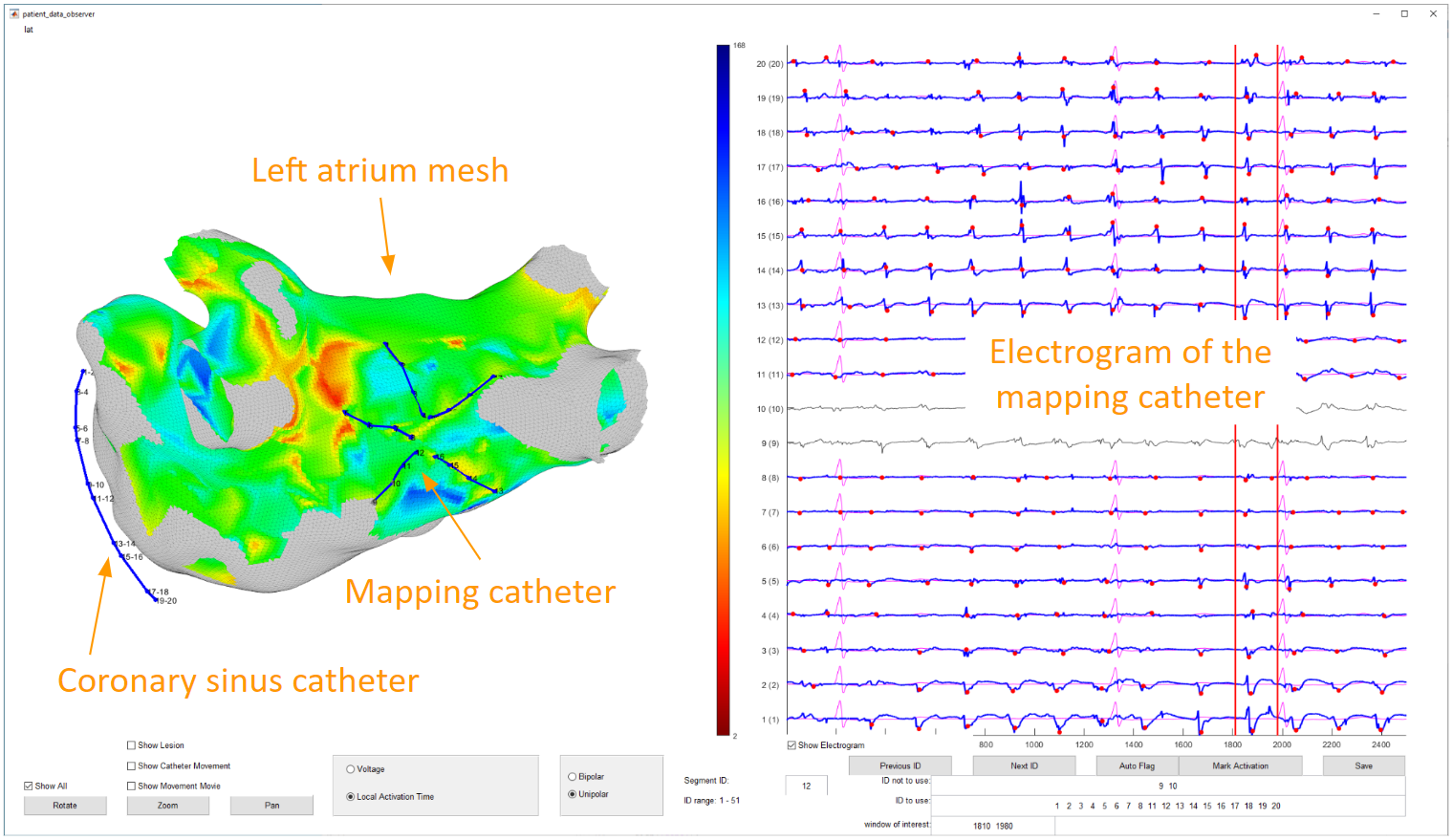}
\caption{User interface for screening patient data.}
\label{fig:patient_data_observer}
\end{figure}

\newpage

\section{Voltage map}
Voltage map displays electrogram peak-to-peak voltage values on the left atrium mesh as a color coded map. The colors of the voltage map are based on a pre-specified cutoff threshold, where areas above the threshold are marked in magenta, and areas below the threshold are considered low voltage areas. As shown in Figure \ref{fig:voltage_map}, voltage values need to be interpolated from the electrode locations to the vertices of the left atrium triangular mesh.

\begin{figure}[!ht]
\centering
\includegraphics[width=1\textwidth]{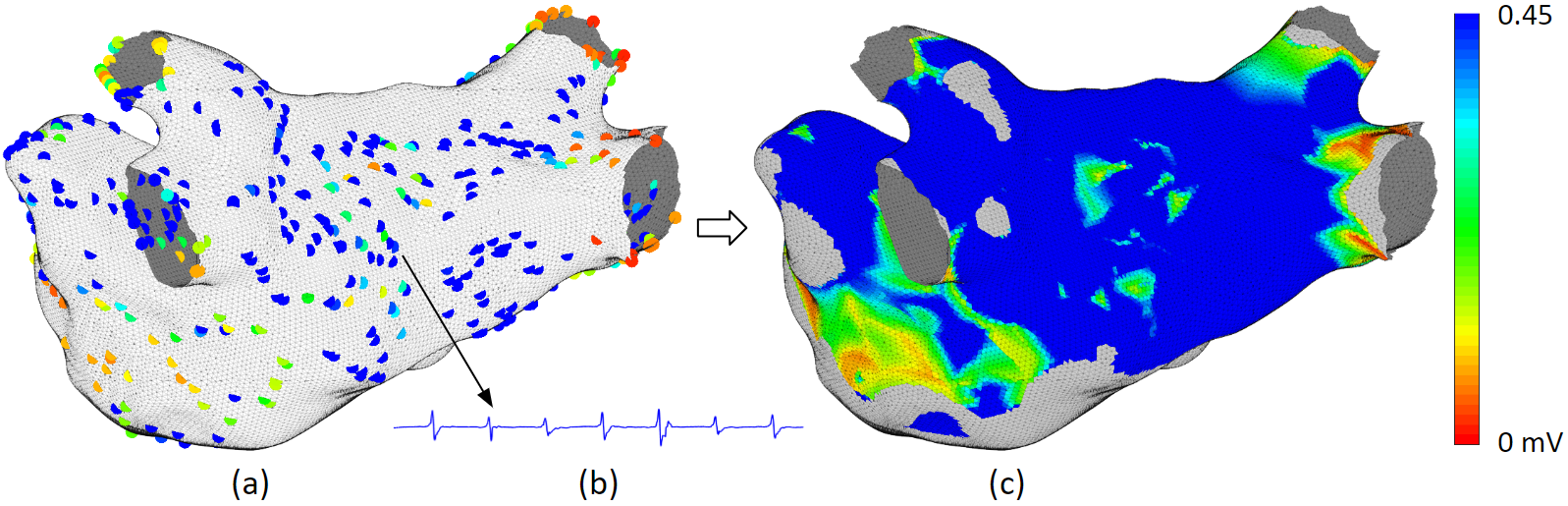}
\caption{Creation of a voltage map.}
\label{fig:voltage_map}
\end{figure}

Low voltage value indicate poor conductivity, which is a main character of scar. Scar can alter or block the heart's electrical activations, resulting in complex activation waves that can lead to atrial fibrillation. Therefore, identifying scar locations is important for figuring out ablation strategies. 

\newpage

\section{Local activation time map}
If we process the electrograms within the window of interest to find out the local activation time values of each electrode locations, then interpolate them to the left atrium mesh vertices, we will obtain a local activation time map. 

The local activation time is found at the time instance of maximum absolute slope computed with Equation \ref{eq:LAT_bipolar} if bipolar electrogram is used, or found at the time instance of maximum negative slop computed with Equation \ref{eq:LAT_unipolar} if unipolar electrogram is used \cite{Cantwell}.

\begin{equation}
\label{eq:LAT_bipolar}
LAT = \underset{t}{argmax}\left ( \left | \frac{dV_{bipolar}(t)}{dt} \right | \right )
\end{equation}

\begin{equation}
\label{eq:LAT_unipolar}
LAT = \underset{t}{argmax}\left ( - \frac{dV_{unipolar}(t)}{dt} \right )
\end{equation}

Figure \ref{fig:clinical_lat_map} shows a clinical example of a local activation time map. In the map, red represents early and blue represents late activation. We can see that there are a lot of red or pink dots near the red area, those are ablation lesions (red means a deep lesion, pink means a shallower lesion). This patient has left atrium tachycardia, and this local activation time map shows the tachycardia focal source location by displaying it in red. After ablated the red area, tachycardia terminated.

\begin{figure}[!ht]
\centering
\includegraphics[width = 1\textwidth]{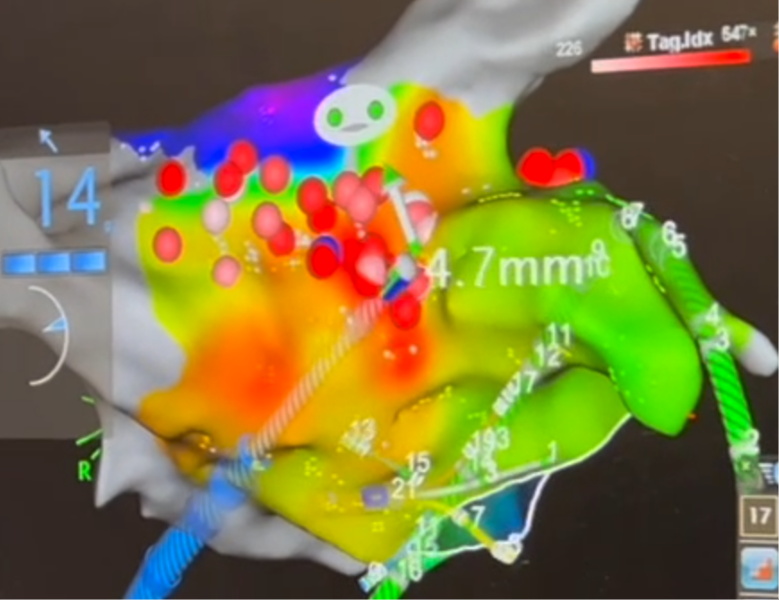}
\caption{Local activation time map. Red represents early and blue represents late activation.}
\label{fig:clinical_lat_map}
\end{figure}

Local activation time map is sensitive to the choice of the time window of interest. Different time window of interest results in different coloring of the map. To better illustrate what happen, a 2D simulations is run and processed into 2 local activation time maps each with different time window of interest. The simulation movie snapshots are shown in Figure \ref{fig:2d_focal_movie}.

\begin{figure}[!ht]
\centering
\includegraphics[width = 1\textwidth]{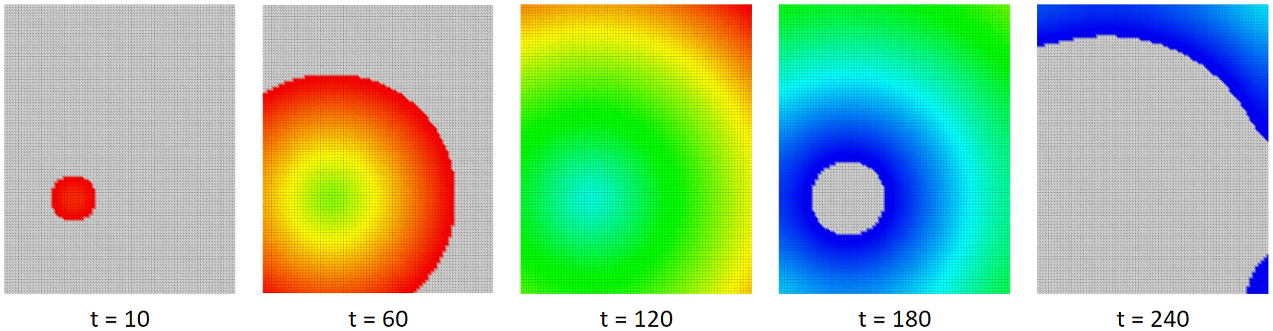}
\caption{Focal arrhythmia simulation movie snapshots at different times.}
\label{fig:2d_focal_movie}
\end{figure}

Different choice of the time window of interest result in different visuals of the local activation time map. With a good choice of the time window of interest, the focal source will be marked in red as shown in Figure \ref{fig:focal_source_detection}(a1), indicating the earliest activation location with respect to the time window of interest, which coincidentally also the earliest of the entire activation wave. However, if the time window of interest starts at the time when the focal source had already spread out a bit, the earliest activation locations, red region in (a2), no longer represents the focal origin. 

\begin{figure}[!ht]
\centering
\includegraphics[width = 1\textwidth]{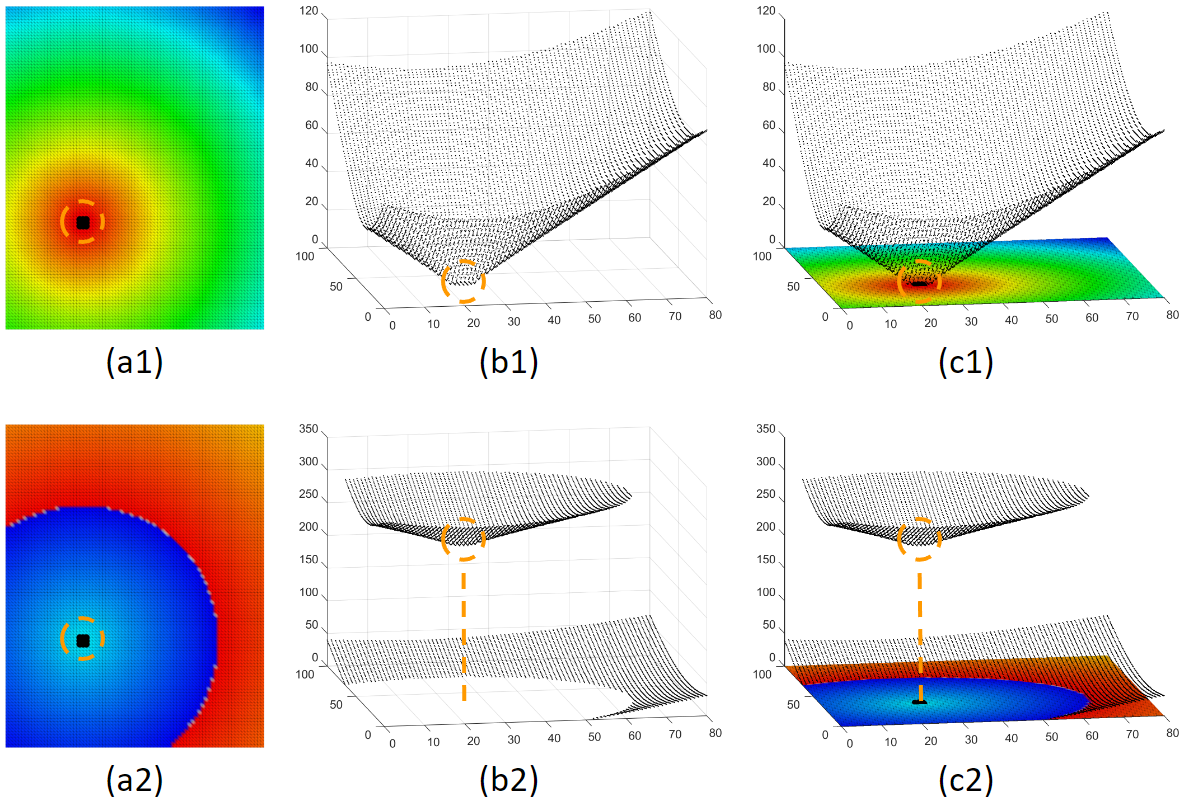}
\caption{Different choice of the time window of interest result in different visuals of the local activation time map. It can be difficult to visually identify the focal origin (marked with orange dashed circle), but numerically it is easy to find: it is the local minima as shown in (b1), (c1), (b2), and (c2).}
\label{fig:focal_source_detection}
\end{figure}

It can be confusing to visually identify focal source location, because physicians were trained to look to red areas to find focal sources. However, it is not a problem for a computer program, all it needs is to find the local minima. This shows that our system can make what is invisible visible.

\newpage

\section{Useful maps for ablation guidance}
Having multiple maps of different features is helpful for identifying arrhythmia source locations. If several maps point to the same location, then we will have high confidence that ablating that location can terminate at least the regional arrhythmia. As shown in Figure \ref{fig:useful_maps}, the region of the black circled has been identified as potential ablation targets in 4 different maps.

\begin{figure}[!ht]
\centering
\includegraphics[width = 1\textwidth]{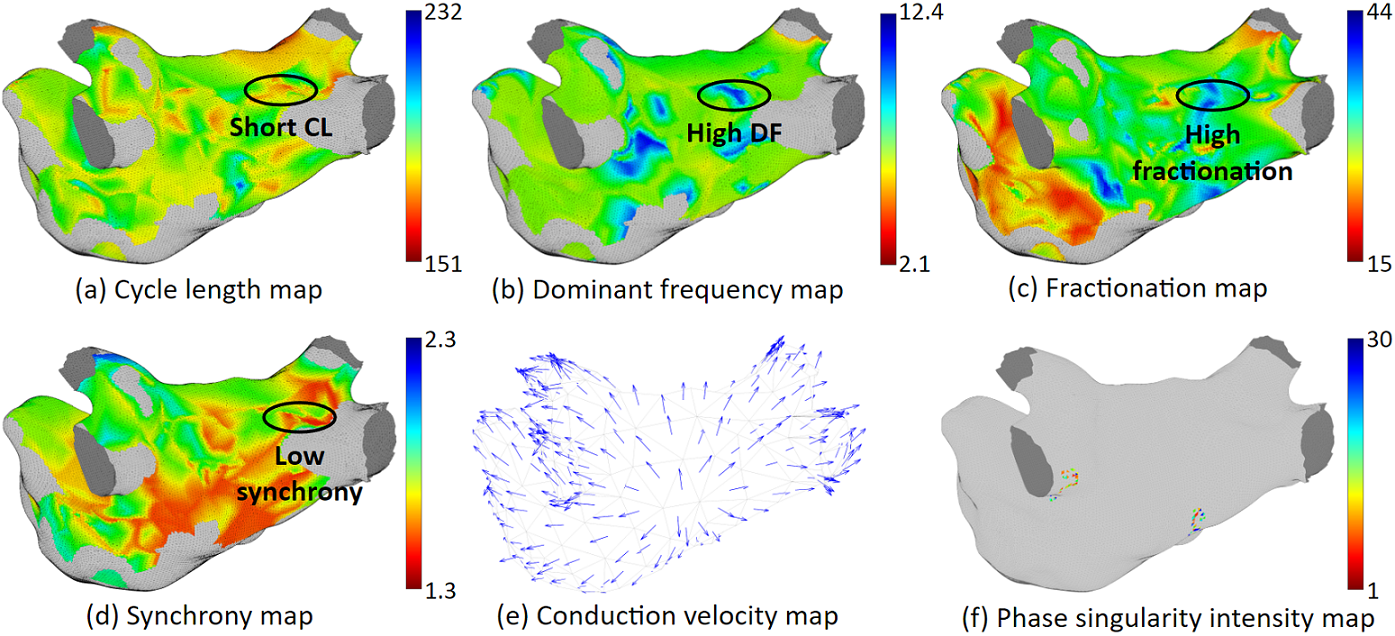}
\caption{Various useful maps our system can provide for ablation guidance.}
\label{fig:useful_maps}
\end{figure}

\subsection{Cycle length map}
Cycle length is a useful metric for locating arrhythmia sources. It is the time in between two consecutive activations as shown in Figure \ref{fig:cl}. Short cycle length indicates fast heart rate which is one of the symptoms of tachycardia, flutter, and atrial fibrillation. 

\begin{figure}[!ht]
\centering
\includegraphics[width = 1\textwidth]{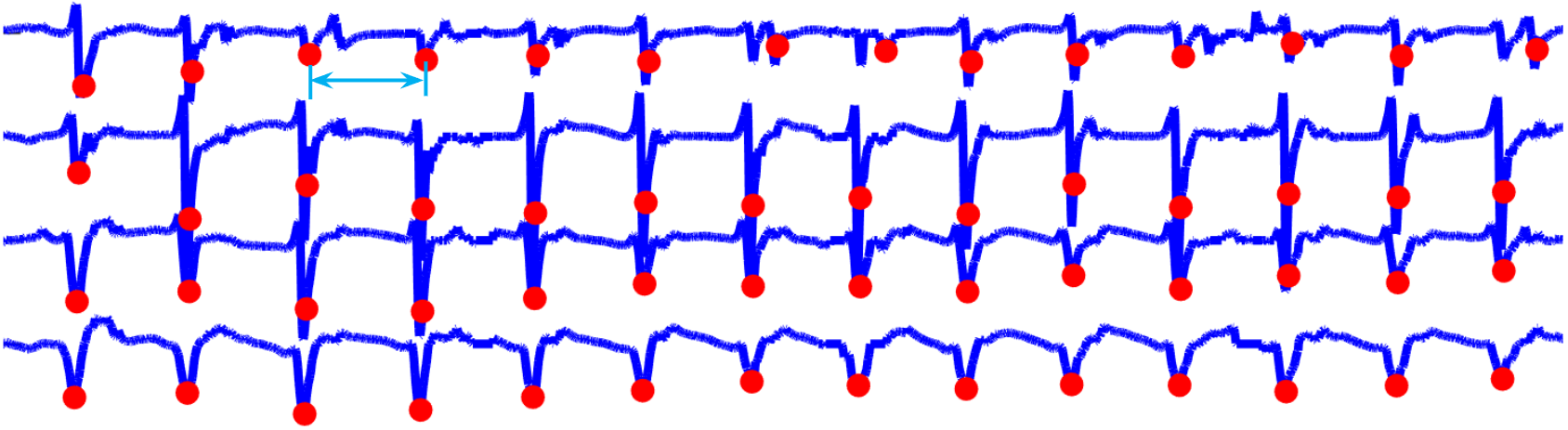}
\caption{Cycle length is the time in between two consecutive activations.}
\label{fig:cl}
\end{figure}

\subsection{Dominant frequency map}
Dominant frequency is closely related to cycle length, it is roughly the reciprocal of the cycle length. Figure \ref{fig:dominant_frequency} shows an example. Because the electrograms are not always clean, sometimes we can get accurate cycle length, other times it may be better to look at the dominant frequency.

\begin{figure}[!ht]
\centering
\includegraphics[width = 1\textwidth]{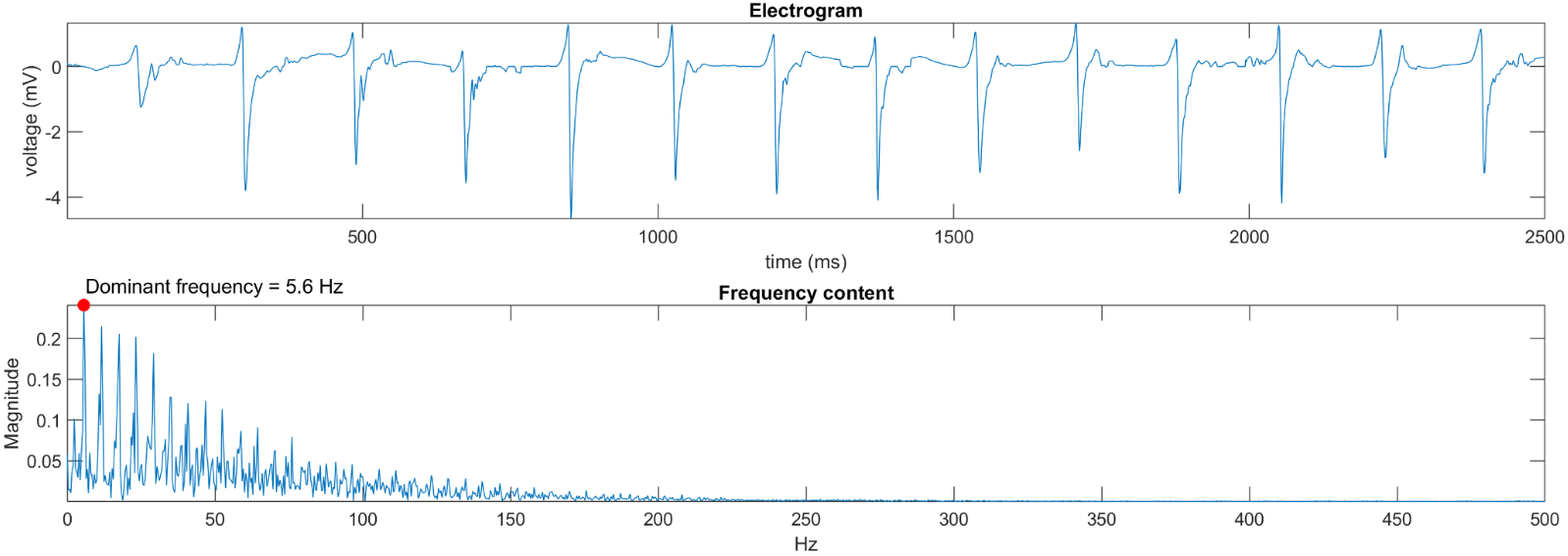}
\caption{Dominant frequency of an electrogram.}
\label{fig:dominant_frequency}
\end{figure}

\subsection{Fractionation map}
For a fractionated electrogram, multiple activations clustered together. It can be caused by complex scars or multiple activation wave collisions. To compute fractionation, first, smooth the electrogram with a moving average with window length of 10. Then, find peaks on the negative derivative of the electrogram. Next, remove peaks of small magnitude. Lastly, count the number of peaks as the fractionation. Areas with high fractionation are ablation targets. However, this feature is less reliable, better used in conjunction with other features.

\begin{figure}[!ht]
\centering
\includegraphics[width = 1\textwidth]{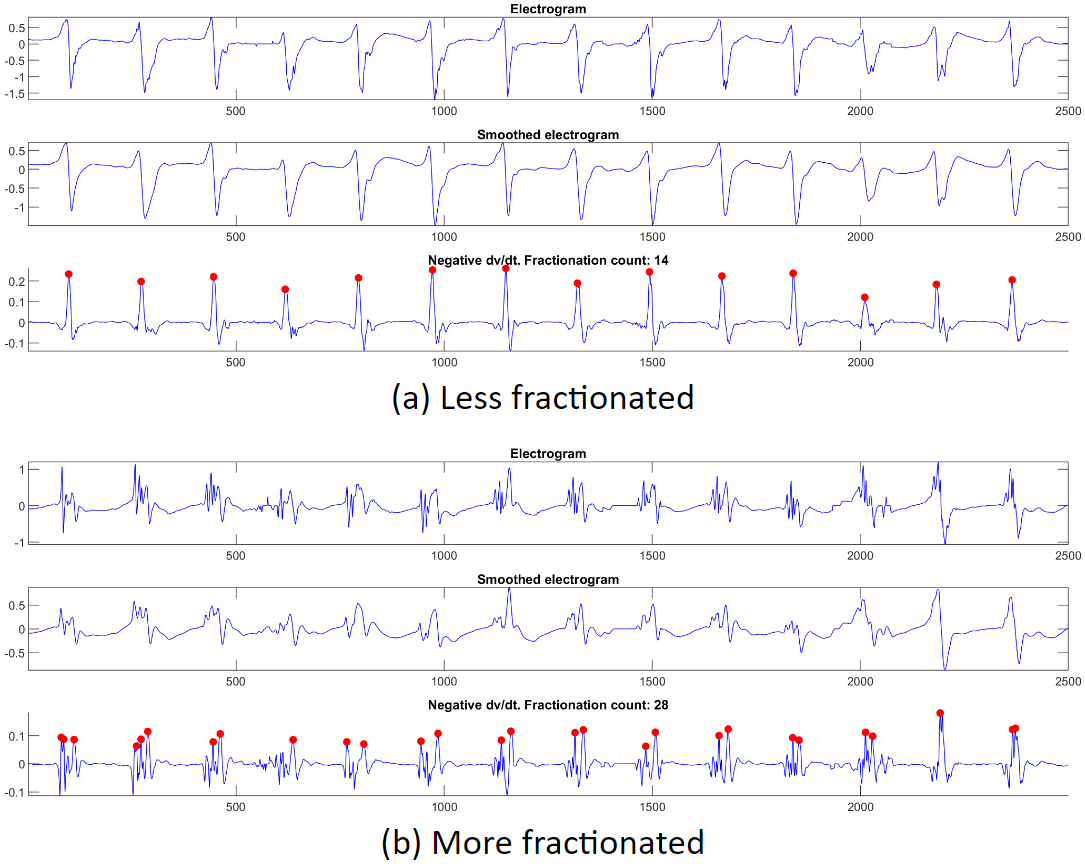}
\caption{Examples of electrogram fractionation.}
\label{fig:egm_fractionation}
\end{figure}

\subsection{Synchrony map}
Synchrony is a measurement presenting how simultaneously the region is activating. To compute synchrony, first, take the absolute value of the electrogram. Than, subtract a noise threshold from the absolute values. Synchrony is the ratio of the length of times below average and above average. Usually a plane wave will have a higher synchrony than a spiral wave. And the rotating centers of the spiral waves are arrhythmia sources.

\begin{figure}[!ht]
\centering
\includegraphics[width = 1\textwidth]{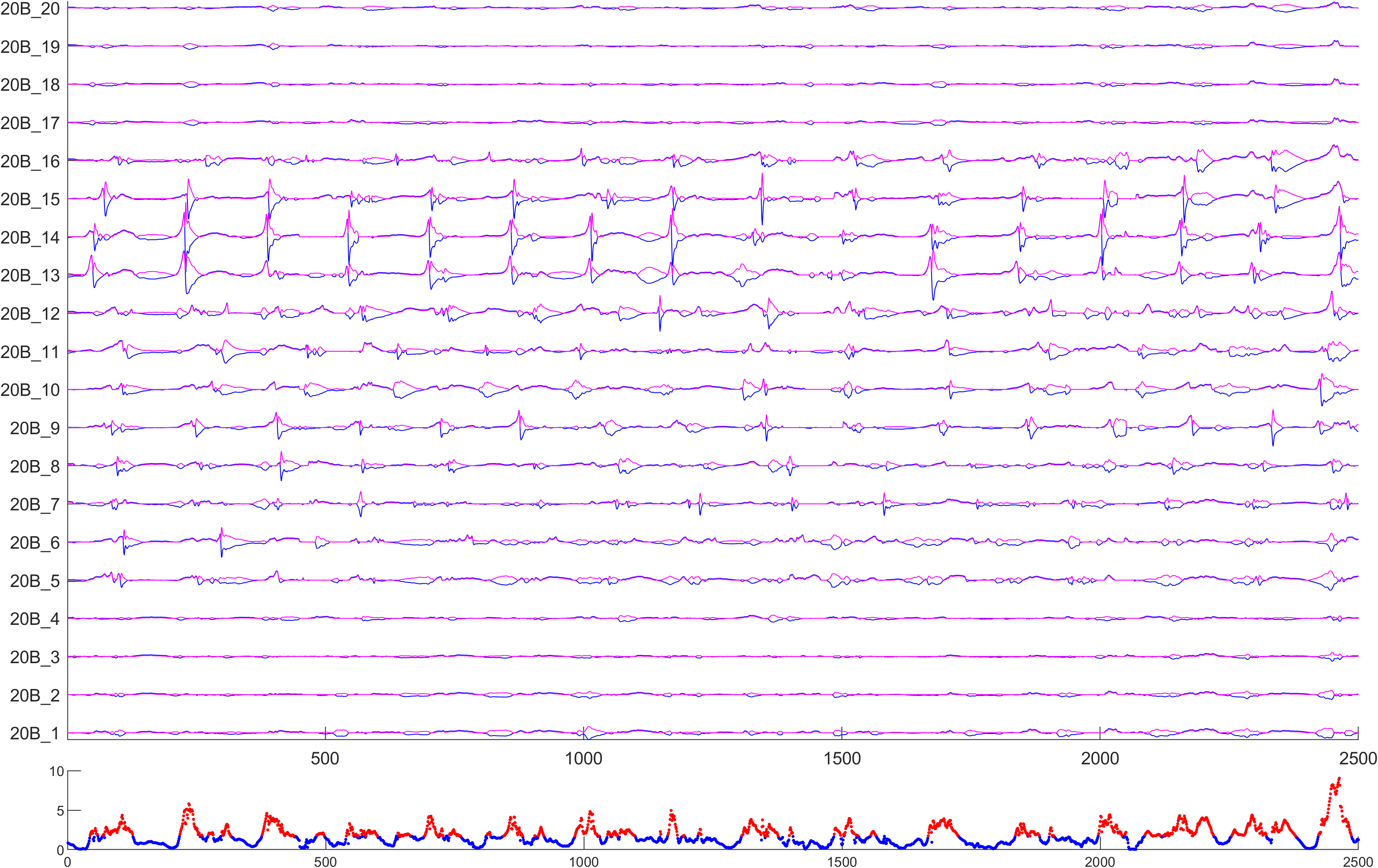}
\caption{The synchrony value of this Pentaray recording region is 1.29. }
\label{fig:synchrony=1.29_s50}
\end{figure}

\begin{figure}[!ht]
\centering
\includegraphics[width = 1\textwidth]{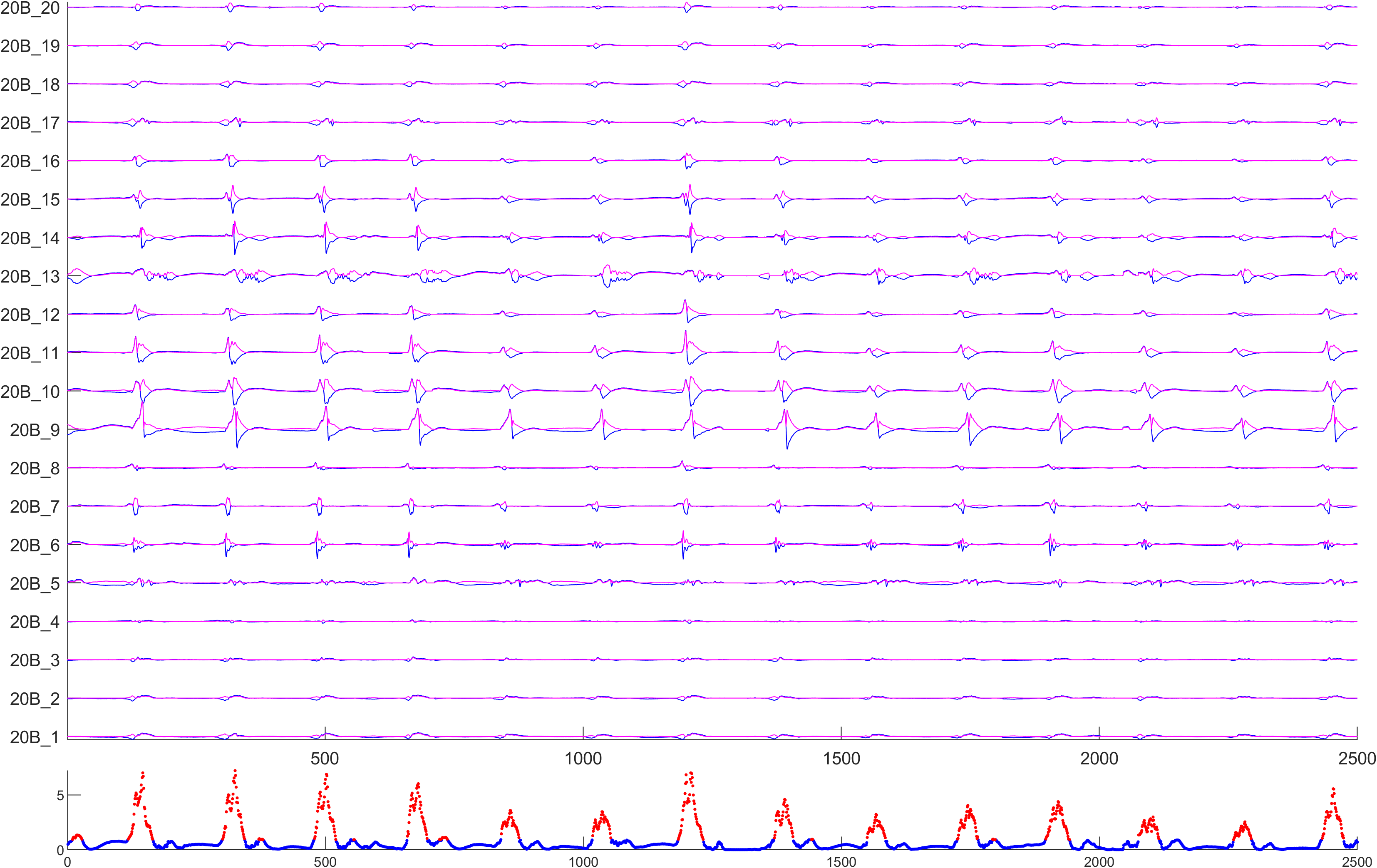}
\caption{The synchrony value of this Pentaray recording region is 3.03.}
\label{fig:synchrony=3.03_s26}
\end{figure}

\subsection{Conduction velocity vector map}
We can compute conduction velocity on the mesh triangles. There are 4 scenarios of how an activation wave can travel through a triangle region as shown in Figure \ref{fig:cv_calculation}. For scenarios (a) (b) and (c), we have one set of equations, and for scenario (d) we have another set of equations for computing the conduction velocity. Conduction velocity direction indicates activation wave direction. Its magnitude indicates if the underlying tissue is healthy or not.

\begin{figure}[!ht]
\centering
\includegraphics[width = 1\textwidth]{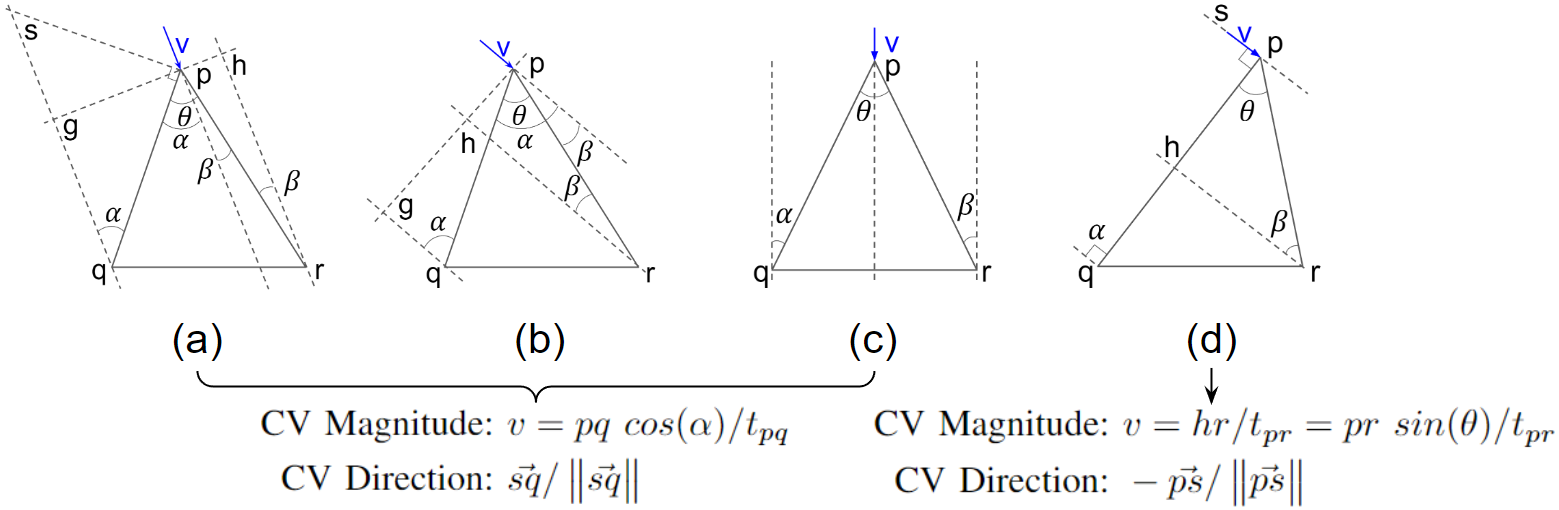}
\caption{How to calculate conduction velocity.}
\label{fig:cv_calculation}
\end{figure}

\subsection{Phase singularity intensity map}
Phase singularity is the center of a rotating wave. Each time frame of the activation movie is a colored map of the phase of activations. On one of the mesh triangles, we can discretize the phase values into one of three values (red, green, or blue). If a triangle contains all three values, that triangle contains a phase singularity. More details can be found in Appendix \ref{app:ps_detection}.

\chapter{\MakeUppercase{Left atrium model validation on patient data}}
\label{chapter: validation}

In previous chapters, we presented our fiber-independent left atrium model, showed its accuracy in producing arrhythmias, and explained the maths. Then we described the clinical patient data and how we process them. In this chapter, we will implement all the methods and algorithms to process patient data and create patient-specific models, then we will validate our model's simulation with the patient data. 

We tested our model with 15 patient data, to show that our model is capable of accurately reproducing patient arrhythmias. Among these patient, 8 of them have sinus rhythm maps, and the other 7 of them have tachycardia maps. 

\section{Data collection}
At the start of the catheter ablation procedure, the physician captures an electroanatomical map with a roving mapping catheter from the Carto3 System at the Hospital of the University of Pennsylvania, it associates the 3D atrium triangular mesh with electrical activities. The mesh mapping fill threshold was 5 mm, and electrogram recording filters were set at 2 to 240 Hz for unipolar electrograms, 16-500 Hz for bipolar electrograms, and 0.5-200 Hz for surface electrogram. Each map has about 1,500 electrogram recordings spread across the endocardium. Each electrogram records 2.5 seconds unipolar and bipolar signals at 1 kHz. 

Some of the electrograms contained noise and needed to be excluded: 

\begin{itemize}
\item Electrode distance to the nearest mesh vertex is greater than 8 mm. This is an empirical threshold we decided to use. The electrode was most likely not in contact with atrium tissue beyond this threshold.
\item Maximum voltage is less than 0.45 mV. These electrodes are either too far way from tissue or in contact with scar tissue \cite{He2019}. Neither will provide clean electrogram.
\item Complex and fractionated electrogram as shown in Figure \ref{fig:delete_fractionated_egm}. Fractionated electrograms consist of multiple high frequency components with low amplitudes and long duration, makes it difficult to find the accurate activation time.
\end{itemize}

\begin{figure}[htb]
\centerline{\includegraphics[width = 0.8\textwidth]{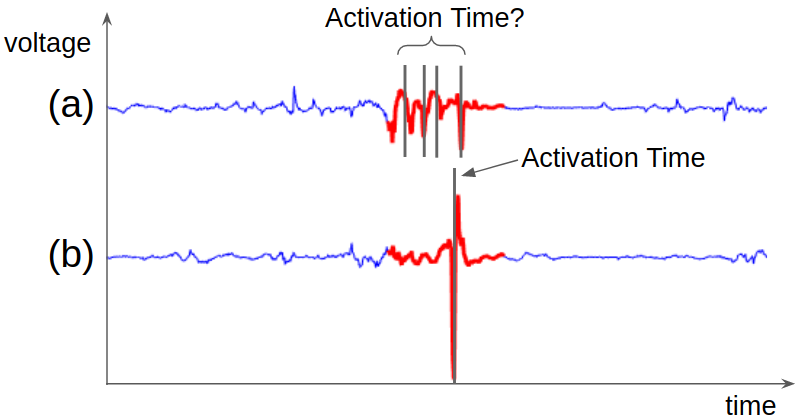}}
\caption{(a) Electrogram is too complex and fractionated, making it difficult to find the activation time. (b) Good electrogram. The activation time is easy to identify.}
\label{fig:delete_fractionated_egm}
\end{figure}

The amount of raw electrogram recordings captured and filtered are shown in Table \ref{tb:number_of_electrode_sample}. The percentage of good electrogram recordings is low for tachycardia maps because they contain more fractionated electrograms.

\begin{table}[htb]
\begin{center}
\caption{Number of Electrode Recordings}
\begin{tabular}{ c c c c c } 
\hline
ID & Rhythm & \# Electrode & \# Used & \% Used \\ \hline
1 & Sinus Rhythm & 976 & 557 & 57.1 \\
2 & Sinus Rhythm & 3263 & 1361 & 41.7 \\
3 & Sinus Rhythm & 3156 & 1788 & 56.7 \\
4 & Sinus Rhythm & 1488 & 663 & 44.6 \\
5 & Sinus Rhythm & 2477 & 1655 & 66.8 \\
6 & Sinus Rhythm & 2905 & 2079 & 71.6 \\
7 & Sinus Rhythm & 1744 & 861 & 49.4 \\
8 & Sinus Rhythm & 1801 & 1106 & 61.4 \\
\hline
\multicolumn{2}{ c }{Average} & 2226 & 1259 & 56.1 \\ \hline
\hline
9 & Tachycardia & 278 & 73 & 26.3 \\
10 & Tachycardia & 958 & 326 & 34.0 \\
11 & Tachycardia & 822 & 197 & 24.0 \\
12 & Tachycardia & 484 & 215 & 44.4 \\
13 & Tachycardia & 1227 & 198 & 16.1 \\
14 & Tachycardia & 714 & 130 & 18.2 \\
15 & Tachycardia & 1469 & 425 & 28.9 \\
\hline
\multicolumn{2}{ c }{Average} & 850 & 223 & 27.4 \\ \hline
\end{tabular}
\label{tb:number_of_electrode_sample}
\end{center}
\end{table}

Then for each patient data, we followed the method described in Section \ref{sec:tune diffusion} to tune a heart model. To compare patient data and our model generated data, we calculate the absolute local activation time error according to Equation \ref{eq:lat_diff}, accuracy according to Equation \ref{eq:lat_accuracy}, and we also calculate the root mean square error and correlation between these two sets of data.

\begin{equation}
\label{eq:lat_diff}
\textit{LAT Err} = \frac{1}{N}\sum_{n=1}^{N}\left | LAT_{simulation,n}-LAT_{patient,n} \right |
\end{equation}

\begin{equation}
\label{eq:lat_accuracy}
Accuracy = \left ( 1-\frac{\textit{LAT Err}}{range(LAT_{patient})}\right ) \times 100\%
\end{equation}

\newpage

\section{Model accuracy on patient sinus rhythm data}
The patient maps and simulation maps are shown in Figure \ref{fig:performance_summary_lat_map_sr}. We can see that the model reproduced patient maps quite accurately. Figure \ref{fig:performance_summary_lat_plot_sr} are activation time scatter plots of patient data vs model generated data. For a 100\% accurate model, all the blue dots will be on the red line. We can see that the blue dots are very near the red line, and the correlation values are close to 1, which indicates high similarity between the two sets of data.

\begin{figure}[!ht]
\centerline{\includegraphics[width = 1\textwidth]{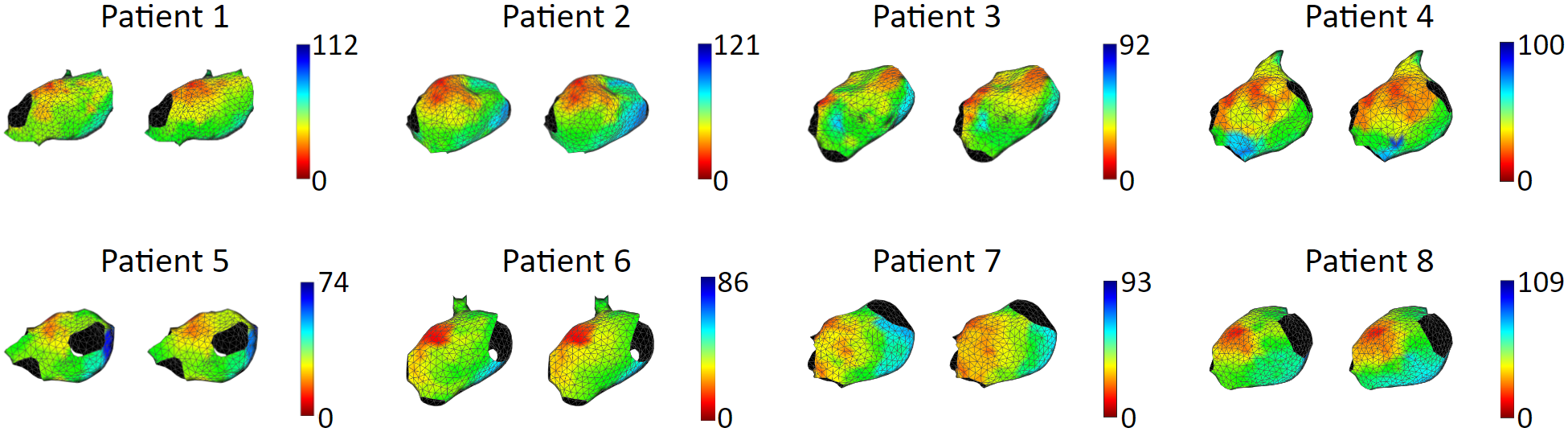}}
\caption{Patient v.s. Simulation LAT Map. Patient \ac{LAT} map is on the left, simulation \ac{LAT} map is on the right. Red represent early activation, blue represent late activation. The color bar is in unit ms.}
\label{fig:performance_summary_lat_map_sr}
\end{figure}

\begin{figure}[!ht]
\centerline{\includegraphics[width = 1\textwidth]{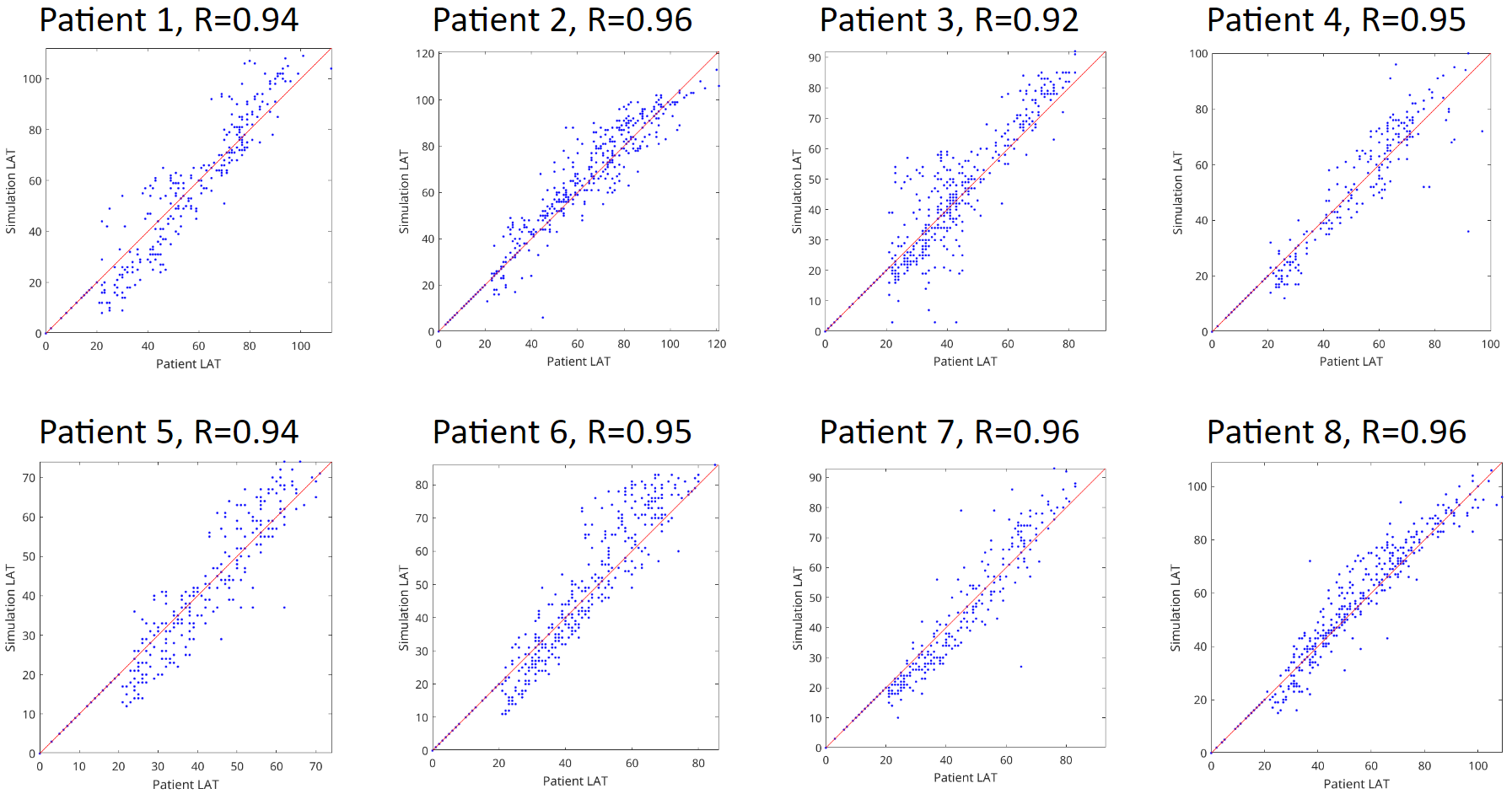}}
\caption{Patient v.s. Simulation LAT Plot. $X$ axis is patient $Y$ axis is simulated \ac{LAT}. The red line is $Y = X$. R: correlation.}
\label{fig:performance_summary_lat_plot_sr}
\end{figure}

Quantitatively, Table \ref{tb:performance_summary_sinus_rhythm} shows the results of the 8 sinus rhythm maps. We achieved an average activation time error of 5.47 ms with 0.95 correlation between patient data and simulation data. 

\begin{table}[!ht]
\begin{center}
\resizebox{0.4\textwidth}{!}{
\begin{tabular}{ c | c c c c} 
\hline
Patient ID & LAT Err & RMSE & Corr & Acc \\
\hline
1 & 7.66 & 9.76 & 0.94 & 93.16 \\ 
2 & 5.74 & 8.19 & 0.96 & 95.25 \\
3 & 5.86 & 8.64 & 0.92 & 92.85 \\ 
4 & 5.22 & 7.99 & 0.95 & 94.62 \\
5 & 4.42 & 6.13 & 0.94 & 93.77 \\
6 & 5.65 & 7.81 & 0.95 & 93.36 \\
7 & 4.48 & 6.61 & 0.96 & 94.60 \\
8 & 4.69 & 6.81 & 0.96 & 95.70 \\
\hline
Average & 5.47 & 7.74 & 0.95 & 94.16 \\ \hline
\end{tabular}
}
\end{center}
\caption{Performance summary. LAT Err: local activation time error, defined in Equation \ref{eq:lat_diff}, unit: ms. RMSE: root-mean-square error, unit: ms. Corr: correlation. Acc: accuracy, defined in Equation \ref{eq:lat_accuracy}.}
\label{tb:performance_summary_sinus_rhythm}
\end{table}

Sinus rhythm is the healthy rhythm, thus it is not a disease that needs the arrhythmia ablation procedure. However, achieving a high accuracy in reproducing patient's sinus rhythm map is helpful for the ablation procedure. 

Usually, at the beginning of the ablation procedure, even if the patient was undergoing arrhythmia, cardioversion will be applied to restore normal heartbeats, because voltage maps are more accurate when patient is in sinus rhythm. Voltage map helps physician identify scar regions. This is specially helpful for redo patients who previously had ablations but some lesions (scar caused by ablation) regrew and became conductive again which led to arrhythmia recurrence. It is crucial to accurately map these scar locations for terminating arrhythmias.

Not all patient can be cardioverted to sinus rhythm, and during the ablation procedure, there are times that it is necessary to induce arrhythmia to find out the arrhythmia source locations. Therefore, a heart model also needs to be able to accurately reproduce patient's arrhythmia. The next section shows our model's accuracy on tachycardia.

\newpage

\section{Model accuracy on patient tachycardia data}
The patient maps and simulation maps are shown in Figure \ref{fig:performance_summary_lat_map_at}. Figure \ref{fig:performance_summary_lat_plot_at} are activation time scatter plots of patient data vs model generated data. We can see that the model performs well for the majority patient, but performance is not good for patient 13. We assume all patient arrhythmias are focal sources, therefore, our models generate focal arrhythmias. On that patient 13's LAT map (the most lower left corner of Figure \ref{fig:performance_summary_lat_map_at}), there seems to be two rotors. Rotor arrhythmia is significantly different than focal arrhythmia in activation patterns, this may be the reason. It would have higher accuracy if our model produced rotor arrhythmia simulation for patient 13. But we have not developed a good method to accurately initiate a rotor arrhythmia according to a patient LAT map.

\begin{figure}[!ht]
\centerline{\includegraphics[width = 1\textwidth]{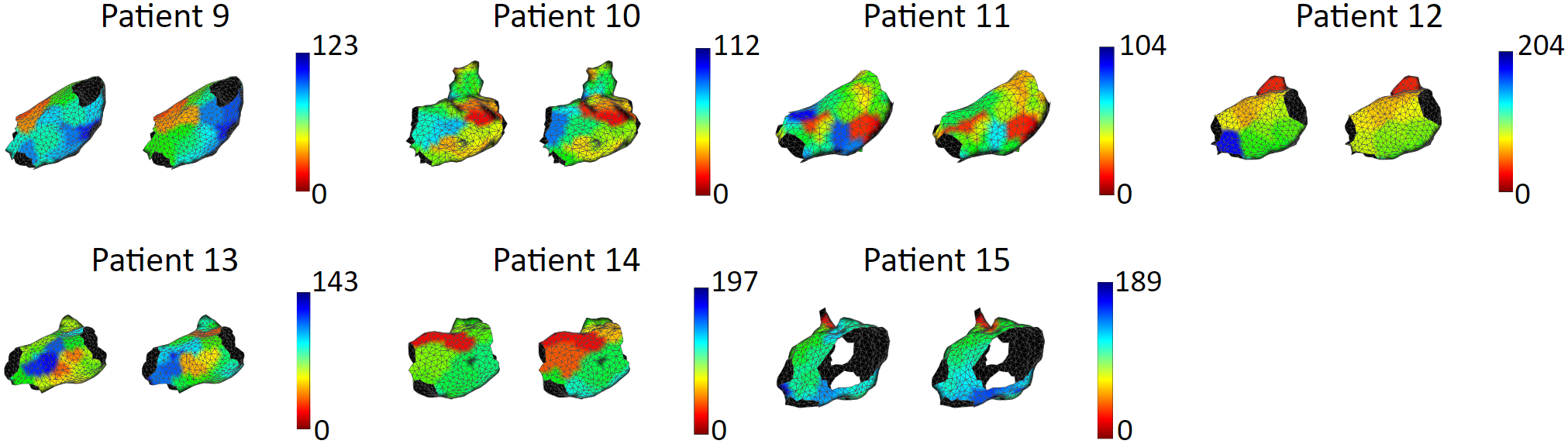}}
\caption{Patient v.s. Simulation LAT Map. Patient \ac{LAT} map is on the left, simulation \ac{LAT} map is on the right. Red represent early activation, blue represent late activation. The color bar is in unit ms. }
\label{fig:performance_summary_lat_map_at}
\end{figure}

\begin{figure}[!ht]
\centerline{\includegraphics[width = 1\textwidth]{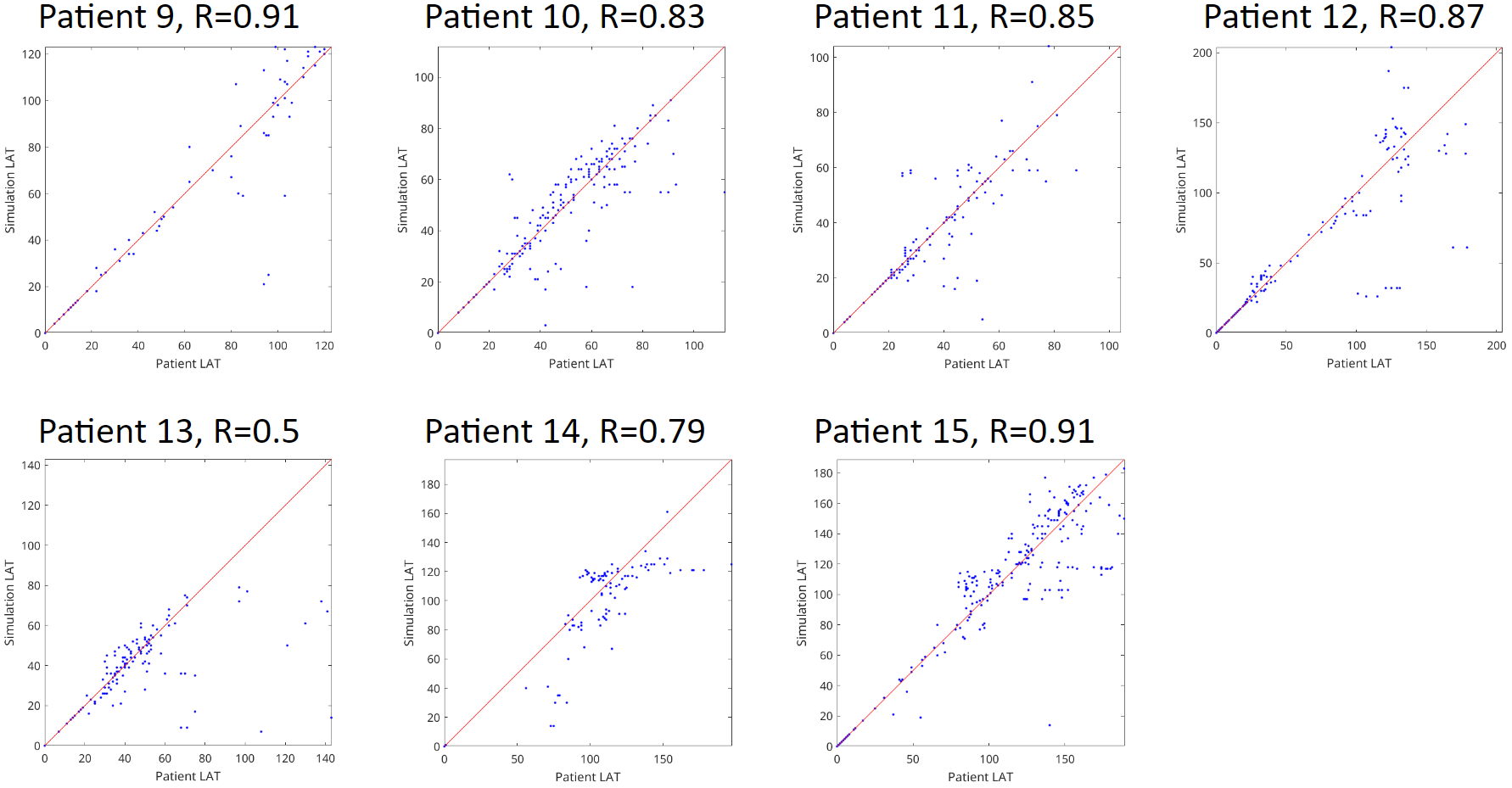}}
\caption{Patient v.s. Simulation LAT Plot. $X$ axis is patient $Y$ axis is simulated \ac{LAT}. The red line is $Y = X$. R: correlation.}
\label{fig:performance_summary_lat_plot_at}
\end{figure}

Quantitatively, Table \ref{tb:performance_summary_tachycardia} shows the results of the 7 tachycardia maps. We achieved an average activation time error of 10.97 ms with 0.81 correlation between patient data and simulation data.

\begin{table*}[!ht]
\begin{center}
\resizebox{0.4\textwidth}{!}{
\begin{tabular}{ c | c c c c } 
\hline
Patient ID & LAT Err & RMSE & Corr & Acc \\
\hline
9 & 8.34 & 16.40 & 0.91 & 93.05 \\
10 & 6.87 & 11.82 & 0.83 & 93.87 \\
11 & 5.49 & 10.11 & 0.85 & 93.76 \\
12 & 13.77 & 27.96 & 0.87 & 92.31 \\
13 & 10.89 & 23.42 & 0.50 & 92.39 \\
14 & 17.71 & 23.45 & 0.79 & 91.01 \\
15 & 13.70 & 21.49 & 0.91 & 92.75 \\
\hline
Average & 10.97 & 19.24 & 0.81 & 92.73 \\ \hline
\end{tabular}
}
\end{center}
\caption{Performance Comparison. LAT Err: local activation time error, defined in Equation \ref{eq:lat_diff}, unit: ms. RMSE: root-mean-square error, unit: ms. Corr: correlation. Acc: accuracy, defined in Equation \ref{eq:lat_accuracy}.}
\label{tb:performance_summary_tachycardia}
\end{table*}

\newpage

\section{Discussion}
The method we implemented for tuning the diffusion coefficients allowed every location of the atrium has individual diffusion coefficient values. That is to compensate for the heterogeneity in tissue conductivity caused by fiber organizations. We will refer to the model tuned in this method "Model I".

As we discovered in Section \ref{sec:cancellation effect}, the heterogeneity caused by fiber organization was not large. This naturally lead to a question of whether we need to tune the diffusion coefficient individually. Different patient's left atrium has different tissue conduction properties, therefore it is necessary to tune the heart model's diffusion coefficients to reflect that. However, what if we tune the diffusion coefficient uniformly, which means assigning the same diffusion coefficient value to all locations of the left atrium? We will refer to the model tuned in this method "Model II".

For atrial fibrillation, the shape of the action potential can greatly impact activation wave dynamics. This is less of an issue for sinus rhythm and tachycardia. To get more insight of the effect of the action potential parameters, we tuned heart models with different $\tau$ values: $\tau_{in} = 0.3$, $\tau_{out} = 3$, $\tau_{open} = 120$, $\tau_{close} = 50$, $v_{gate} = 0.13$ (instead of $\tau_{in} = 0.3$, $\tau_{out} = 6$, $\tau_{open} = 120$, $\tau_{close} = 150$, $v_{gate} = 0.13$). Here the diffusion coefficients were individually tuned, and we will refer to the model tuned in this method "Model III".

The resulting performance is shown in Table \ref{tb:performance_summary}. Compare to Model I, performance of Model II is worse: the average \ac{LAT} error increases, root-mean-square error increases, correlation decreases, and accuracy decreases. Moreover, patient with arrhythmia often has scars in the left atrium. The scars can be caused by the arrhythmia or by ablations. Scars do not conduct well, or have low diffusion coefficient values, while healthy tissue regions have high diffusion coefficient values. Therefore, a heart model with uniform diffusion coefficient value will not perform well if the left atrium contains large regions of scar.

Compare Model III to Model I, the average performance of sinus rhythm decreases 12.98\% ((6.18-5.47)/5.47), but the average performance of tachycardia increases 20.69\% ((10.97-8.7)/10.97). The reason may be that the original parameters are more accurate for sinus rhythm maps while these parameters are more accurate for tachycardia maps.

\begin{table*}[ht]
\begin{center}
\caption{Performance Comparison}
\resizebox{1\textwidth}{!}{
\begin{tabular}{ c c | c c c c | c c c c | c c c c } 
\hline
\multicolumn{2}{c}{} & \multicolumn{4}{|c}{Model I: $d$ tuned individually} & \multicolumn{4}{|c}{Model II: $d$ tuned uniformly} & \multicolumn{4}{|c}{Model III: different $\tau$s}\\
\hline
ID & Rhythm & LAT Err & RMSE & Corr & Acc & LAT Err & RMSE & Corr & Acc & LAT Err & RMSE & Corr & Acc\\
\hline
1 & Sinus Rhythm & 7.66 & 9.76 & 0.94 & 93.16 & 8.73 & 11.06 & 0.91 & 92.21 & 8.45 & 10.73 & 0.92 & 92.45 \\ 
2 & Sinus Rhythm & 5.74 & 8.19 & 0.96 & 95.25 & 10.71 & 14.40 & 0.88 & 91.15 & 9.28 & 12.05 & 0.93 & 92.33 \\
3 & Sinus Rhythm & 5.86 & 8.64 & 0.92 & 92.85 & 7.15 & 10.08 & 0.85 & 91.28 & 5.41 & 7.61 & 0.92 & 93.4 \\ 
4 & Sinus Rhythm & 5.22 & 7.99 & 0.95 & 94.62 & 5.50 & 8.55 & 0.93 & 94.33 & 4.89 & 7.04 & 0.95 & 94.96 \\
5 & Sinus Rhythm & 4.42 & 6.13 & 0.94 & 93.77 & 5.74 & 7.79 & 0.89 & 91.92 & 5.17 & 6.90 & 0.93 & 92.72 \\
6 & Sinus Rhythm & 5.65 & 7.81 & 0.95 & 93.36 & 6.19 & 8.19 & 0.91 & 92.72 & 5.52 & 7.24 & 0.94 & 93.50 \\
7 & Sinus Rhythm & 4.48 & 6.61 & 0.96 & 94.60 & 4.61 & 6.76 & 0.95 & 94.45 & 4.04 & 5.94 & 0.96 & 95.13 \\
8 & Sinus Rhythm & 4.69 & 6.81 & 0.96 & 95.70 & 7.50 & 9.97 & 0.91 & 93.12 & 6.67 & 8.49 & 0.95 & 93.88 \\
\hline
\multicolumn{2}{ c| }{Average} & 5.47 & 7.74 & 0.95 & 94.16 & 7.02 & 9.60 & 0.90 & 92.65 & 6.18 & 8.25 & 0.94 & 93.55 \\ \hline
\hline
9 & Tachycardia & 8.34 & 16.40 & 0.91 & 93.05 & 12.91 & 17.49 & 0.89 & 89.24 & 5.98 & 9.66 & 0.97 & 95.01 \\
10 & Tachycardia & 6.87 & 11.82 & 0.83 & 93.87 & 12.22 & 16.85 & 0.67 & 89.09 & 6.88 & 10.79 & 0.87 & 93.86 \\
11 & Tachycardia & 5.49 & 10.11 & 0.85 & 93.76 & 9.95 & 15.24 & 0.62 & 88.69 & 5.65 & 11.23 & 0.84 & 93.58 \\
12 & Tachycardia & 13.77 & 27.96 & 0.87 & 92.31 & 22.37 & 31.12 & 0.82 & 87.50 & 10.07 & 15.59 & 0.96 & 94.38 \\
13 & Tachycardia & 10.89 & 23.42 & 0.50 & 92.39 & 13.09 & 21.68 & 0.58 & 90.85 & 8.66 & 17.35 & 0.74 & 93.94 \\
14 & Tachycardia & 17.71 & 23.45 & 0.79 & 91.01 & 17.81 & 24.20 & 0.69 & 90.96 & 13.11 & 17.23 & 0.84 & 93.34 \\
15 & Tachycardia & 13.70 & 21.49 & 0.91 & 92.75 & 22.22 & 32.18 & 0.84 & 88.24 & 10.53 & 17.56 & 0.94 & 94.43 \\
\hline
\multicolumn{2}{ c| }{Average} & 10.97 & 19.24 & 0.81 & 92.73 & 15.80 & 22.68 & 0.73 & 89.22 & 8.70 & 14.20 & 0.88 & 94.08 \\ \hline
\end{tabular}
}
\label{tb:performance_summary}
\end{center}
\begin{flushleft}
LAT Err: local activation time error, unit: ms. RMSE: root-mean-square error, unit: ms. Corr: correlation. Acc: accuracy.
\end{flushleft}
\end{table*}

\chapter{\MakeUppercase{Conclusion}}
To summarize, we developed an integrated computational heart model for clinical atrial arrhythmia ablation. We were the first to integrate a heart model that can run in real-time to a clinical mapping system and do not require data outside of the clinical operating room. 

We investigated the fiber effects on activation patterns, which was the first rigorous investigation in the field of left atrium modeling. 

We developed a fast tuning method for patient-specific heart modeling. Our approach tunes parameters locally with continuous values, rather than one value for healthy tissue and one value for scar. Our heart model can run in real-time, the tuning can be done within 15 seconds, and simulation for 1 heartbeat takes 5 seconds on a personal computer, instead of hours or days on a supercomputer.

Our model achieved high accuracy of simulating arrhythmias. Local activation time accuracy is 96\% for focal arrhythmia, and 93\% for rotor arrhythmia.

We processed and refined clinical data which improved electroanatomical map accuracy, developed custom maps for assisting ablation, and validated our system with patient data.

\end{mainf}

\begin{append}
\chapter{\MakeUppercase{Publications}}
\label{app:publication}
\begin{itemize}[leftmargin=0pt, itemsep=-10pt]
\item \underline{Jiyue He}, Arkady Pertsov, John Bullinga, Rahul Mangharam, “Individualization of atrial tachycardia models for clinical applications: Performance of fiber-independent model”, IEEE Transactions on Biomedical Engineering. 2023.
\item \underline{Jiyue He}, Arkady Pertsov, John Bullinga, Rahul Mangharam, “Tachycardia activation pattern predictivity of a fiber-independent left atrium model”, Cardiac Physiome Society. 2023.
\item \underline{Jiyue He}, Arkady Pertsov, Rahul Mangharam, “Real-time atrial tachycardia ablation guidance with a left atrium model”, Heart Rhythm Society. 2023.
\item \underline{Jiyue He}, Arkady Pertsov, Elizabeth Cherry, Flavio Fenton, Caroline Roney, Steven Niederer, Zirui Zang, Rahul Mangharam, “Fiber organization has little effect on electrical activation patterns during focal arrhythmias in the left atrium”, IEEE Transactions on Biomedical Engineering. 2022.
\item \underline{Jiyue He}, Arkady Pertsov, Sanjay Dixit, Katie Walsh, Eric Toolan, Rahul Mangharam, “Patient-specific heart model towards atrial fibrillation”, Proceedings of the ACM/IEEE 12th International Conference on Cyber-Physical Systems. 2021.
\item \underline{Jiyue He}, \underline{Kuk Jin Jang}, Katie Walsh, Jackson Liang, Sanjay Dixit, Rahul Mangharam, “Electroanatomic mapping to determine scar regions in patients with atrial fibrillation”, 41st Annual International Conference of the IEEE Engineering in Medicine and Biology Society. 2019.
\end{itemize}

IEEE Transactions on Biomedical Engineering is ranked No. 3 by Google Scholar among biomedical technology journals. The "Individualization of atrial tachycardia ..." paper and "Fiber organization ..." paper are awarded featured articles. Heart Rhythm Society is the largest international scientific organization on cardiac rhythm disorders.

\chapter{\MakeUppercase{Fiber registration}}
\label{app:fiber_registration}
The Cartesian nodes are created using the endocardial mesh, note that it is different from the epicardial mesh, therefore, we need to register epicardial fiber onto endocardial mesh as shown in Figure \ref{fig:fiber registration}: first, implement non-rigid iterative closest point morphing \cite{Amberg2007}, that matches the epicardium shape to the endocardium, then perform a coordinate transformation on fiber vectors. 

\begin{figure}[!ht]
\centering
\includegraphics[width = 1\textwidth]{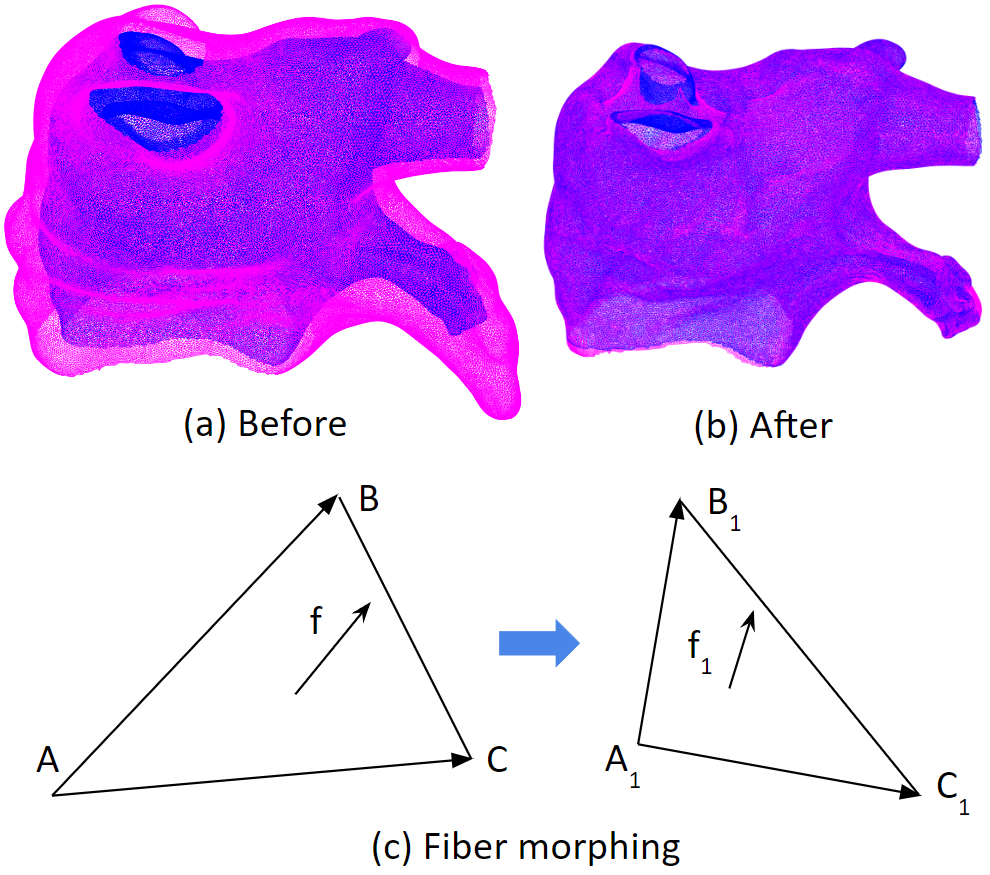}
\caption{Fiber registration. $LA_1$ endocardial mesh (blue) and epicardial mesh (magenta) are shown. (a) and (b) are the before and after mesh morphing. (c) shows how to transform fiber from the original mesh (magenta mesh in (a)) to the morphed mesh (magenta mesh in (b)): Given xyz coordinates of $A$, $B$, $C$, $A_1$, $B_1$, $C_1$, and the fiber (3D vector $f$), calculate $f_1$. We have $f = a \textbf{AB} + b \textbf{AC}$, by solving for $a$ and $b$, we can obtain $f_1 = a \textbf{A}_1\textbf{B}_1 + b \textbf{A}_1\textbf{C}_1$, then make $f_1$ unit length, it will be the morphed fiber. (Note that $f = a \textbf{AB} + b \textbf{AC}$ will give 3 equations, since vectors $\textbf{AB}$, $\textbf{AC}$ and $f$ are co-plane, any 2 of the 3 equations will give the same $a$ and $b$.)}
\label{fig:fiber registration}
\end{figure}

We can see that the morphed mesh matches the target well, on average, the distance between the nearest vertex pairs between the morphed mesh and the target mesh is: 1.4 +/- 0.5 mm. Then we need to morph the fiber data. The final step is to register the morphed fiber to the target mesh, this can be done by copying the fibers that are nearest to the target mesh's triangles.

Since there are two layers of fibers, the Cartesian nodes also need to be categorized into two layers. As shown in Figure \ref{fig:endo_epi_node}. Each node was attributed to either the epicardial or endocardial layer, based on its location with respect to the mesh. Note that although there are two normal directions (opposite to each other) for a triangle, we can easily determine which points out from the mesh: it is a property of a triangular mesh that the three vertices of a triangle are in a particular sequence, and the outward normal vector thus can be found according to the sequence and the right-hand-rule. Then the endocardial and epicardial fibers were assigned to these nodes via minimum distance mapping. 

\begin{figure}[!ht]
\centering
\includegraphics[width = 1\textwidth]{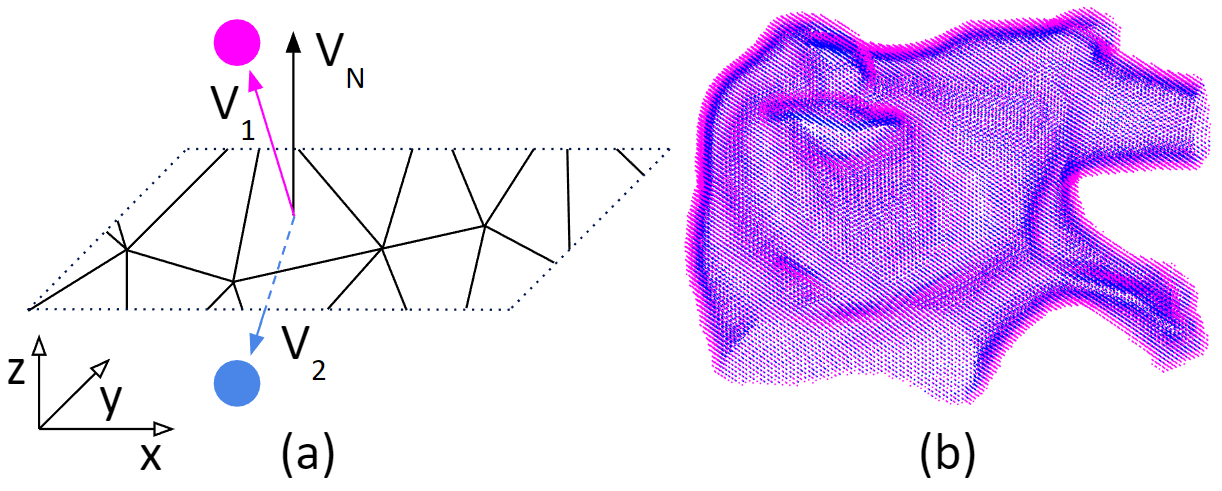}
\caption{Labeling of endocardial and epicardial nodes. (a) $V_1$ and $V_2$ are vectors pointing from the nearest triangle center to the respective node. $V_N$ is the surface normal vector. Because the angle between $V_1$ and $V_N$ is less than 90\degree, the magenta node was labeled as an epicardial node; If the angle is larger than 90\degree, the blue node was labeled as an endocardial node. If the angle is equal to 90\degree, then randomly assign it to endocardial or epicardial node. (b) All nodes labeled as endocardial (blue) or epicardial (magenta) nodes.}
\label{fig:endo_epi_node}
\end{figure}

\chapter{\MakeUppercase{Spatial and temporal resolutions}}
\label{app:resolution}
\begin{figure}[!ht]
\centering
\includegraphics[width = 0.8\textwidth]{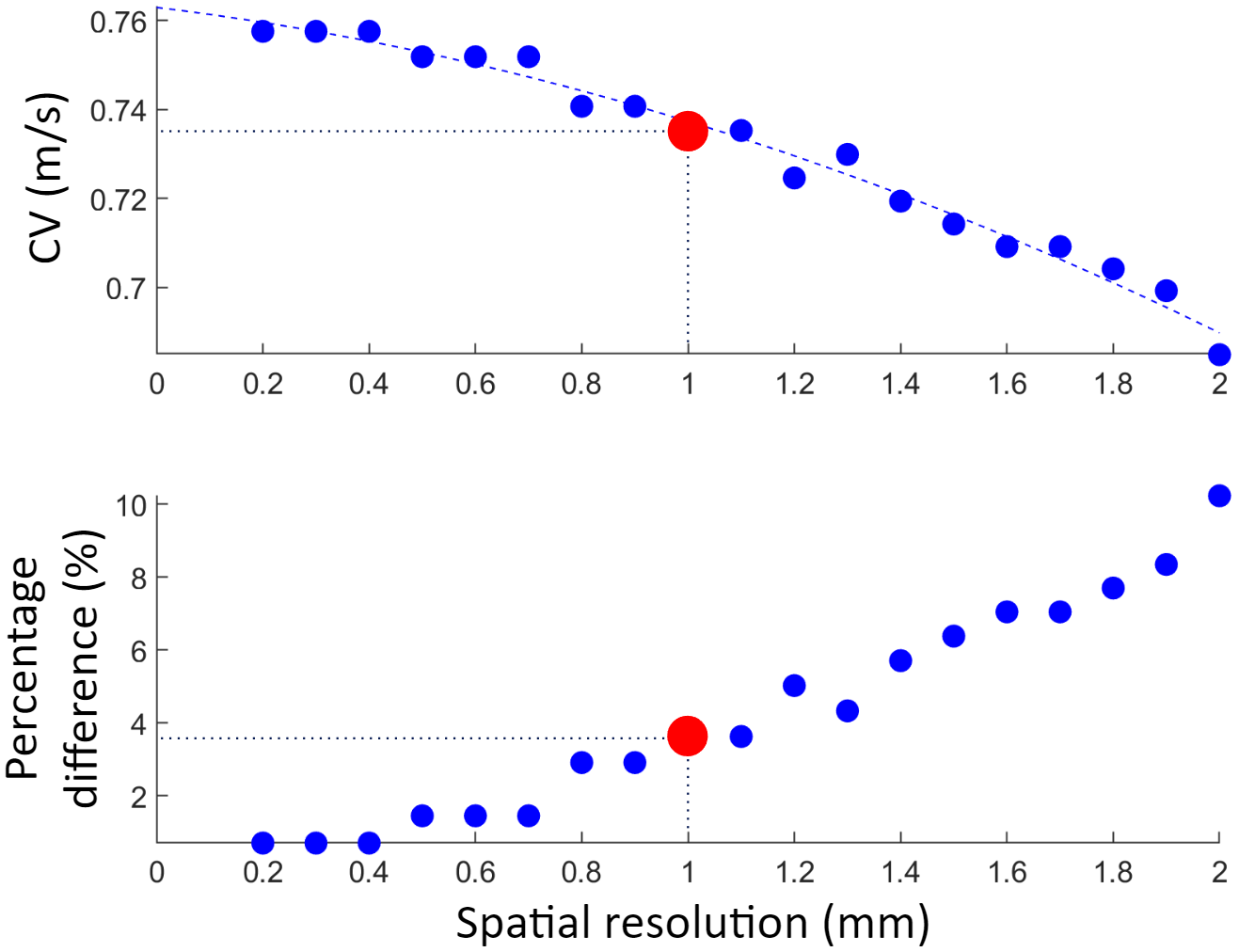}
\caption{Top: Conduction velocity (CV) values (the dots) of implementing different resolutions. The dashed blue line is a quadratic fit of the dots. Bottom: \ac{CV} percentage differences compare to the asymptotic value, which is the value of the dashed blue line at 0 mm spatial resolution. For spatial resolution of 1 mm and temporal resolution of 0.01 ms (the big red dot), the \ac{CV} percentage difference is 3.6\%, much smaller than the adequate 10\%.}
\label{fig:delta vs conduction velocity}
\end{figure}

Simulations are ran on a slab of 100 mm $\times$ 4 mm $\times$ 4 mm. (The long length of the slab is to help increase \ac{CV} computational accuracy.) The heart model parameters are chosen such that the \ac{CV} values are close to the physical values \cite{Harrild2000}. Assume isotropic conduction, or no fiber organization. Spatial resolutions are set to 0.2, 0.3, ..., 2.0 mm. Temporal resolution $dt$ is set to 0.01 ms. Results are summarized in Figure \ref{fig:delta vs conduction velocity}. For spatial resolution of 1 mm and temporal resolution of 0.01 ms, the accuracy of \ac{CV} is adequate with a deviation from the asymptotic value of 3.6\%, which is much smaller than the usual required 10\%.

\chapter{\MakeUppercase{Phase singularity detection}}
\label{app:ps_detection}
\ac{PS} refers to a pivot point around which an activation wave rotates. To identify \ac{PS}, the phase values of the three vertices of a triangle on the left atrium mesh are analyzed. If specific criteria are met, the triangle is determined to contain a \ac{PS}. The phase values of a vertex is a time sequence that linearly increases from 0 to 2$\pi$. It starts at the beginning of an activation and ends at the beginning of the subsequent activation. For each time instance, we transform the phase value $p$ into one of three colors:

\begin{itemize} 
\setlength\itemsep{0.5em}
  \item If ($p \geq 0$ and $p < \frac{1}{6}2\pi$) or ($p > \frac{5}{6}2\pi$ and $p \leq 2\pi$), assign red, or $[1\ 0\ 0]$ in RGB representation.
  \item If $p \geq \frac{1}{6} 2\pi$ and $p<\frac{1}{2} 2\pi$, assign green, or $[0\ 1\ 0]$.
  \item If $p \geq \frac{1}{2} 2\pi$ and $p \leq \frac{5}{6} 2\pi$, assign blue, or $[0\ 0\ 1]$.
\end{itemize}

For a triangle, we can create a $3\times3$ matrix that each row is the RGB representation of a vertex's color. This matrix can then be transformed so that different combinations of vertex colors are converted into different numerical values:

\begin{equation}
\begin{split}
\label{eq:matrix to number}
&[3^2 \ 3^1 \ 3^0] \begin{bmatrix} c_{11} & c_{12} & c_{13} \\ c_{21} & c_{22} & c_{23} \\ c_{31} & c_{32} & c_{33} \end{bmatrix} \begin{bmatrix} 0 \\ 1 \\ 2 \end{bmatrix} \\
=& 9c_{12} + 18c_{13} + 3c_{22} + 6c_{23} + c_{32} + 2c_{33} 
\end{split}
\end{equation}

If a triangle contains a \ac{PS}, then the numerical value will be one of these: 7, 11, 21, 5, 19, and 15. And these six scenarios are depicted in Figure \ref{fig:ps}. 

\begin{figure}[!ht]
\centering
\includegraphics[width = 0.5\textwidth]{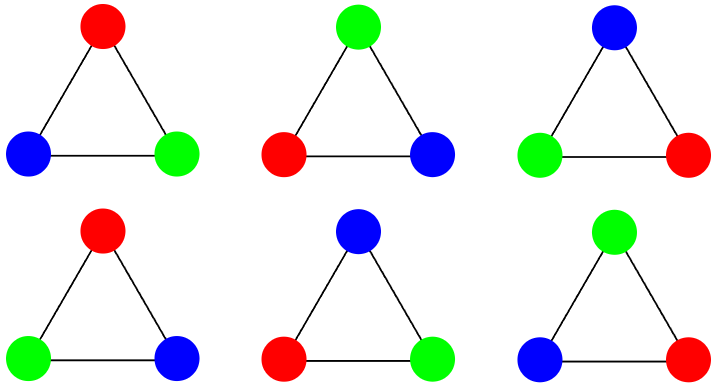}
\caption{The six scenarios that a triangle contains a phase singularity.}
\label{fig:ps}
\end{figure}

Performing \ac{PS} detection for every time instance on all triangles of the mesh will result in \ac{PS} trajectories, representing the movement of the \ac{PS} over time.

Note that it is a property of the triangular mesh that the sequence of the three vertices of a triangle follows the right-hand-rule: right hand four fingers curl following the vertices sequence, then the thumb will represent the face normal that points outwards of the atrium mesh. This ensures the detected \ac{PS} rotational direction (clockwise or counterclockwise) is correct.

\chapter{\MakeUppercase{Additional figures}}

\begin{figure}[!ht]
\centering
\includegraphics[width = 1\textwidth]{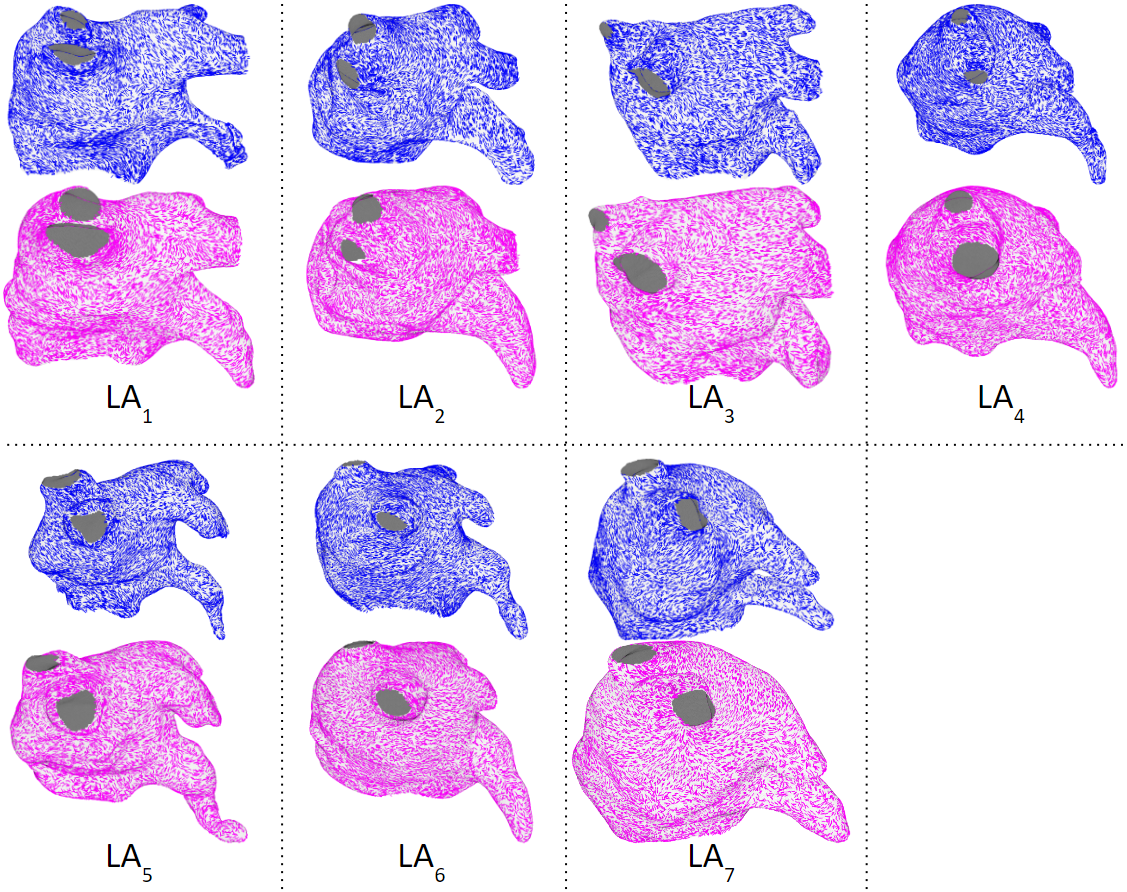}
\caption{Fiber data of the 7 atria. The upper blue ones are the endocardium fibers, and the lower magenta ones are the epicardium fibers.}
\label{fig:fiber_data_7_atria}
\end{figure}

\newpage

\begin{figure}[!ht]
\centering
\includegraphics[width = 1\textwidth]{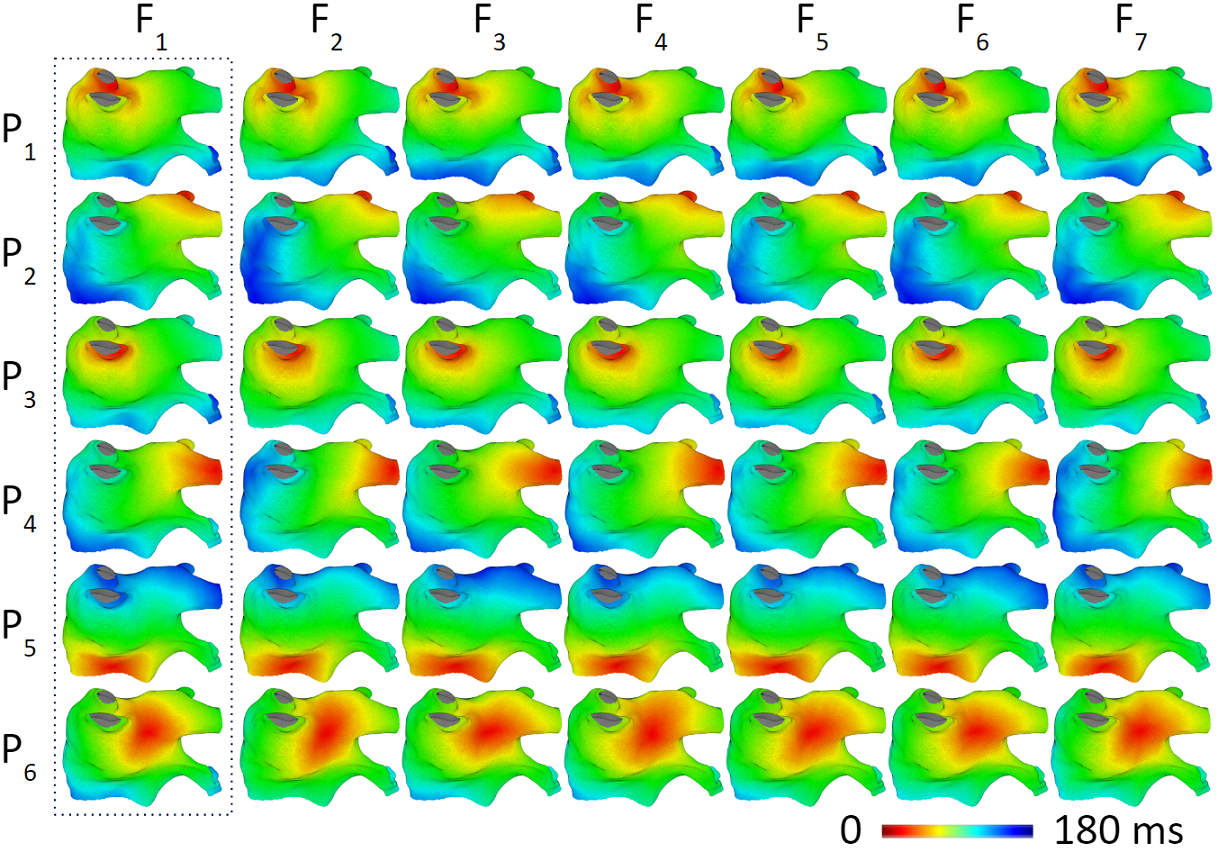}
\caption{\ac{LAT} maps produced by pacing from six different locations in the ground-truth (column F1) and chimeric models (columns F2 to F7). Color scales are normalized to each individual map. The maps in each row are similar to each other, demonstrating that despite significant differences in fiber patterns the activation pattern is not significantly affected.}
\label{fig:lat maps}
\end{figure}

\newpage

\begin{figure}[!ht]
\centering
\includegraphics[width = 1\textwidth]{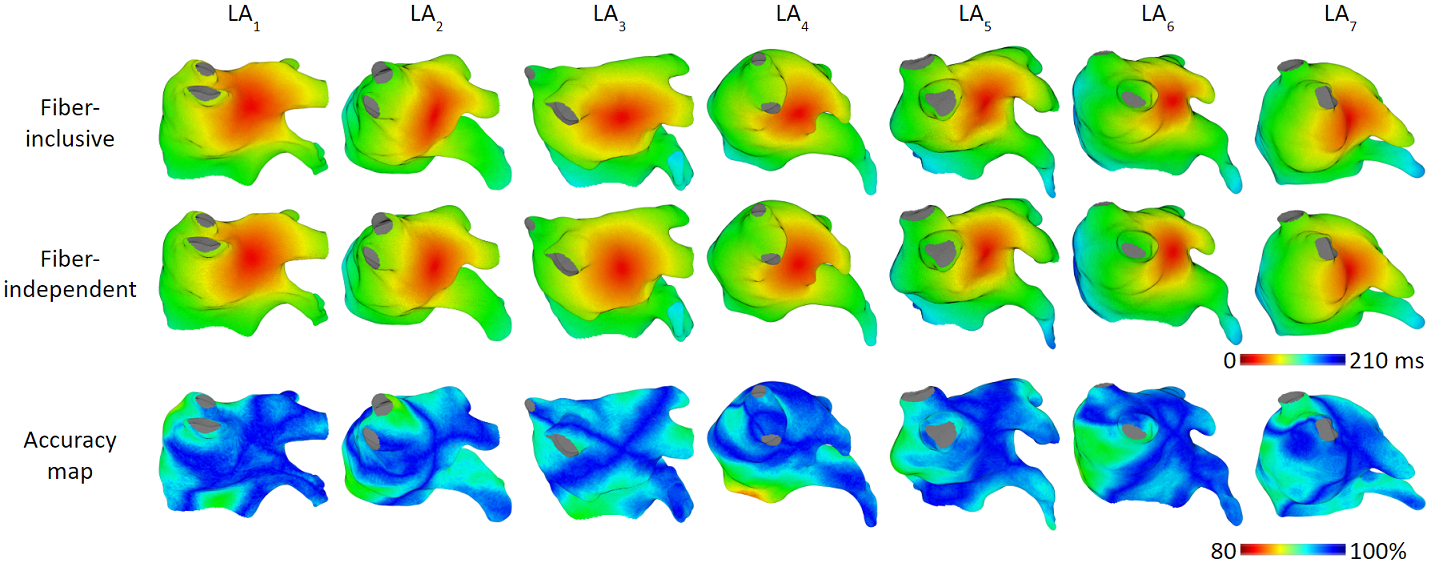}
\caption{Focal arrhythmia comparison: A total of 42 focal arrhythmias are generated (7 left atria each with 6 focal pacing scenarios), here shows results of focal pacing at location P$_6$. Row 1 and 2 show local activation time maps of fiber-inclusive and fiber-independent models respectively. We can see that the red regions in row 2 are more rounded, because the model assumes isotropic conduction; while red regions in row 1 are skewed by fiber orientations. Row 3 shows the accuracy maps calculated according to Equation \eqref{eq:accuracy}, with 100\% meaning no difference between the two models.}
\label{fig:focal lat map}
\end{figure}

\newpage

\begin{figure}[!ht]
\centering
\includegraphics[width = 1\textwidth]{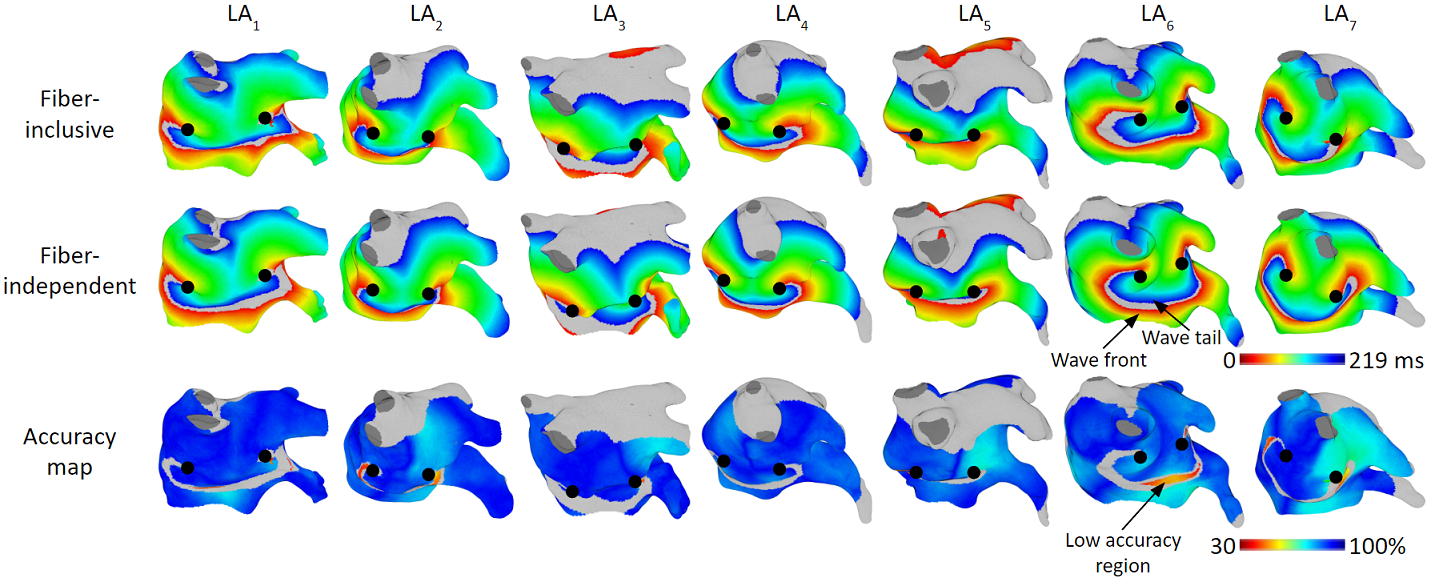}
\caption{Rotor arrhythmia comparison. A pair of rotors are generated on each left atrium. These rotors were made stable by placing small non-conducting patches as anchors (marked with black dots) at the rotation center, which allowed us to compare the activation patterns of the rotor arrhythmias between models. Row 1 and 2 show local activation time maps of fiber-inclusive and fiber-independent models respectively. Row 3 shows the accuracy maps calculated according to Equation \eqref{eq:accuracy}, with 100\% meaning no difference between the two models.}
\label{fig:rotor lat map}
\end{figure}

\end{append}

\begin{bibliof}
\bibliography{reference}
\end{bibliof}
\end{document}